\documentclass[a4paper,11pt]{article}
\usepackage{jheppub} 
\usepackage{lineno}
\nolinenumbers
\usepackage{mystuff}
\pgfplotsset{compat=1.18}
\usepackage{graphicx} 
\usepackage{xcolor}

\definecolor{amethyst}{rgb}{0.54, 0.17, 0.89}
\definecolor{coral}{rgb}{1.0, 0.3, 0.4}

\newcommand{\dket}[1]{\lvert #1 \rangle\!\rangle}
\newcommand{\dbra}[1]{\langle\!\langle #1 \rvert}
\newcommand{\dbraket}[2]{\langle\!\langle #1 \vert #2 \rangle\!\rangle}
\newcommand{\dnorm}[1]{\langle\!\langle #1 \vert #1 \rangle\!\rangle}
\newcommand{\dketbra}[2]{\lvert #1 \rangle\!\rangle\langle\!\langle #2 \rvert}

\newcommand{\Hext}[1]{(\Ha_{\mathrm{ext}})_{#1}}
\newcommand{\Hmatt}[1]{\Ha_{\mathrm{matt}}^{#1}}

\title{
Topological entanglement entropy in 3D gravity
}

\author[a,b,c]{Vijay Balasubramanian,}
\author[a,d]{and Charlie Cummings}
\affiliation[a]{David Rittenhouse Laboratory, University of Pennsylvania,  209 S.33rd Street, Philadelphia, Pennsylvania 19104, USA}
\affiliation[b]{Santa Fe Institute, 1399 Hyde Park Road, Santa Fe, NM 87501, USA}
\affiliation[c]{Theoretische Natuurkunde, Vrije Universiteit Brussel, Pleinlaan 2, B-1050 Brussels, Belgium}
\affiliation[d]{Kavli Institute for Theoretical Physics,
University of California, Santa Barbara, Kohn Hall, Lagoon Rd, Santa Barbara, CA 93106}

\emailAdd{charlie5@sas.upenn.edu}

\abstract{
We derive an entropy formula for a recently proposed tensor network model of states in 3D gravity which satisfy the constraint equations of diffeomorphism invariance, and including the effects of matter. After accounting for the invertibility of the spacetime metric, we argue that this formula matches the quantum extremal surface formula (for small but finite $G_N$), and is not restricted to time symmetric spacetimes. We derive the result via canonical quantization without relying on path integrals, and  discuss an extension to compact bulk subregions.
}

\begin{document}
\maketitle
\flushbottom



\section{Introduction}

One of the sharpest probes of quantum gravity in AdS/CFT has been the Ryu--Takayanagi formula \cite{Ryu_2006}, together with its generalizations to include quantum corrections \cite{Faulkner:2013ana} and time dependent spacetimes (the HRT formula) \cite{Hubeny:2007xt}, culminating in the quantum extremal surface (QES) formula \cite{Engelhardt:2014gca}. The QES formula computes the entropy of a boundary subregion $R$ as the generalized entropy of a bulk surface homologous to $R$, extremized and then minimized over all such surfaces. Beyond just computing entropies, this formula has also been crucial for understanding the holographic map between the bulk and boundary theories. For example, the bulk region between $R$ and its quantum extremal surface, called the entanglement wedge of $R$, is the part of the bulk which can be reconstructed from the boundary subregion itself \cite{Jafferis_2016,Dong:2016eik,Faulkner:2017vdd}. The QES formula also constrains the entanglement structure of holographic states \cite{Hayden:2011ag}. Applied to evaporating black holes, it even reproduces the Page curve consistent with unitarity \cite{Penington:2019npb,Almheiri:2019psf}. Given how much of the holographic dictionary rests on the QES formula, it is important to understand how it comes about within the gravitational theory itself.

The QES formula has been derived from the gravitational path integral by using the replica trick \cite{Lewkowycz_2013}. Briefly, this derivation proceeds by computing $\Tr(\rho_R^n)$ of a state $\rho_R$ for a boundary subregion $R$ using the gravitational path integral, taking a derivative with respect to $n$, and taking the limit $n\to 1$. Assuming that the dominant saddle preserves the replica symmetry in this limit, the von Neumann entropy of $\rho_R$ localizes to become the area of the surface which is fixed under the replica symmetry (which is extremal by the equations of motion), divided by $4G_N$. Similar arguments extend this derivation to include quantum corrections \cite{Faulkner:2013ana} and time dependent spacetimes \cite{Dong:2016hjy}. This line of reasoning can also be extended to derive the replica wormholes responsible for the Page curve \cite{Penington:2019kki,Almheiri:2019qdq}. However, the gravitational path integral is not fully understood: the sum over topologies and the integration contour over metrics are not defined from first principles, and the path integral sometimes appears to compute averaged rather than exact quantities \cite{Maloney:2007ud,Saad:2019lba,Marolf:2020xie,Belin:2020hea,Balasubramanian:2022lnw,Balasubramanian:2025zey}. Even setting these issues aside, the path integral is a tool for computing amplitudes, whereas questions about which states exist in quantum gravity and which operators act on those states are most naturally posed in the canonical formalism. It would therefore be valuable to know whether the QES formula can be derived directly within the canonical formalism of gravity, starting from explicit gauge invariant states on a Cauchy slice.

Tensor networks are a frequently used toy model for this canonical point of view \cite{Swingle_2012,Pastawski_2015,Yang:2015uoa,Hayden_2016}. A tensor network is a map from a bulk Hilbert space to a boundary Hilbert space, and in this sense it can be interpreted as a toy model of the translation between a canonically quantized bulk theory and its holographic dual. This interpretation comes with some well known subtleties. For instance, a tensor network lives on a single Cauchy slice and has no intrinsic dynamics, so it is only a model of a time symmetric slice in its standard implementation. This means that a standard tensor network can only hope to simulate the Ryu--Takayanagi surface (possibly with quantum corrections), for which the covariant extremization of the HRT/QES formula plays no role. Furthermore, the lattice on which the network is drawn breaks the diffeomorphism invariance, which is the gauge symmetry of gravity. Separately, the states prepared by random tensor networks \cite{Hayden_2016}, constructed to mimic the error correcting properties \cite{Pastawski_2015,Harlow:2016vwg} of gravitational entanglement, are known to have a flat R\'enyi spectrum, in contrast to semiclassical states of gravity \cite{Dong:2018seb}. 

A number of proposals have addressed one or another of these issues, both from a tensor network \cite{Balasubramanian:2025rcr,Balasubramanian:2026ymu,Akers:2024wab,Dong2024,Akers:2024ixq,Swingle_2012,Pastawski_2015,Hayden_2016,Dong:2018seb,Donnelly:2016qqt,Qi:2022lbd,Singh_2010,Colafranceschi:2020ern,Basteiro:2022xvu,Basteiro:2024cuh,Basteiro:2024crz,Basteiro:2022zur,Caputa:2020fbc,Frenkel:2024smt,Singh:2017tet,Colafranceschi:2022dig,Chirco:2017wgl,Chirco:2021chk,Yang:2015uoa,Sahu:2025upe,Chen:2024unp,Hung:2019bnq,Chen:2022wvy,Cheng:2023kxh,Hung:2024gma,Harlow:2016vwg} and gauge theory \cite{Wong:2025asd,Wong:2025kpz,Ball:2024hqe,Donnelly:2020teo,Jiang:2020cqo,Donnelly:2018ppr,Wong:2017pdm,Blommaert:2018oue,Blommaert:2018rsf,Wong:2022eiu,Mertens:2022ujr,Mertens:2025ydx,Mertens:2022aou,Chua:2023ios} perspective. In this paper, we will focus on the tensor network model we proposed in \cite{Balasubramanian:2025rcr,Balasubramanian:2026ymu}, which can be thought of as equivalent to the canonical quantization of a topological field theory with gauge group $G$. When $G$ is a finite group, this model is the same as Kitaev's quantum double model \cite{Kitaev:1997wr}, or equivalently a string net \cite{Levin_2005,kirillov2011stringnet}, and the gravitational properties of such models have been studied in \cite{Akers:2024wab,Delcamp:2016eya,Fliss:2023dze}. The tensor networks of \cite{Balasubramanian:2025rcr,Balasubramanian:2026ymu}, however, allow $G$ to be any semisimple Lie group.\footnote{More generally, any transformable group, meaning a unimodular group of type I \cite{Balasubramanian:2025rcr}.} When $G = \SL(2,\R)$, the networks prepare states of the large level limit of $\SL(2,\R) \times \SL(2,\R)$ Chern--Simons theory, and by this theory's relation with general relativity in AdS$_3$ \cite{Witten:1988hc}, we can think of them as preparing a version of 3D gravity with small $G_N$ and possibly non-invertible metrics. It is then natural to ask whether a QES-like formula can be derived within this formalism, without referring to the path integral at all.

Towards this goal, in \cite{Balasubramanian:2026ymu} we derived an entropy formula for a boundary subregion $R$ in this model. In that paper, we focused on tensor networks without matter, which we review in Sec.~\ref{sec:entropy_nomatter} (see there for details). We found that
\begin{align}
	S(\rho_R) = H[\,p_\pi] + \langle \hat{A}\rangle_\rho + S_{\mathrm{bulk}}(\rho_R) \,. \label{eq:SrhoR_intro}
\end{align}
Here, $\rho_R$ is the reduced state of the subregion, which is a mixture over superselection sectors labeled by unitary irreducible representations $\pi \in \widehat{G}$ of the gauge group, $H[\,p_\pi]$ is the differential entropy of the probability distribution over these sectors, $\hat{A}$ is a state independent operator supported on the cut which separates $R$ from its complement, and $S_{\mathrm{bulk}}(\rho_R)$ is the average over sectors of the entropy of the state within each. While suggestively similar to the QES formula in some ways, there is no minimization or extremization over surfaces in \eqref{eq:SrhoR_intro}. This is not necessarily surprising: without matter, every cut separating $R$ from $\overline{R}$ is homotopic to every other, so there is exactly one gauge invariant area operator separating $R$ and its complement, so there is nothing for a minimization to choose between. How seriously, then, should we take the similarities between \eqref{eq:SrhoR_intro} and the QES formula?

In this paper, we make progress towards answering this question by extending the analysis of \cite{Balasubramanian:2026ymu} in two ways. First, we incorporate the effects of matter into the tensor network, and we consider two possibilities for doing so. If the state of the matter is fixed, we find in Sec.~\ref{sec:entropy_matter_norandom} that each topological equivalence class of cut through the network, relative to the matter legs, defines a notion of entropy between the two sides of the cut. This is somewhat analogous to a version of the generalized entropy formula without the final minimization step. If we instead take the matter states to be sufficiently random, in a precise sense defined in Sec.~\ref{sec:entropy_matter_random}, we find that the entropy of a boundary subregion is computed by the minimal (topological equivalence class of) cut through the network. Defining the random ensemble requires some care, because for non-compact $G$ the matter Hilbert spaces are infinite dimensional, so the Haar ensemble of random tensor networks does not exist, but we make this notion of averaging precise in Sec.~\ref{sec:entropy_matter_random}.

We then explain how we expect these results to change if the invertibility of the metric is incorporated into the tensor networks. We work out the consequences of this proposal in Sec.~\ref{sec:examples}; conditioned on this proposal, we find that the spectrum of the area operator $\hat{A}$ is $\ell/4G_N$, to leading order in $G_N$. Here, $\ell$ is the length of the geodesic in the homotopy class of the cut, in the hyperbolic metric which the flat $\SL(2,\R)$ connections on the slice determine, and $G_N$ is Newton's constant. For a boundary anchored cut on a time symmetric slice, this is the Ryu--Takayanagi formula.  We find that the same statement holds for rotating BTZ black holes \cite{Banados:1992wn}, where the cut is the closed curve separating the two asymptotic boundaries, and that the area eigenvalue is $2\pi r_+/4G_N$, the Bekenstein--Hawking entropy. This holds even though the tensor network does not necessarily live on the same Cauchy slice as the bifurcation surface: as we explain in Sec.~\ref{sec:continuum}, the network lives on the maximal volume slice, which pulls away from the horizon under boundary time evolution. In both cases, then, the area operator is measuring the length of the surface which is \emph{extremal in the spacetime}, rather than the length of a minimal curve on the slice which the network tessellates.
This is possible because the representation theoretic data which simultaneously defines the tensor networks and the length of the bifurcation surface are constants of motion under time evolution by local boundary Hamiltonians $H = H_L + H_R$, so changing the boundary time that the maximal volume Cauchy slice is anchored to does not change the entanglement entropy shared between the two sides of the black hole. Indeed, because the entanglement entropy only depends on the spectrum of, e.g., the reduced state $\rho_R$ of the right boundary, $\rho_R$ and $\exp( -iH_R t) \rho_R \exp(iH_R t)$ must have the same entropy. We therefore propose that, assuming the invertibility of the metric can be incorporated as expected, our results amount to a derivation of the quantum extremal surface formula, and not only of the Ryu--Takayanagi formula reproduced by standard tensor networks. 

We then turn to a possible generalization of our formula to compact regions in the bulk. This is interesting for a few reasons. First, the tensor network seems to supply a definition of a compact bulk subregion which is consistent with the gauge symmetry of the model: because the dressed matter legs are the only bulk features of the lattice which survive the graphical moves which allow us to change the lattice defining the tensor network, a co-dimension zero subregion of the Cauchy slice can be specified by its matter content. In other words, we define a compact subregion in the bulk as a subset $a$ of the matter legs, together with the edge modes (see Sec.~\ref{sec:background}) which make that content gauge invariant on its own, in the spirit of \cite{Delcamp:2016eya,Balasubramanian:2023dpj,Ciambelli:2021nmv}. 
Second, in Sec.~\ref{sec:finite_regions} we find that a region defined in this way satisfies the same entropy formula as a boundary anchored one. 
If the matter outside $a$ is held fixed, the entropy is the generalized entropy of $a$ itself, with the area term evaluated on the boundary of the region, as proposed in \cite{Jensen2023}; if the exterior matter is taken to be random, it is instead the minimum of the generalized entropy over all regions of the network which contain $a$, which is the structure of the generalized entanglement wedge of Bousso and Penington \cite{Bousso:2022hlz,Bousso:2023sya}. 
It is interesting to speculate that a formula of this kind, which assigns a generalized entropy to a gauge invariantly defined bulk region without reference to the asymptotic boundary, could be a path toward a local version of the holographic principle.

\subsection{Outline}

The rest of the paper is organized as follows. 
In Sec.~\ref{sec:background}, we review the topological tensor network model we proposed in \cite{Balasubramanian:2025rcr}, as well as the entropy of matter-free topological tensor networks we computed in \cite{Balasubramanian:2026ymu}. 
In Sec.~\ref{sec:entropy_matter_norandom}, we compute the entropy of a topological tensor network with matter along a fixed cut through the network. 
In Sec.~\ref{sec:entropy_matter_random}, we show that averaging the matter state of the topological tensor network using the Gaussian ensemble (see this section for details) leads to a minimization of the entropy over possible cuts through the network, as in the quantum extremal surface formula.
In Sec.~\ref{sec:continuum}, we discuss the possible connections between our results and 3D gravity.
In Sec.~\ref{sec:examples}, we use these connections to argue that our entropy formula is the analog of the full quantum extremal surface formula, not just the Ryu--Takayanagi formula.
In Sec.~\ref{sec:finite_regions}, we discuss a proposed generalization of our formula to define the entropy of a compact subregion in the bulk.
We conclude with a discussion in Sec.~\ref{sec:discussion}.

\section{Background} \label{sec:background}

\subsection{The model}\label{sec:themodel}

We begin by reviewing the construction of topological tensor networks \cite{Balasubramanian:2025rcr}, and the results of \cite{Balasubramanian:2026ymu} concerning their entanglement structure for tensor networks without matter. We refer the reader to \cite{Balasubramanian:2025rcr,Balasubramanian:2026ymu} for details and proofs. Throughout, $G$ is a transformable Lie group,\footnote{That is, a unimodular group of type I \cite{Balasubramanian:2025rcr}. Every semisimple Lie group, compact or not, is transformable, and $G = \SL(2,\R)$ is the case relevant to 3D gravity.} which may be continuous and/or non-compact, and $\widehat{G}$ is its unitary dual, which parameterizes the space of unitary irreducible representations $V_\pi$ of $G$. In other words, on each vector space $V_\pi$, there is a family of unitary operators $\pi(g)$ which satisfy $\pi(g)\pi(h) = \pi(gh)$, and have matrix elements $\pi(g)_{ab}$. 

\subsubsection{The physical Hilbert space}
Let $\Sigma$ be an orientable spatial surface, possibly with boundary, and let $\Lambda = (V,E,P)$ be a tessellation of $\Sigma$ with vertices $V$, oriented edges $E$, and plaquettes $P$. Vertices which lie on $\partial \Sigma$ are called boundary vertices, and the edges which end on them are called boundary legs. The kinematic Hilbert space assigns a copy of $L^2(G)$ to every edge,
\begin{align}
	\Ha(\Lambda) = \bigotimes_{\ell \in E} L^2(G)_\ell \,,
\end{align}
and each factor decomposes over the unitary dual as
\begin{align}
	L^2(G) = \int^{\oplus}_{\widehat{G}} d\mu(\pi) \, V_\pi \otimes V_\pi^* \,,
	\label{eq:peterweylnoncompact}
\end{align}
where $d\mu(\pi)$ is the Plancherel measure of $G$. When $G$ is compact, this is called the Peter--Weyl theorem, the integral is a sum, and $\mu(\pi) = d_\pi/\mathrm{Vol}(G)$ is the dimension of the representation divided by the volume of the group. When $G$ is non-compact, every $V_\pi$ appearing in \eqref{eq:peterweylnoncompact} is infinite dimensional, $\widehat{G}$ contains continuous families of representations, and the Plancherel measure is a nontrivial function on $\widehat{G}$. Nevertheless, this measure always exists, is intrinsic to $G$, and is finite on compact subsets of the unitary dual \cite{HarishChandra1952,HarishChandra1954complex,HarishChandra1976}. 

There are two natural bases of $L^2(G)$ that we will use: the group basis $\ket{g}$, with $\braket{g}{h} = \delta(g^{-1}h)$, and the representation basis $\ket{\pi,ab}$, with
\begin{align}
	\braket{\pi,ab}{\omega,mn} = \delta(\pi,\omega) \delta_{am}\delta_{bn}\,, \qquad \braket{g}{\pi,ab} = \pi(g)_{ab} \,.\label{eq:repbasisoverlap}
\end{align}
Here, $\delta(\pi,\omega)$ is the delta function normalized with respect to the Plancherel measure. The Plancherel measure is thus the unique measure on $\widehat{G}$ for which this change of basis is unitary, and its non-uniformity over $\widehat{G}$ will be the origin of an ``area term'' in the entropy associated to subregions of $\Sigma$.

Next, we define two families of operators that act on $\Ha(\Lambda)$. For each vertex $v$ and $h \in G$, the \emph{electric operator} $A_v(h)$ multiplies the group element on every edge incident to $v$ by $h$: on the left for edges flowing out of $v$ and on the right (by $h^{-1}$) for edges flowing in.\footnote{This convention associates the $V_\pi$ factor of \eqref{eq:peterweylnoncompact} on an edge with its outflowing end and the $V_\pi^*$ factor with its inflowing end, and we will use it throughout.} This implements a gauge transformation at $v$, which we can think of as a translation of the vertex within $\Sigma$. For each plaquette $p$ with a base point $v \in \partial p$, the \emph{magnetic operator} $B_{(v,p)}(h)$ projects onto configurations whose holonomy around $p$, starting at $v$, is equal to $h$. Gauge invariance and flatness (the two constraint equations that define the physical Hilbert space) are then imposed by the ``projectors''
\begin{align}
	\Pi_A^{(v)} = \int_G dg \, A_v(g) \,, \qquad \Pi_B^{(p)} = B_{(v,p)}(e) \,, \qquad \Pi = \prod_{v \in V_{\mathrm{bulk}}} \Pi_A^{(v)} \prod_{p \in P} \Pi_B^{(p)} \,, \label{eq:Pidef}
\end{align}
where $e$ is the identity element of $G$. Neither $\Pi_A^{(v)}$ nor $\Pi_B^{(p)}$ squares to itself for a general transformable group $G$: $(\Pi^{(v)}_A)^2$ diverges as $\mathrm{Vol}(G)$, and $(\Pi_B^{(p)})^2$ as $\delta(e)$. However, both have finite matrix elements between bounded, compactly supported wave functions, and this is enough to define the physical Hilbert space. Let $\Ha_{\mathrm{null}} \subset \Ha(\Lambda)$ be the kernel of $\Pi$, and define the equivalence classes
\begin{align}
	\dket{\psi} \equiv [\ket{\psi} \sim \ket{\psi} + \ket{\chi}] & \, \text{ for all }  \ket{\psi} \in \Ha(\Lambda) \,\text{ and } \, \ket{\chi} \in \Ha_{\mathrm{null}}\,.\label{eq:electricnullstates}
\end{align}
We then equip this space of equivalence classes with the co-invariant inner product
\begin{align}
	\dbraket{\psi}{\sigma} = \bra{\psi} \Pi \ket{\sigma} \,. \label{eq:ImpInnerProduct}
\end{align}
This definition is independent of a choice of representative for $\dket{\psi},\dket{\sigma}$, so it is well-defined.
Completing the space of equivalence classes with respect to \eqref{eq:ImpInnerProduct} gives the physical Hilbert space defining our tensor networks, which we will denote $\Ha_{\mathrm{phys}}(\Sigma)$ for reasons explained below. Note that we do not define the physical Hilbert space by projecting with $\Pi$ and then use the kinematic inner product, since $\bra{\psi}\Pi^\dagger \Pi\ket{\sigma}$ diverges; the quotient by null states is what allows the construction to go through for non-compact $G$. Physical operators are those which map $\Ha_{\mathrm{null}}$ to itself, and therefore descend to the quotient.  A detailed exposition of the above construction can be found in \cite{Balasubramanian:2025rcr,Balasubramanian:2026ymu}. 

\subsubsection{Boundary conditions}
The constraints in \eqref{eq:Pidef} are imposed at every bulk vertex and every plaquette, but not at the boundary vertices. This defines what we call \emph{open} boundary conditions with $n$ marked points \cite{Cirac:2011oss,Cong:2017ffh,Cheipesh:2018imk,Balasubramanian:2026ymu}: we choose $n$ marked points on each boundary component of $\Sigma$, take a tessellation whose boundary vertices sit at the marked points, and drop the electric constraint there. In the continuum, these are the $n$ points at which a charged line operator of the theory is allowed to end.\footnote{A magnetically charged line operator ends on the boundary segments between adjacent boundary vertices, and a more general dyonic ``ribbon operator'' (see Appendix~\ref{app:ribbons}) ends on an adjacent vertex--boundary-segment pair.} We write $\Ha_{\mathrm{phys}}(\Sigma^{(n)})$ when we wish to emphasize the dependence on $n$. Tuning a boundary Hamiltonian at the marked points to criticality and sending $n \to \infty$ produces a conformal boundary condition, in which a full two-dimensional CFT lives on $\partial\Sigma$ \cite{Kong:2019byq,Kong:2019cuu}. This is the boundary condition for $\partial \Sigma$ of asymptotically AdS gravity \cite{Brown1986} which leads to the famous ``boundary gravitons''. We will nonetheless work at finite $n$, because the contribution to the entropy we are interested in, the area term, turns out to be independent of $n$ altogether: it is fixed by the gauge group and by the boundary condition at the cut which factorizes the Hilbert space (as we explain in Sec.~\ref{sec:factorization}), and not by the boundary condition on $\partial \Sigma$ \cite{Balasubramanian:2026ymu}.

\subsubsection{Lattice independence}
The physical Hilbert space only depends on the surface $\Sigma$ and its boundary conditions, and does not depend on the tessellation $\Lambda$. This follows from two graphical moves, shown in Figs.~\ref{fig:move1} and \ref{fig:move2}, and their associated unitaries. Move 1 splits a bulk vertex into two bulk vertices joined by a new leg carrying the state $\ket{e}$, where $e$ is the identity element of the group, and move 2 splits a plaquette into two plaquettes separated by a new leg carrying $\ket{1} = \int dg\,\ket{g}$. Both moves define unitary maps $\Delta_{1,2}: \Ha_{\mathrm{phys}}(\Lambda) \to \Ha_{\mathrm{phys}}(\Lambda')$ between the physical Hilbert spaces of the two tessellations \cite{Balasubramanian:2025rcr}, and any two tessellations of $\Sigma$ with the same boundary vertices are related by a sequence of them. This justifies the notation $\Ha_{\mathrm{phys}}(\Sigma)$. Two consequences of lattice independence will be used repeatedly below. First, we are free to present any state in whichever tessellation is most convenient, and we will treat the choice of tessellation as a gauge choice. Second, moves 1 and 2 are local, and so they can be performed on any sublattice of $\Lambda$ without affecting the rest of the lattice.

\begin{figure}
	\centering
	\begin{tikzpicture}
		\def\sep{3}
		\def\nn{5}
		\def\scale{2}
		\def\hh{0.75}
		\node at (-\sep,0) {\begin{tikzpicture}[scale=\scale]
				
				\filldraw (0,0) circle (0.025);
				\foreach \xx in {1,...,\nn}{
					\draw[thick,->-=0.6] (0,0) -- ({cos(360*\xx/\nn)},{sin(360*\xx/\nn)});
				}
				\node at (1,-0.75) {$\Lambda$};
		\end{tikzpicture}};
		\node[scale=2] at (0,0) {$\Leftrightarrow $};
		\node at (\sep,0) {\begin{tikzpicture}[scale=\scale]
				\filldraw (0,0) circle (0.025);
				\filldraw (\hh,0) circle (0.025);
				\foreach \xx in {1,...,3}{
					\draw[thick,->-=0.6] (\hh,0) -- ({\hh+cos(360*(\xx-1)/\nn)},{sin(360*(\xx-1)/\nn)});
				}
				\foreach \xx in {4,...,\nn}{
					\draw[thick,->-=0.6] (0,0) -- ({cos(360*(\xx-1)/\nn)},{sin(360*(\xx-1)/\nn)});
				}
				\draw[thick,->-=0.5] (0,0) -- (\hh,0);
				\node[anchor=north] at (\hh/2,0) {$\ket{e}$};
				\node at (1,-0.75) {$\Lambda'$};
		\end{tikzpicture}};
		
	\end{tikzpicture}
	\caption{An example of move 1, which can be performed to add or remove a bulk vertex/leg from the graph $\Lambda$. In other words, we can split or combine bulk vertices. 
	}
	\label{fig:move1}
\end{figure}
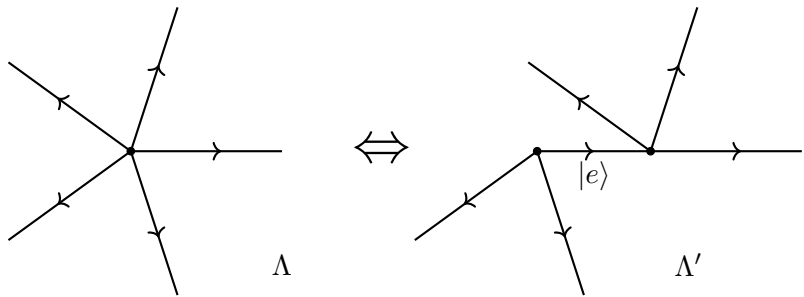
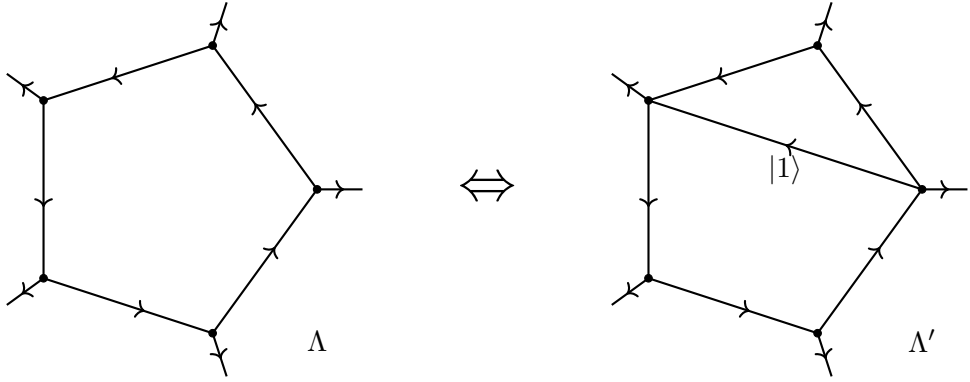
\begin{figure}
	\centering
	\begin{tikzpicture}
		\def\sep{4}
		\def\nn{5}
		\def\rr{1.3}
		\def\scale{2}
		\def\hh{0.75}
		\node at (-\sep,0) {\begin{tikzpicture}[scale=\scale]
				
				\foreach \xx in {1,...,\nn}{
					\filldraw ({cos(360*(\xx-1)/\nn)},{sin(360*(\xx-1)/\nn)}) circle (0.025);
					\draw[thick,->-=0.6] ({cos(360*(\xx-1)/\nn)},{sin(360*(\xx-1)/\nn)}) -- ({\rr*cos(360*(\xx-1)/\nn)},{\rr*sin(360*(\xx-1)/\nn)});
					\draw[thick,->-=0.6] ({cos(360*(\xx-1)/\nn)},{sin(360*(\xx-1)/\nn)}) -- ({cos(360*\xx/\nn)},{sin(360*\xx/\nn)});
				}
				\node at (1,-1) {$\Lambda$};
		\end{tikzpicture}};
		\node[scale=2] at (0,0) {$\Leftrightarrow $};
		\node at (\sep,0)  {\begin{tikzpicture}[scale=\scale]
				
				\draw[thick,->-=0.5] (1,0) -- ({cos(360*(2)/\nn)},{sin(360*(2)/\nn)});
				
				\foreach \xx in {1,...,\nn}{
					\filldraw ({cos(360*(\xx-1)/\nn)},{sin(360*(\xx-1)/\nn)}) circle (0.025);
					\draw[thick,->-=0.6] ({cos(360*(\xx-1)/\nn)},{sin(360*(\xx-1)/\nn)}) -- ({\rr*cos(360*(\xx-1)/\nn)},{\rr*sin(360*(\xx-1)/\nn)});
					\draw[thick,->-=0.6] ({cos(360*(\xx-1)/\nn)},{sin(360*(\xx-1)/\nn)}) -- ({cos(360*\xx/\nn)},{sin(360*\xx/\nn)});
				}
				\node[anchor=north] at ({0.5+cos(360*(2)/\nn)/2},{sin(360*(2)/\nn)/2}) {$\ket{1}$};
				\node at (1,-1) {$\Lambda'$};
		\end{tikzpicture}};
		
	\end{tikzpicture}
	\caption{An example of move 2, which can add or remove a leg/plaquette from the graph $\Lambda$. In other words, we can split or fuse two bulk plaquettes together. 
	}
	\label{fig:move2}
\end{figure}

Because $\Ha_{\mathrm{phys}}(\Sigma)$ is independent of the tessellation $\Lambda$, it is a topological invariant of $\Sigma$, and so it is the Hilbert space of a three-dimensional topological field theory. When $G$ is a finite group, this construction is Kitaev's quantum double model \cite{Kitaev:1997wr}, and the TQFT is Dijkgraaf--Witten theory \cite{Dijkgraaf:1989pz,Hu:2012wx}. When the legs are labeled instead by integrable representations of the quantum group $G_k$, it is a Levin--Wen string net \cite{Levin_2005,kirillov2011stringnet}, and the TQFT is the doubled $G_k \times G_{-k}$ Chern--Simons theory. Our construction allows every representation in the support of the Plancherel measure on every leg, so we can think of the TQFT associated with  $\Ha_{\mathrm{phys}}(\Sigma)$ as the large level limit of the doubled Chern--Simons theory. We return to this in Sec.~\ref{sec:continuum}.

\subsubsection{Matter}
Matter is incorporated by attaching additional \emph{matter legs} to the tessellation.\footnote{A recent construction \cite{Bourne:2026jbx} suggests that we can think of a tensor network with $n$ matter legs as preparing a state with $n$ particles. The complete matter Hilbert space for a minimally coupled scalar field, then, may be a ``Fock space'' of these particles, i.e., a direct sum over tensor networks over the number of matter legs. In this paper, however, we will focus on states with a fixed number of matter legs.} A matter leg is an auxiliary Hilbert space $\Hmatt{}$ which must support an action $A^{\mathrm{matt}}_v(h)$, $B^{\mathrm{matt}}_{(v,p)}(h)$ of the electric and magnetic operators, so it is attached to both a bulk vertex $v$ and a bulk plaquette $p$. Using moves 1 and 2, we can always arrange that distinct matter legs are attached to distinct pairs $(v,p)$. The kinematic Hilbert space and constraint operators become
\begin{align}
	\Ha(\Lambda) &= \bigotimes_{\ell \in E} L^2(G)_\ell \bigotimes_{(v,p) \in V \times P} \Hmatt{(v,p)} \,, \label{eq:HLambdamatter}\\
	A^{\mathrm{tot}}_v(h) &= A_v(h) A_v^{\mathrm{matt}}(h) \,,\\
	B^{\mathrm{tot}}_{(v,p)}(h) &= \int dg\, B_{(v,p)}(hg^{-1})B^{\mathrm{matt}}_{(v,p)}(g) \,,
\end{align}
with $\Pi_A^{(v)}$, $\Pi_B^{(p)}$ and $\Pi$ defined from the total operators exactly as in \eqref{eq:Pidef}, and the physical Hilbert space is again the quotient by the null states of $\Pi$. A vertex/plaquette pair without matter corresponds to $\Hmatt{(v,p)} = \C$, $A^{\mathrm{matt}}_v = 1$, $B^{\mathrm{matt}}_{(v,p)}(h) = \delta(h)$. The multiplication of the electric operators and convolution of the magnetic ones is what keeps the theory topological away from the matter legs \cite{Akers:2024wab}. In particular, moves 1 and 2 can still be applied to any part of the lattice which does not contain a matter leg, so a lattice with matter can be reduced until each matter leg, together with the vertex and plaquette it is dressed to, is joined to the rest of the lattice by a single bulk leg. We call this structure a \emph{lollipop} (Fig.~\ref{fig:reduced}) \cite{Akers:2024wab}. The lollipops are the only features of the bulk lattice which survive the quotient by the graphical moves, a fact we will exploit in Sec.~\ref{sec:finite_regions} to define compact bulk regions in a gauge invariant way.

For the rest of this section, unless otherwise specified, we will work with tensor networks without matter legs. We return to the case of tensor networks with matter, the main purpose of this paper, in Sec.~\ref{sec:entropy_matter_norandom}.

\subsubsection{Example: the disk}
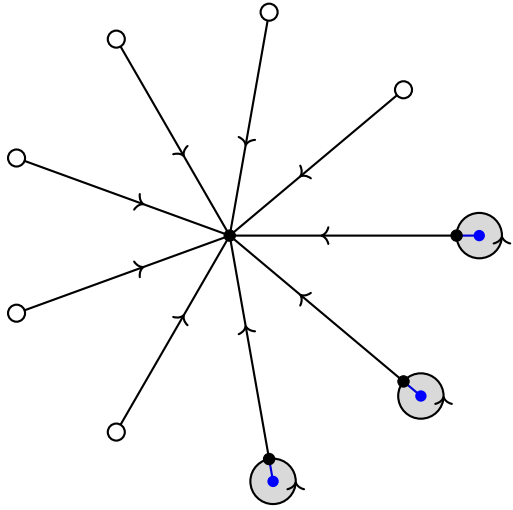
\begin{figure}
	\centering
	\begin{tikzpicture}[scale=3]
		\def\nn{6}
		\def\mm{3}
		\def\rr{1.1}
		
		\foreach\xx in {1,...,\nn}{
			\draw[thick,->-=0.6]  ({cos(360*(\xx)/(\nn+\mm))},{sin(360*(\xx)/(\nn+\mm))}) -- (0,0);
			\filldraw[thick,fill=white] ({cos(360*(\xx)/(\nn+\mm))},{sin(360*(\xx)/(\nn+\mm))}) circle ({0.025*1.5});
		}
		\foreach\xx in {1,...,\mm}{
			\draw[thick,->-=0.6] ({cos(360*(\xx+\nn)/(\nn+\mm))},{sin(360*(\xx+\nn)/(\nn+\mm))}) -- (0,0);
			
			\draw[blue,thick] ({cos(360*(\xx+\nn)/(\nn+\mm))},{sin(360*(\xx+\nn)/(\nn+\mm))}) -- ({\rr*cos(360*(\xx+\nn)/(\nn+\mm))},{\rr*sin(360*(\xx+\nn)/(\nn+\mm))});
			
			\filldraw ({cos(360*(\xx+\nn)/(\nn+\mm))},{sin(360*(\xx+\nn)/(\nn+\mm))}) circle (0.025);
			
			\filldraw[thick,->-=0,fill opacity=0.15] ({\rr*cos(360*(\xx+\nn)/(\nn+\mm))},{\rr*sin(360*(\xx+\nn)/(\nn+\mm))}) circle ({\rr-1});
			
			\fill[blue] ({\rr*cos(360*(\xx+\nn)/(\nn+\mm))},{\rr*sin(360*(\xx+\nn)/(\nn+\mm))}) circle (0.025);
		}
		\filldraw (0,0) circle (0.025);
	\end{tikzpicture}
	\caption{Reduced lattice of the disk with six marked points on the boundary and three legs carrying matter. The boundary vertices are shown in white, and the out-of-plane legs that carry matter are shown in blue. The matter legs are attached to a bulk vertex and plaquette (shown in gray), which we call a ``lollipop''. This lollipop is connected to the central vertex of the reduced lattice through another bulk leg. }
	\label{fig:reduced}
\end{figure}

For most of this paper, we will take $\Sigma$ to be the disk $\Sigma_{0,1}$ with open boundary conditions, i.e., $n$ marked points on the boundary of the disk. There are two tessellations we will primarily use to describe the physical Hilbert space. The \emph{reduced lattice} of Fig.~\ref{fig:reduced} has $n$ boundary legs, no plaquettes, and every boundary leg feeds into a single bulk vertex. In the case of a tensor network with $m$ matter legs, the reduced lattice will also have $m$ lollipop factors feeding into the bulk vertex. Returning to the matter-free case, the magnetic constraint is empty and there is a single electric constraint, imposed at the lone bulk vertex. Orienting all boundary legs inwards, so that the electric operator acts on the $V_\pi^*$ factor of each leg and leaves the $V_\pi$ factor untouched, and writing $\vec{\pi} = \pi_1 \otimes \cdots \otimes \pi_n$, the physical Hilbert space is
\begin{align}
	\Ha_{\mathrm{phys}}\left(\Sigma_{0,1}^{(n)}\right) = \int_{\widehat{G}}^{\oplus} d\mu(\vec{\pi}) V_{\vec{\pi}} \otimes \Pi_A [ V_{\vec{\pi}}^*] \,. \label{eq:Hphys}
\end{align}
The subspace $\Pi_A[V_{\vec{\pi}}^*]$ is the space of intertwiners from $V_{\vec{\pi}}$ to the trivial representation. For non-compact $G$, the trivial representation is not in the support of the Plancherel measure and these states are not normalizable in the kinematic inner product. But they are normalizable in the co-invariant inner product \eqref{eq:ImpInnerProduct}, so we will continue to call them intertwiners. Since the intertwiner space is fixed by group theory, all of the independent data in the state sits in the $V_\pi$ factors at the boundary vertices: in this sense, on the disk, the model is holographic.

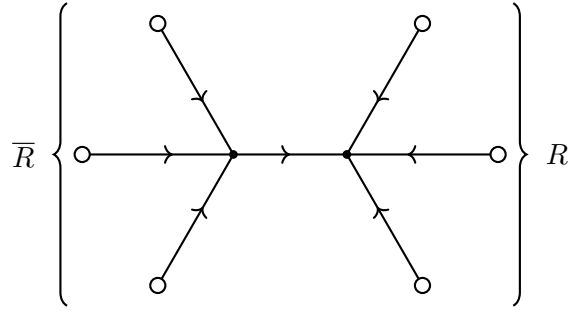
\begin{figure}
	\centering
	\begin{tikzpicture}[scale=2]
		\def\sep{3}
		\def\nn{6}
		\def\hh{0.75}
		
		\filldraw (0,0) circle (0.025);
		\filldraw (\hh,0) circle (0.025);
		\foreach \xx in {1,...,3}{
			\draw[thick,->-=0.6]  ({\hh+cos(360*(\xx-2)/\nn)},{sin(360*(\xx-2)/\nn)}) -- (\hh,0);
			\filldraw[thick,fill=white] ({\hh+cos(360*(\xx-2)/\nn)},{sin(360*(\xx-2)/\nn)}) circle (0.05);
		}
		\foreach \xx in {4,...,\nn}{
			\draw[thick,->-=0.6]  ({cos(360*(\xx-2)/\nn)},{sin(360*(\xx-2)/\nn)}) -- (0,0);
			\filldraw[thick,fill=white] ({cos(360*(\xx-2)/\nn)},{sin(360*(\xx-2)/\nn)}) circle (0.05);
		}
		\draw[thick,->-=0.5] (0,0) -- (\hh,0);
		
		\draw [thick,decorate,decoration={brace,amplitude=5pt}]
		(-1-0.1,-1) -- (-1-0.1,1) node[midway,xshift=-1.5em]{$\overline{R}$};
		
		\draw [thick,decorate,decoration={brace,amplitude=5pt,mirror}]
		(1+\hh+0.1,-1) -- (1+\hh+0.1,1) node[midway,xshift=1.5em]{$R$};
		
	\end{tikzpicture}
	\caption{The bowtie lattice $\Lambda_b$ between the boundary legs $R$ and $\overline{R}$ for a tensor network without matter legs.}
	\label{fig:bowtie}
\end{figure}

The second tessellation is adapted to a bipartition of the boundary vertices into a subregion $R$ and its complement $\overline{R}$. The \emph{bowtie lattice} $\Lambda_b$ of Fig.~\ref{fig:bowtie} is obtained from the reduced lattice by move 1: the central vertex is split into two, with the $R$ legs attached to one and the $\overline{R}$ legs to the other, and the new leg joining them is the \emph{corner leg}. In the continuum, operators supported on the corner leg are line operators supported on a boundary anchored curve $\gamma$ with $\Sigma \setminus \gamma = \Sigma_R \sqcup \Sigma_{\overline{R}}$, which separates the $R$ boundary vertices from the $\overline{R}$ boundary vertices (Fig.~\ref{fig:gluing_disks}). Without matter, $\gamma$ is unique up to homotopy on the disk. Defining the auxiliary Hilbert spaces
\begin{align}
	\Ha_R(\omega) = \int_{\widehat{G}}^{\oplus} d\mu(\vec{\pi}) V_{\vec{\pi}} \otimes \Pi_A[V_{\vec{\pi}}^* \otimes V_\omega^*] \,, \label{eq:HRomega}
\end{align}
and similarly $\Ha_{\overline{R}}(\omega)$, the bowtie presentation of the same physical Hilbert space reads
\begin{align}
	\Ha_{\mathrm{phys}}(\Sigma) = \int_{\widehat{G}}^{\oplus} d\mu(\pi)\, \Pi_A[\Ha_R(\pi) \otimes \Ha_{\overline{R}}(\overline{\pi})] \,. \label{eq:fixedareadecompositionintro}
\end{align}
Here, $\Ha_R(\pi)$ catalogs the fusion channels from the representations on the $R$ boundary legs to an intermediate representation $\overline{\pi}$ on the corner leg, $\Ha_{\overline{R}}(\overline{\pi})$ does the same for fusion channels from $\overline{R}$ to the $\pi$ representation, and by Schur's lemma there is a unique channel from each pair $(\pi,\overline{\pi})$ to the trivial representation, which is what the $\Pi_A$ in \eqref{eq:fixedareadecompositionintro} enforces. The direct integral over $\pi$ is the statement that the physical Hilbert space does \emph{not} factorize between $R$ and $\overline{R}$: the two sides are correlated through the representation $\pi$ carried by the corner leg. We will see in Sec.~\ref{sec:continuum} that \eqref{eq:fixedareadecompositionintro} is the tensor network version of the decomposition of the gravitational Hilbert space into fixed area states \cite{Dong:2018seb,Dong:2022ilf}, with $\pi$ playing the role of the area and the Plancherel measure supplying their density. 

\subsection{The factorization map}\label{sec:factorization}

The reduced state of a subregion $R$ is ordinarily defined by a partial trace, $\rho_R = \tr_{\overline{R}}\ketbra{\psi}$, which presupposes a factorization $\Ha = \Ha_R \otimes \Ha_{\overline{R}}$. The direct integral in \eqref{eq:fixedareadecompositionintro} shows that $\Ha_{\mathrm{phys}}(\Sigma)$ admits no such factorization across the boundary vertices. Following \cite{Balasubramanian:2026ymu},\footnote{See also \cite{Casini:2013rba,Wong:2025asd,Wong:2025kpz,Ball:2024hqe,Donnelly:2020teo,Jiang:2020cqo,Donnelly:2018ppr,Wong:2017pdm,Blommaert:2018oue,Blommaert:2018rsf} for other work on factorization in gauge theories.} we circumvent this by constructing a linear map
\begin{align}
	V : \Ha_{\mathrm{phys}}(\Sigma) \to \Hext{R} \otimes \Hext{\overline{R}} \label{eq:Vdef}
\end{align}
which embeds the physical Hilbert space into a factorizing one. The two factors are the \emph{extended Hilbert spaces} of $R$ and $\overline{R}$, and they carry additional degrees of freedom, the \emph{edge modes}, beyond those of the physical Hilbert space. The map is built on the bowtie lattice $\Lambda_b$ of Fig.~\ref{fig:bowtie}, and it cuts open the corner leg.

\subsubsection{The group basis}
\begin{figure}
	\centering
	\begin{tikzpicture}[scale=2]
		\def\sep{0.5}
		\def\nn{6}
		\def\hh{2}
		
		\filldraw (0,0) circle (0.025);
		\filldraw (\hh,0) circle (0.025);
		\foreach \xx in {1,...,3}{
			\draw[thick,->-=0.6]  ({\hh+cos(360*(\xx-2)/\nn)},{sin(360*(\xx-2)/\nn)}) -- (\hh,0);
			\filldraw[thick,fill=white] ({\hh+cos(360*(\xx-2)/\nn)},{sin(360*(\xx-2)/\nn)}) circle (0.05);
		}
		\foreach \xx in {4,...,\nn}{
			\draw[thick,->-=0.6]  ({cos(360*(\xx-2)/\nn)},{sin(360*(\xx-2)/\nn)}) -- (0,0);
			\filldraw[thick,fill=white] ({cos(360*(\xx-2)/\nn)},{sin(360*(\xx-2)/\nn)}) circle (0.05);
		}
		\draw[thick,->-=0.5] (0,0) -- (\hh/2 - \sep/2,0) node[midway,anchor=north] {$k$};
		\draw[thick,->-=0.5] (\hh/2 + \sep/2,0) -- (\hh,0) node[midway,anchor=north] {$h$};
		\filldraw[thick,fill=black] (\hh/2 - \sep/2,0) circle (0.05);
		\filldraw[thick,fill=black] (\hh/2 + \sep/2,0) circle (0.05);
		
		\node at (0,-1.1) {$\dket{\vec{g}_{\overline{R}},k}_{\overline{R}}$};
		\node at (\hh,-1.1) {$\dket{h,\vec{g}_{R}}_{R}$};
		
		\draw [thick,decorate,decoration={brace,amplitude=5pt}]
		(-1-0.1,-1) -- (-1-0.1,1) node[midway,xshift=-1.5em]{$\overline{R}$};
		
		\draw [thick,decorate,decoration={brace,amplitude=5pt,mirror}]
		(1+\hh+0.1,-1) -- (1+\hh+0.1,1) node[midway,xshift=1.5em]{$R$};
		
	\end{tikzpicture}
	
	\caption{The state after the Gauss constraint between $R$ and $\overline{R}$ is lifted. The newly exposed legs, with black dots, are the edge modes. Each subgraph represents the entanglement wedge for $\overline{R},R$, respectively. Left $G$ multiplication on the edge modes is a gauge/global symmetry for the $\overline{R}$, $R$, edge modes, respectively. Right $G$ multiplication on the edge modes is a global/gauge symmetry for the $\overline{R}$, $R$ edge modes, respectively. The edge mode legs are therefore crucial for each subregion being gauge-invariantly defined, while simultaneously allowing the Hilbert space between the subregions to factorize because the other multiplication is lifted to a global symmetry.}
	\label{fig:factorizedgraph}
\end{figure}
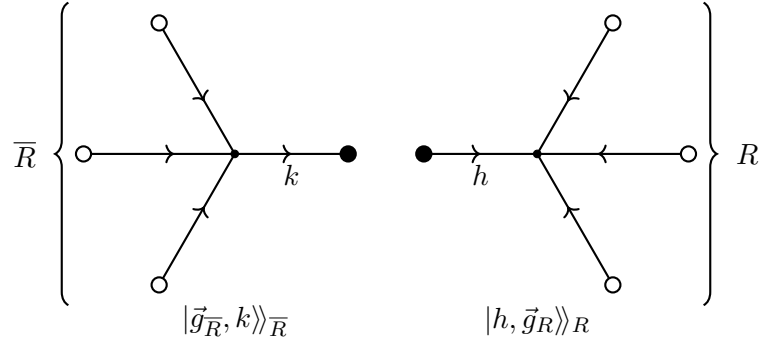

Any state in the bowtie presentation has an expansion
\begin{align}
	\dket{\psi} = \int d[\vec{g}_R,\vec{g}_{\overline{R}},h] \,\psi(\vec{g}_R, \vec{g}_{\overline{R}}, h) \dket{\vec{g}_{\overline{R}},h,\vec{g}_R} \,,
\end{align}
where $\vec{g}_R$ and $\vec{g}_{\overline{R}}$ are the group elements on the two sets of boundary legs, and $h$ is the group element on the corner leg. The factorization map is defined by
\begin{align}
	V\dket{\psi} = \int d[\vec{g}_R,\vec{g}_{\overline{R}}, h, k ]\,\psi(\vec{g}_R, \vec{g}_{\overline{R}}, kh) \dket{\vec{g}_{\overline{R}},k}_{\overline{R}} \dket{h,\vec{g}_R}_R\,. \label{eq:Vgroupbasisfactorized}
\end{align}
The states $\dket{h,\vec{g}_R}_R$ span $\Hext{R}$: they are co-invariant states of the partial lattice $\Lambda_R$ consisting of the $R$ boundary legs, the $R$ bulk vertex, and one copy of the corner leg, with the Gauss law still imposed at the bulk vertex (Fig.~\ref{fig:factorizedgraph}). The same holds for $\dket{\vec{g}_{\overline{R}},k}_{\overline{R}}$ and $\Lambda_{\overline{R}}$. Since $V$ is defined on the bowtie presentation, which exists for every state of the disk, it is defined on all of $\Ha_{\mathrm{phys}}(\Sigma)$.

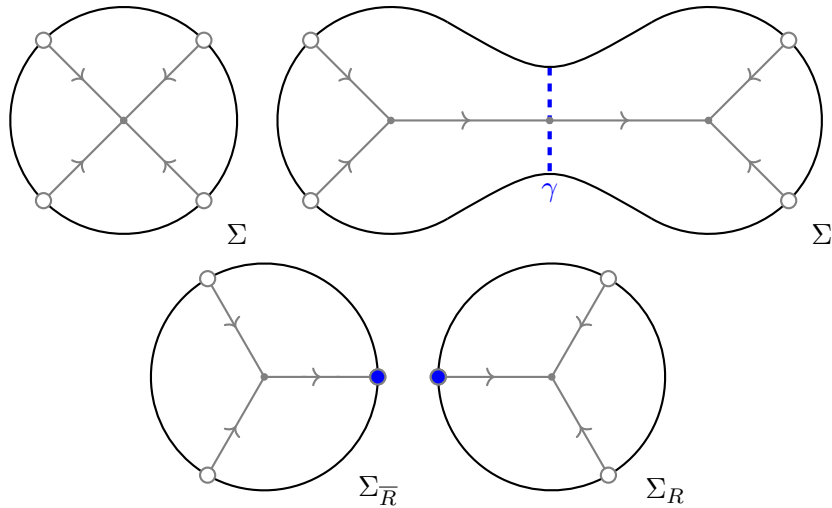
\begin{figure}
	\centering
	\newcommand{\legdisk}[4]{%
		\pgfmathsetmacro{\gap}{asin(#3/#2)}
		\foreach \i in {0,...,\numexpr#1-1\relax}{%
			\pgfmathsetmacro{\ang}{#4 + \i*360/#1}%
			\pgfmathsetmacro{\angnext}{#4 + (\i+1)*360/#1}%
			\draw[thick]
			(\ang+\gap:#2) arc[start angle=\ang+\gap, end angle=\angnext-\gap, radius=#2];
			
			\draw[color=black!50, thick,->-=0.5] (\ang:#2) -- (0,0);
			\filldraw[black!50,thick,fill=white] (\ang:#2) circle (0.1);
			
		}%
		\fill[color=black!50] (0,0) circle (0.05);%
	}
	
	\begin{tikzpicture}
		\legdisk{4}{1.5}{0.1}{45}
		\node at (1.5,-1.5) {$\Sigma$};
	\end{tikzpicture}
	\begin{tikzpicture}
		\def\R{1.5}      
		\def\r{0.1}     
		\def\cx{2.1}     
		\def\na{60}      
		\pgfmathsetmacro{\gap}{asin(\r/\R)}
		\coordinate (L)  at (-\cx,0);
		\coordinate (Rc) at (\cx,0);
		\coordinate (M)  at (0,0);
		\coordinate (NTL) at ([shift={(L)}]\na:\R);
		\coordinate (NTR) at ([shift={(Rc)}]180-\na:\R);
		\coordinate (NBL) at ([shift={(L)}]-\na:\R);
		\coordinate (NBR) at ([shift={(Rc)}]\na-180:\R);
		
		\draw[ultra thick,dashed,blue] (0,0.7) -- (0,-0.7);
		\node[anchor=north,blue] at (0,-0.7) {$\gamma$};
		
		\draw[thick] ([shift={(L)}]135+\gap:\R) arc[start angle=135+\gap, end angle=225-\gap, radius=\R];
		\draw[thick] ([shift={(L)}]135-\gap:\R) arc[start angle=135-\gap, end angle=\na, radius=\R];
		\draw[thick] ([shift={(L)}]225+\gap:\R) arc[start angle=225+\gap, end angle=360-\na, radius=\R];
		\draw[thick] ([shift={(Rc)}]45-\gap:\R) arc[start angle=45-\gap, end angle=-45+\gap, radius=\R];
		\draw[thick] ([shift={(Rc)}]180-\na:\R) arc[start angle=180-\na, end angle=45+\gap, radius=\R];
		\draw[thick] ([shift={(Rc)}]\na-180:\R) arc[start angle=\na-180, end angle=-45-\gap, radius=\R];
		
		\draw[thick] (NTL) to[out=-90+\na, in=-90-\na, looseness=1.5] (NTR);
		\draw[thick] (NBL) to[out=90-\na, in=90+\na, looseness=1.5] (NBR);

		
		\draw[color=black!50, thick, ->-=0.5] ([shift={(L)}]135:\R) -- (L);
		\draw[color=black!50, thick, ->-=0.5] ([shift={(L)}]225:\R) -- (L);
		\draw[color=black!50, thick, ->-=0.5] ([shift={(Rc)}]45:\R) -- (Rc);
		\draw[color=black!50, thick, ->-=0.5] ([shift={(Rc)}]-45:\R) -- (Rc);
		\draw[color=black!50, thick, ->-=0.5] (L) -- (M);
		\draw[color=black!50, thick, ->-=0.5] (M) -- (Rc);
		\foreach \c/\a in {L/135, L/225, Rc/45, Rc/-45}{
			\filldraw[thick,color=black!50, fill=white] ([shift={(\c)}]\a:\R) circle (0.1);
		}
		\foreach \p in {L, Rc, M}{ \fill[color=black!50] (\p) circle (0.05); }
		
		\foreach \c/\a in {L/135, L/225, Rc/45, Rc/-45}{
		}
		
		\node at (\cx+1.5,-1.5) {$\Sigma$};
		
	\end{tikzpicture}
	
	\begin{tikzpicture}
		\begin{scope}[shift={(-1.9,0)}]
			\legdisk{3}{1.5}{0.1}{0}
			\fill[fill=white] (0.75,0) circle (0.2);
			\draw[thick,color=black!50,->-=0.5] (0.5,0) -- (1,0);
			\filldraw[black!50,fill=blue,thick] (1.5,0) circle (0.1);
			\node at (1.5,-1.5) {$\Sigma_{\overline{R}}$};
		\end{scope}
		\begin{scope}[shift={(1.9,0)}]
			\legdisk{3}{1.5}{0.1}{180}
			\filldraw[black!50,fill=blue,thick] (-1.5,0) circle (0.1);
			\node at (1.5,-1.5) {$\Sigma_{R}$};
		\end{scope}
	\end{tikzpicture}
	
	\caption{Top: Two equivalent presentations of the disk Hilbert space with four marked points, in the reduced lattice gauge and the bowtie lattice gauge, respectively. The physical Hilbert space $\Ha_{\mathrm{phys}}(\Sigma)$ is identical between them. As in Fig.~\ref{fig:factorizedgraph}, the boundary points for $\overline{R}$ are on the left, and the boundary points for $R$ are on the right.  The boundary anchored curve $\gamma$ which separates $\overline{R}$ from $R$ is shown in blue on the right.
		Bottom: The two disks of the extended Hilbert spaces $\Hext{\overline{R}}$ and $\Hext{R}$, respectively, after we cut the disk open along $\gamma$. The degrees of freedom that live on the blue dot descend from the curve $\gamma$, and are the observer edge modes. The gauge fixing edge modes feed into the bulk vertex of each partial disk.
	}
	\label{fig:gluing_disks}
\end{figure}

We can think of the factorization map $V$ as cutting the tensor network open along the corner leg. This leaves behind a finite boundary within each partial disk, as in Fig.~\ref{fig:gluing_disks}: in the continuum, the cut is the boundary anchored curve $\gamma$ of Sec.~\ref{sec:themodel} which separates $R$ from $\overline{R}$, and each of $\Sigma_R$ and $\Sigma_{\overline{R}}$ acquires $\gamma$ as a new boundary component. Like any boundary, the cut must be given a boundary condition, and the same one on both sides, so that the cut is invisible after regluing. For the factorization map to be independent of the precise details of the lattice near the cut $\gamma$, this must be a \emph{topological} boundary condition \cite{Kitaev:2011dxc,Kong:2014xyz,Kapustin:2010hk,Beigi:2010htr}. Briefly, a topological boundary condition is characterized by the set of bulk line operators which can end on it without leaving an excitation behind.

As in \cite{Balasubramanian:2026ymu}, in this paper we will impose the \emph{rough} boundary condition at the cut for concreteness. The rough boundary condition is the one on which the pure charges, labeled by $\pi \in \widehat{G}$, condense \cite{Bravyi:1998sy,Beigi:2010htr,Kitaev:2011dxc,Balasubramanian:2026ymu}. On the lattice, it is obtained by dropping the electric constraint at the boundary vertices while keeping the magnetic constraint on every plaquette. This is exactly what cutting the corner leg does, since it exposes one new vertex on each side of the cut at which the Gauss law is not imposed; the cut is treated as a new pair of marked points, one for $\Sigma_R$ and one for $\Sigma_{\overline{R}}$. In the language of gauge theory, it is a Dirichlet boundary condition for the gauge field, under which $G$ survives as a global symmetry of the boundary degrees of freedom \cite{Wen:2023otf,Wen:2024udn}. Unlike the boundary conditions at $\partial \Sigma$, the possible topological boundary conditions are determined solely by the bulk theory \cite{Kong:2014xyz,KONG2021115607}, and a different choice (for instance the smooth boundary condition, on which the pure fluxes condense) would lead to a factorization map with an alternative edge mode structure, labeled by whichever anyons the cut can then support \cite{Wong:2017pdm,Donnelly:2018ppr}. We will discuss this point more in Sec.~\ref{sec:entropy_bcs} after explaining the rough boundary factorization map and the entropy it defines in detail. Furthermore, we explain which topological boundary condition leads to the quantum extremal surface formula in 3D gravity in Sec.~\ref{sec:quantumgroups}.

Let us now understand the structure of the factorization map $V$ in more detail. The corner leg introduced into the extended Hilbert space plays two roles at once. A gauge transformation at the bulk vertex of $\Lambda_R$ acts on $h$ by right multiplication, so the corner leg is what allows $\Hext{R}$ to be defined gauge invariantly: it is the leg into which the $R$ boundary legs fuse. Left multiplication on $h$, however, is not gauged in $\Hext{R}$; it is a global symmetry of the extended Hilbert space, which we call the \emph{corner symmetry}. In the representation basis \eqref{eq:peterweylnoncompact} of the corner leg, right multiplication acts on $V_\pi^*$ and left multiplication acts on $V_\pi$. We call the degrees of freedom in $V_\pi^*$, which are contracted into the $R$ vertex by the Gauss law, the \emph{gauge fixing edge modes}. The degrees of freedom in $V_\pi$, which are left free, are called the \emph{observer edge modes}.\footnote{For $\Hext{\overline{R}}$ the orientation of the corner leg is reversed, so the roles of left and right multiplication are exchanged. Splitting the edge modes into the two indices of a matrix element in this way is expected to be a general feature of gravitational theories, whose soliton charges are always in the adjoint representation \cite{McNamara:2021cuo,Harlow:2018tng}, and matches the role of corner symmetry in continuum gauge theory and gravity \cite{Donnelly_2016}.} The observer edge modes of $R$ and $\overline{R}$ are not independent: writing $A_\eth(g)$ for the simultaneous transformation $h \to gh$, $k \to kg^{-1}$, a change of variables in \eqref{eq:Vgroupbasisfactorized} shows that
\begin{align}
	A_\eth(g) V \dket{\psi} = V\dket{\psi} \,, \label{eq:cornersymmetry}
\end{align}
so the diagonal corner symmetry is a gauge symmetry of the image of $V$, even though each of the individual left multiplications acts nontrivially. Geometrically, $A_\eth$ moves the cut $\gamma$ within $\Sigma$ in a correlated way on both sides, leaving the glued surface invariant.

The action of $V$ on a basis element makes the structure of the edge modes explicit:
\begin{align}
	V\dket{\vec{g}_{\overline{R}}, g, \vec{g}_R}
	&= \int d[h, k] \delta(g^{-1}kh) \dket{\vec{g}_{\overline{R}}, k}\dket{h, \vec{g}_R}
	\\&= \int dh \,\dket{\vec{g}_{\overline{R}},h^{-1}} \dket{hg, \vec{g}_R} \label{eq:Vgroupbasis}
	\\&= \left(\int dh A_{\eth}(h) \right) \dket{\vec{g}_{\overline{R}},e} \dket{g,\vec{g}_R} \,.
\end{align}
The integral over $h$ is the Gauss law projector $[\Pi_A]_\eth$ for a new bulk vertex, the \emph{corner vertex} $\eth$, which glues the two copies of the corner leg back together (the black vertices of Fig.~\ref{fig:factorizedgraph}). So $V$ can be understood as follows: introduce an auxiliary leg in the state $\ket{e}$, and impose the Gauss law at the vertex joining it to the corner leg. The choice $\ket{e}$ is the unique wave function for the observer edge modes for which the two halves glue back to the original state with no relative ``kink'' at $\eth$. Any other choice, say $\ket{g}$, would leave behind an electric flux through the corner vertex after regluing, which is to say a physical charge inserted on the cut. In this sense, $V$ is the unique factorization map which introduces no defect at the cut, a property we will use to fix the remaining ambiguities below.

Finally, we note that $V$ is only \emph{projectively} an isometry. Suppressing the boundary legs,
\begin{align}
	\bra{g} V^\dagger V \ket{h} = \int d[m\,n] \braket{m}{n}\braket{mg}{nh} = \left(\int_G dm \right) \braket{g}{h} \,, \label{eq:groupbasisVoldiverge}
\end{align}
so $V^\dagger V = \mathrm{Vol}(G)\,\Id$. For compact $G$ we can normalize the Haar measure so that $\mathrm{Vol}(G) = 1$ and there is nothing to discuss. For non-compact $G$ the divergence is real, and dealing with it is the subject of Sec.~\ref{sec:algebras}.

\subsubsection{The representation basis}
Before we explain how to cure this divergence, it will be useful to investigate the action of the factorization map in another basis. The same map $V$ acts on the representation basis of the corner leg as follows. Repeatedly inserting the resolution of the identity $\Id_{L^2(G)} = \int d\mu(\pi)\sum_{ij}\ketbra{\pi,ij}$ into \eqref{eq:Vgroupbasis} gives
\begin{align}
	V \dket{\vec{g}_{\overline{R}} ; \pi,ab ; \vec{g}_R} = \sum_c \dket{\vec{g}_{\overline{R}} ; \pi,ac}_{\overline{R}}  \dket{\pi,cb; \vec{g}_R}_R \,, \label{eq:Vcoinvariantrepbasis}
\end{align}
that is, matrix multiplication on the corner leg. The sum over $c$ is gauge invariant because the $b$ index is contracted into the $R$ vertex while the $c$ index, which belongs to the observer edge modes, remains free \cite{Balasubramanian:2025rcr}. It is useful to split the basis state of the corner leg as $\ket{\pi,ab} = \ket{\pi,a}\ket{b_\pi}$, into a factor which carries the continuous label $\pi$ and one which does not, normalized as
\begin{align}
	\braket{\pi,a}{\omega,c} &= \delta(\pi,\omega) \delta_{ac} \,, \label{eq:fakesplitdelta}\\
	\braket{b_\pi}{d_\omega} &= \delta_{\pi\omega} \delta_{bd} \label{eq:fakesplitkronecker}\,,
\end{align}
where $\delta_{\pi\omega}$ is a Kronecker delta. The assignment of the continuous label to one factor is a convention, chosen (with the opposite convention on the $\overline{R}$ corner leg) so that the gauge fixing edge modes of both $R$ and $\overline{R}$ are unit normalized. Then \eqref{eq:Vcoinvariantrepbasis} takes the form
\begin{align}
	V \dket{\vec{g}_{\overline{R}} ; \pi,ab ; \vec{g}_R} &= \dket{\vec{g}_{\overline{R}} ; a_\pi }_{\overline{R}} \ket{\chi_\pi}' \dket{b_\pi ; \vec{g}_R}_R \,, \label{eq:Vrepresentationbasis}
\end{align}
where the middle factor is proportional to the character of the representation $\pi$,
\begin{align}
	\ket{\chi_\pi}' \propto \ket{\chi_\pi} = \sum_c \ket{\pi,c} \ket{c_\pi} = \int dg \,\chi_\pi(g) \ket{g} \,, \qquad \chi_\pi(g) = \tr[\pi(g)] \,, \label{eq:characterfndef}
\end{align}
and the prime records that $\ket{\chi_\pi}'$ carries the ``double delta'' normalization inherited from \eqref{eq:fakesplitdelta} rather than that of the $L^1$ function $\chi_\pi$.\footnote{For non-compact $G$ the character is a distribution, and $\chi_\pi(g)$ should be read as the locally integrable function representing it \cite{HarishChandra1952,HarishChandra1976}.} Equation \eqref{eq:Vrepresentationbasis} is a useful way to think about what $V$ does intuitively. Within each superselection sector $\pi$, it introduces a maximally entangled state of the observer edge modes of $R$ and $\overline{R}$. That entangled state is the character, as is made clear by the Bell pair structure of \eqref{eq:characterfndef}. This is forced by gauge invariance: with two legs meeting at the corner vertex, the intertwiner space $\Pi_A[V_\pi^* \otimes V_\pi]$ is one dimensional by Schur's lemma, and the character spans it. The entanglement structure of the edge modes within a fixed sector is therefore unique, and the only freedom left in $V$ is the normalization of $\ket{\chi_\pi}'$ sector by sector. The volume divergence \eqref{eq:groupbasisVoldiverge} lives entirely in this normalization: every state $\dket{b_\pi;\vec{g}_R}_R$ is normalizable, so $V^\dagger V$ diverges only because $\braket{\chi_\pi}{\chi_\pi}'$ does. Both facts, the residual freedom and the divergent norm, are taken up in Sec.~\ref{sec:reducedstate} and Sec.~\ref{sec:algebras}.

Reading \eqref{eq:Vrepresentationbasis} as a statement about Hilbert spaces, the extended Hilbert space decomposes as
\begin{align}
	\Hext{R} = \int_{\widehat{G}}^{\oplus} d\mu(\pi) \, \Ha_R(\pi) \otimes V_\pi \,, \label{eq:Hextdecomp}
\end{align}
with $\Ha_R(\pi)$ the auxiliary space \eqref{eq:HRomega} spanned by the boundary legs together with the gauge fixing edge modes, and $V_\pi$ contains the observer edge modes. Comparing with \eqref{eq:fixedareadecompositionintro}, the factorization map has replaced the single intertwiner space shared between $R$ and $\overline{R}$ by a pair of observer edge mode spaces $V_\pi$ and $V^*_\pi$, one on each side, and it is the entanglement between them that carries the information the intertwiner used to carry. 

\subsection{The reduced state} \label{sec:reducedstate}

With the factorization map in hand, the reduced state of the subregion $R$ is
\begin{align}
	\rho_R = \Tr_{\overline{R}}\left[V \dket{\psi}\dbra{\psi} V^\dagger\right] \,, \label{eq:rhoRdef}
\end{align}
where $\Tr_{\overline{R}}$ is the trace over the extended Hilbert space $\Hext{\overline{R}}$. A complete basis of $\Hext{\overline{R}}$ consists of the states $\dket{\vec{g}_{\overline{R}};a_\pi}_{\overline{R}}$ together with the $V_\pi^*$ half of the observer edge mode state $\ket{\chi_\pi}'$, so the first step is the partial trace of the character over one of its two tensor factors,
\begin{align}
	\tr_{V^*_\pi}[\ket{\chi_\pi}'
	\bra{\chi_\omega}'] &\equiv \delta(\pi,\omega) \Pi_\pi \,, \label{eq:partialtracecharacters}
\end{align}
where $\delta(\pi,\omega)$ is the Plancherel-normalized delta function, and $\Pi_\pi$ is a Plancherel-normalized projector which is characterized by its algebra
\begin{align}
	\Pi_\pi \Pi_\omega = \delta(\pi,\omega) \Pi_\pi \,. \label{eq:Piproj}
\end{align}
Taking the trace of both sides of \eqref{eq:partialtracecharacters} gives the norm of the observer edge mode state,
\begin{align}
	\braket{\chi_\pi}{\chi_\omega}' = \delta(\pi,\omega) \tr(\Pi_\pi) \,,\label{eq:character_delta_overlap}
\end{align}
which shows that the volume divergence \eqref{eq:groupbasisVoldiverge} of $V^\dagger V$ is the same thing as the statement that $\Pi_\pi$ is not trace class with respect to the ordinary trace on $V_\pi$. When $G$ is compact, $\Pi_\pi = \Id_{V_\pi}/d_\pi$ is the maximally mixed state on the observer edge modes, and \eqref{eq:rhoRdef} reproduces the familiar structure of the reduced state in lattice gauge theory \cite{Donnelly:2011hn}. For non-compact $G$ the family of operators $\Pi_\pi$ satisfying \eqref{eq:Piproj} is the correct replacement for the maximally mixed state, and the fact that its trace diverges means that we need to implement a renormalization procedure to define the trace of $\rho_R$.

To complete the computation of the reduced state $\rho_R$, the remainder of the partial trace is taken over the boundary legs. Leaving the details to \cite{Balasubramanian:2026ymu}, we find that
\begin{align}
	\rho_R = \int d\mu(\pi) \, q_\pi \, \rho_\pi \otimes \Pi_\pi \,. \label{eq:rhoRfirst}
\end{align}
Here, $q_\pi$ is a Plancherel-normalized probability distribution which satisfies
\begin{align}
	\int d\mu(\pi) \, q_\pi = \dnorm{\psi}\,, \label{eq:qpi_normalized}
\end{align}
so it integrates to one against the Plancherel measure exactly when the global state is normalized in the co-invariant inner product. Furthermore, each $\rho_\pi$ is a normalized partial density matrix on $\Ha_R(\pi)$ (defined in \eqref{eq:Hextdecomp}) with $\tr_{\Ha_R(\pi)}[\rho_\pi] = 1$.\footnote{To avoid confusion with the other traces which we will define in the next subsection, $\tr_{\Ha_R(\pi)}$ is the ordinary trace within the Hilbert space $\Ha_R(\pi)$, which is unique because the operator algebra acting on $\Ha_R(\pi)$ has no center.}
The reduced state $\rho_R$ is thus a classical mixture over superselection sectors, labeled by the representations $\pi \in \widehat{G}$. Within the sector $\pi$, the boundary legs and gauge fixing edge modes are in the state $\rho_\pi$, and the observer edge modes are in the generalized maximally mixed state $\Pi_\pi$. Furthermore, $q_\pi$ is the probability density, with respect to the Plancherel measure, for the state $\rho_R$ to be found in the $\pi$ sector. We will confirm in Sec.~\ref{sec:algebras} that $\rho_R$ is indeed a normalized state.

\subsubsection{Central operators and the uniqueness of the factorization map} \label{sec:Arcenter}
The entanglement structure within each sector was fixed by Schur's lemma, but the relative weighting of the sectors was not. To make this precise, consider the operator on $\Ha_{\mathrm{phys}}(\Sigma)$ defined by
\begin{align}
	\hat{C} \dket{\vec{g}_{\overline{R}} ; \pi,ab ; \vec{g}_R} = C(\pi) \dket{\vec{g}_{\overline{R}} ; \pi,ab ; \vec{g}_R} \,, \label{eq:Crepbasis}
\end{align}
for a bounded function $C: \widehat{G} \to \C$. In the group basis, $\hat{C}$ acts by convolution on the corner leg with the class function $C(g) = \int d\mu(\pi)\, C(\pi)\, \chi_\pi(g)$,
\begin{align}
	\hat{C} \dket{\vec{g}_{\overline{R}} ; h ; \vec{g}_R}
	= \int_G dk\, C(k) \dket{\vec{g}_{\overline{R}} ; kh ; \vec{g}_R}
	= \int_G dk\, C(k) \dket{\vec{g}_{\overline{R}} ; hk ; \vec{g}_R} \,, \label{eq:Cgroupbasis}
\end{align}
and it commutes with the constraints, so it is a physical operator: a gauge invariant line operator supported on the cut $\gamma$. 

Now, let $\mathcal{A}_R$ be the von Neumann algebra generated by the physical operators which act only on the $R$ boundary vertices, that is, the boundary anchored line operators which end on $R$.\footnote{As with any algebra defined by a generating set, we complete it by taking the double commutant, $\mathcal{A}_R \to \mathcal{A}_R''$.} Every element of $\mathcal{A}_R$ commutes with $\hat{C}$, because $\hat{C}$ acts on the corner leg alone. At the same time, we can deform the line operator defining $\hat{C}$ to act only on the $R$ boundary legs. Therefore, $\hat{C} \in \mathcal{Z}(\mathcal{A}_R)$ is central in $\mathcal{A}_R$. Conversely, by the decomposition \eqref{eq:fixedareadecompositionintro}, the only operators commuting with all of $\mathcal{A}_R$ are those proportional to the identity on each $\Ha_R(\pi)$. So the operators $\hat{C}$ exhaust the center of $\mathcal{A}_R$, and $\mathcal{Z}(\mathcal{A}_R) \cong L^\infty(\widehat{G})$ under pointwise multiplication. 
For any such $\hat{C}$, we can define the alternative factorization map $V_C = V\hat{C}$, which acts as
\begin{align}
	V_C \dket{\vec{g}_{\overline{R}} ; \pi,ab ; \vec{g}_R} &=  \dket{\vec{g}_{\overline{R}} ; a_\pi}_{\overline{R}}\, C(\pi)\ket{\chi_\pi}'  \dket{b_\pi; \vec{g}_R}_R \,. \label{eq:VCdef}
\end{align}
$V_C$ is gauge invariant sector by sector, and differs from $V$ only in the normalization assigned to the observer edge mode state in each sector. This is the entire ambiguity in the factorization map, and it has a clear physical meaning. Regluing the two halves of $V_C \dket{\psi}$ returns not $\dket{\psi}$ but $\hat{C}\dket{\psi}$, so the map $V_C$ inserts the line operator $\hat{C}$ on the cut before factorizing. The map $V = V_{\Id}$ is the unique factorization map which introduces no such defect, and this is the sense in which it is preferred \cite{Balasubramanian:2026ymu}. 

\subsubsection{The renormalized trace}\label{sec:algebras}

We now define the trace with respect to which $\rho_R$ is normalized, using techniques from algebraic quantum field theory. We refer the reader to \cite{Balasubramanian:2026ymu} for more details about the techniques utilized in this section.

States such as $\rho_R$ are used to compute expectation values of operators representing physical observables. So when we define the trace $\Tr$ for which $\Tr(\rho_R)=1$, we first need to specify what observables we are allowed to measure using $\rho_R$, and ensure that $\Tr$ is well-defined for these observables. There are three nested algebras to consider,
\begin{align}
	\mathcal{Z}(\mathcal{A}_R) \subset \mathcal{A}_R \subset \mathcal{B}_R \,, \label{eq:nesting}
\end{align}
where $\mathcal{B}_R$ is the algebra of all bounded operators on $\Hext{R}$, $\mathcal{A}_R$ is the (completion of) all the operators acting only on the $R$ vertices, and $\mathcal{Z}(\mathcal{A}_R)$ is the center of $\mathcal{A}_R$.
The point of this subsection is that each algebra carries its own trace, fixed by different physical principles, and that the three traces are related by canonical maps which allow us to translate between them.

\paragraph{The algebra $\mathcal{B}_R$.} The algebra of all bounded operators on a Hilbert space is a type I factor, so its trace is unique up to normalization: it is the sum of diagonal matrix elements. Writing $\ket{\pi;\alpha,i}$ for a basis of $\Hext{R}$ in the decomposition \eqref{eq:Hextdecomp}, with $\alpha$ indexing $\Ha_R(\pi)$ and $i$ indexing $V_\pi$, a general element of $\mathcal{B}_R$ has matrix elements $\mathcal{O}(\pi,\omega)^{\alpha,i}_{\beta,j}$, and
\begin{align}
	\Tr_{\mathcal{B}_R}[\mathcal{O}] = \int d\mu(\pi) \sum_{\alpha,i} \mathcal{O}(\pi,\pi)_{\alpha, i}^{\alpha, i}\,. \label{eq:BRtrace}
\end{align}
This trace is finite on ordinary density matrices of the boundary legs, but applied to \eqref{eq:rhoRfirst} it gives
\begin{align}
	\Tr_{\mathcal{B}_R}[\rho_R] = \int d\mu(\pi)\, q_\pi \tr_{V_\pi}[\Pi_\pi] = \int d\mu(\pi)\, q_\pi \cdot \dim(V_\pi)\, \delta(\pi,\pi) = \infty \,, \label{eq:tracevoldiv}
\end{align}
which is precisely the volume divergence \eqref{eq:groupbasisVoldiverge}. To see this, note that for compact $G$, $\dim(V_\pi)\delta(\pi,\pi) = \mathrm{Vol}(G)$ and there is no problem; for non-compact $G$, however, $\rho_R$ is not a normalizable state of $\mathcal{B}_R$.

\paragraph{The algebra $\mathcal{A}_R$.} The reduced state $\rho_R$ is, however, a state of the smaller algebra $\mathcal{A}_R$. Represented on the extended Hilbert space, $\mathcal{A}_R$ is the commutant of the corner symmetry,
\begin{align}
	\mathcal{A}_R = \mathcal{A}_{\mathrm{edge}}' \,, \label{eq:ARcommutant}
\end{align}
where $\mathcal{A}_{\mathrm{edge}}$ is the algebra generated by the left multiplications $U_R(g)$ on the corner leg. A general operator in $\mathcal{A}_R$ is given by
\begin{align}
	\mathcal{O}_{\mathcal{A}_R} = \int d\mu(\pi) \, \mathcal{O}(\pi) \otimes \Pi_\pi\,, \label{eq:ARgeneral}
\end{align}
with $\mathcal{O}(\pi)$ an operator on $\Ha_R(\pi)$. Note that $\mathcal{O}_{\mathcal{A}_R}$ is block diagonal in the sector label and proportional to $\Pi_\pi$ on the observer edge modes, and that the reduced state \eqref{eq:rhoRfirst} is of this form. A trace on $\mathcal{A}_R$ is
\begin{align}
	\Tr_{\mathcal{A}_R}[\mathcal{O}] = \int d\mu(\pi) \tr_{\Ha_R(\pi)}[\mathcal{O}(\pi)] \,, \label{eq:physicaltrace}
\end{align}
which amounts to declaring $\tr(\Pi_\pi) = 1$ and using the ordinary trace within each sector. This definition is linear, cyclic, and positive, so it is a legitimate trace. By \eqref{eq:qpi_normalized}, if the global state $\dket{\psi}$ is normalized in the co-invariant inner product of the physical Hilbert space, we have that
\begin{align}
	\Tr_{\mathcal{A}_R}[\rho_R] = \int_{\widehat{G}} d\mu(\pi) \, q_\pi = 1 \,.
\end{align}

However, this definition of the trace for $\mathcal{A}_R$ is not unique, because $\mathcal{A}_R$ has a large center. For any positive, bounded function $C(\pi)$, the definition
\begin{align}
	\Tr_{\mathcal{A}_R}^{C} [\mathcal{O}] = \Tr_{\mathcal{A}_R}[\hat{C} \mathcal{O}] = \int d\mu(\pi) \, C(\pi) \tr_{\Ha_R(\pi)}[\mathcal{O}(\pi)] \label{eq:CasMeasure}
\end{align}
is also a trace. A choice of trace on $\mathcal{A}_R$ is the same as a choice of measure on $\widehat{G}$, and nothing in the algebra itself prefers one measure over another \cite{Sorce:2023fdx}. 

Since the ambient trace \eqref{eq:BRtrace} is infinite on $\mathcal{A}_R$, we cannot simply fix this measure by defining $\Tr_{\mathcal{A}_R}$ as the trace from the larger algebra restricted to elements of $\mathcal{A}_R$. Instead, what fixes the trace on $\mathcal{A}_R$ is that the ambiguity in the trace is the same as the ambiguity in the factorization map. If we factorized the Hilbert space with $V_C$ rather than $V$, then the reduced state becomes
\begin{align}
	\rho_R^C = \Tr_{\overline{R}}[V_C \dket{\psi}\dbra{\psi}V_C^\dagger] = \int d\mu(\pi) \, q_\pi \, \rho_\pi \otimes |C(\pi)|^{2}\,\Pi_\pi \,. \label{eq:rhoRC}
\end{align}
Comparing with \eqref{eq:CasMeasure}, the trace which assigns $\rho_R^C$ unit norm is $\Tr^{|C|^{-2}}_{\mathcal{A}_R}$. In other words,
\begin{align}
	V_C \text{ is an isometry} \qquad \Longrightarrow \qquad \Tr_{\mathcal{A}_R} = \Tr^{|C|^{-2}}_{\mathcal{A}_R} \,. \label{eq:VCisometry}
\end{align}
For each factorization map, there is exactly one trace making it an isometry, so choosing a trace is the same as choosing a factorization map. The latter choice was settled in Sec.~\ref{sec:Arcenter}: the defect-free map $V$ corresponds to $C = 1$, and the physical trace on $\mathcal{A}_R$ is \eqref{eq:physicaltrace}, with the sectors integrated against the Plancherel measure. The one thing this does not fix is the overall normalization of the Haar measure, $dg \to \kappa\, dg$, which rescales $\Tr_{\mathcal{A}_R} \to \kappa^{-1} \Tr_{\mathcal{A}_R}$. This final ambiguity cannot be fixed canonically for non-compact $G$, and will enter the entropy of $\rho_R$ only as a state independent constant, to which we return in Sec.~\ref{sec:entropy_nomatter}. 

\paragraph{The algebra $\mathcal{Z}(\mathcal{A}_R)$.} Finally, consider the center of $\mathcal{A}_R$. The trace \eqref{eq:physicaltrace} is infinite on the center, since a central operator acts as the identity on every $\Ha_R(\pi)$ (an infinite dimensional Hilbert space), so the trace on the center must be renormalized separately. Because $\mathcal{Z}(\mathcal{A}_R) \cong L^\infty(\widehat{G})$ is abelian, every trace on it is an integral against a positive measure, so we can write
\begin{align}
	\Tr^\nu_{\mathcal{Z}(\mathcal{A}_R)}[\hat{C}] = \int_{\widehat{G}} d\nu(\pi) \, C(\pi) \,. \label{eq:centertrace}
\end{align}
The defect-free argument, which we used to fix the measure for the trace of $\mathcal{A}_R$, cannot select the measure $d\nu$ because the center \emph{is} the algebra of defects. The principle we used in \cite{Balasubramanian:2026ymu} to fix $d\nu$ is instead the following. By \eqref{eq:Cgroupbasis}, the center acts on the corner leg through class functions, which resolve a group element only up to conjugacy. In contrast, $\mathcal{A}_R$ resolves all of $G$. The Plancherel measure is the measure on $\widehat{G}$ dual to the Haar measure on $G$, which is why it is the right measure for $\mathcal{A}_R$; the measure appropriate to the center is the one dual to the flat measure on the space of conjugacy classes of $G$. Since almost every conjugacy class meets a unique Cartan subgroup (see \cite{Balasubramanian:2026ymu} for more details), the flat measure on conjugacy classes is the Haar measure on each Cartan subgroup (restricted to a Weyl chamber), and its Fourier dual defines a second measure $d\pi$ on $\widehat{G}$. We call $d\pi$ the \emph{microcanonical measure} over representations. The two are related by the density
\begin{align}
	\frac{d\mu(\pi)}{d\pi} \equiv \mu(\pi) \,. \label{eq:cfunction}
\end{align}
For representations in the principal series of $G$, $\mu(\pi) = |c(\pi)|^{-2}$ is Harish-Chandra's $c$-function \cite{HarishChandra1952,HarishChandra1976,Helgason2000GGA}, and for the remaining representations $\mu(\pi)$ is simply the relative weight the two measures assign to each representation. For compact $G$, the microcanonical measure is the counting measure over $\widehat{G}$ divided by the volume of $G$, and $\mu(\pi) = d_\pi$. In general, $\mu(\pi)$ plays the role of a density of states over $\widehat{G}$, in the sense that $d\mu(\pi) = \mu(\pi)\, d\pi$ in the same way that $dN = \rho(E)\,dE$ in statistical mechanics. The preferred trace on the center is then
\begin{align}
	\Tr_{\mathcal{Z}(\mathcal{A}_R)}[\hat{C}] = \int_{\widehat{G}} d\pi \, C(\pi) \,. \label{eq:centertracemicro}
\end{align}
Because $d\mu(\pi)$ and $d\pi$ both descend from the Haar measure on $G$ (the latter through the Weyl integration formula), a rescaling $dg \to \kappa\, dg$ rescales both by $\kappa^{-1}$, and the density $\mu(\pi)$ is independent of the normalization of the Haar measure altogether.\footnote{The relative normalization of these two measures is fixed by demanding the Lie algebra of $G$ and the Lie algebra of the Cartan subgroup that $d\pi$ descends from have the same normalization.}

\paragraph{Operator valued weights.} The three traces defined above were fixed by three different principles, and because of the renormalizations involved at each step of passing down the nesting \eqref{eq:nesting}, they are not simply restrictions of one another. Nevertheless, there is a canonical map at each step which implements this renormalization. This map, called an \emph{operator valued weight} \cite{Haagerup:1979ovwI,Haagerup:1979ovwII,takesaki2003theory}, allows us to determine, say, the contribution of the observables in the center $\mathcal{Z}(\mathcal{A}_R)$ to the entropy of $\rho_R$, thought of as a state on the algebra $\mathcal{A}_R$. An operator valued weight is a positive, bimodular map that is compatible with the traces on each algebra. There is a unique operator valued weight which implements the renormalization between each of the steps of passing down the nesting \eqref{eq:nesting} which is compatible with the physical principles which determined the trace on each algebra.

The operator valued weight $\mathcal{E}: \mathcal{B}_R\to\mathcal{A}_R$, which renormalizes the trace of $\mathcal{A}_R$, is defined as
\begin{align}
	\mathcal{E}(\mathcal{O}) = \int d\mu(\pi) \left(\sum_{\alpha,\beta,i} \mathcal{O}(\pi,\pi)^{\alpha, i}_{\beta, i} \ketbra{\alpha_\pi}{\beta_\pi}\right) \otimes \Pi_\pi \,, \qquad \Tr_{\mathcal{B}_R} = \Tr_{\mathcal{A}_R} \circ\,\, \mathcal{E} \,. \label{eq:ovwdef}
\end{align}
Here, $\ket{\alpha_\pi}$ denotes an orthonormal basis of $\Ha_R(\pi)$, so the object in parentheses can be thought of as an operator on $\Ha_R(\pi)$. Mechanically, $\mathcal{E}$ decoheres $\mathcal{O} \in \mathcal{B}_R$ in the sector label $\pi$, traces over the observer edge modes within each sector, and reinstates $\Pi_\pi$ in the observer edge modes sector.\footnote{This map is also normal, semifinite, and faithful (see \cite{Balasubramanian:2026ymu}), which is why it is unique.} Note that $\mathcal{E}(\Id_{\mathcal{B}_R})$ diverges by exactly the factor in \eqref{eq:tracevoldiv}: the group volume divergence has been relocated from the trace into the map $\mathcal{E}$. 

Applied to any normalized state of $\mathcal{B}_R$, $\mathcal{E}$ produces a state of the form \eqref{eq:rhoRfirst}. This demonstrates that the structure we found for $\rho_R$ is not an artifact of the factorization map, but is the unique trace-compatible restriction of a state of the extended Hilbert space to $\mathcal{A}_R$.\footnote{Equivalently, one could define an operator valued weight directly from $\mathcal{B}(\Ha_{\mathrm{phys}}(\Sigma))$ to $\mathcal{A}_R$ without ever introducing edge modes, closer in spirit to \cite{Dong:2018seb,Casini:2013rba,Klinger:2023auu,Klinger:2023tgi}; the same ambiguities and the same renormalization appear there. In our perspective, the extended Hilbert space is a tool which makes the physical meaning of these steps more transparent.}

One level down, an operator valued weight $\mathcal{F}: \mathcal{A}_R \to \mathcal{Z}(\mathcal{A}_R)$ is determined by a single density $\lambda(\pi)$, and compatibility with \eqref{eq:physicaltrace} and \eqref{eq:centertracemicro} fixes $\lambda(\pi) = d\mu(\pi)/d\pi = \mu(\pi)$:
\begin{align}
	\mathcal{F}(\mathcal{O}) = \int d\mu(\pi) \, \left(\mu(\pi)  \tr_{\Ha_R(\pi)}[\mathcal{O}(\pi)]\right) \, \Id_{\Ha_R(\pi)} \otimes \Pi_\pi \,, \qquad \Tr_{\mathcal{A}_R} = \Tr_{\mathcal{Z}(\mathcal{A}_R)} \circ \,\, \mathcal{F} \,. \label{eq:AtoZ}
\end{align}
Note the additional factor of $\mu(\pi)$ in this definition.
The weight relating $\mathcal{A}_R$ to its center is therefore not just the naive partial trace: it carries an additional factor of the density of states $\mu(\pi)$. Applied to the reduced state,
\begin{align}
	\mathcal{F}(\rho_R) = \int d\mu(\pi) \, \big(\mu(\pi)  q_\pi\big) \, \Id_{\Ha_R(\pi)} \otimes \Pi_\pi \,, \qquad \Tr_{\mathcal{Z}(\mathcal{A}_R)}[\mathcal{F}(\rho_R)] = \int d\pi \, \mu(\pi) q_\pi = 1 \,. \label{eq:TZonrho}
\end{align}
We can think of $p_\pi := \mu(\pi) q_\pi$ as the probability distribution of $\rho_R$ over superselection sectors, normalized against the microcanonical measure. Looking ahead, the mismatch between $\mu(\pi)q_\pi$ and $q_\pi$ in \eqref{eq:rhoRfirst} is where the area operator \eqref{eq:areaoperator} comes from.

\subsection{Entropy without matter legs}\label{sec:entropy_nomatter}

We can now compute the entropy of the reduced state $\rho_R$ of \eqref{eq:rhoRfirst} using the renormalized trace \eqref{eq:physicaltrace}:
\begin{align}
	S(\rho_R) = - \Tr_{\mathcal{A}_R}[\rho_R \ln \rho_R] = - \partial_n\Tr_{\mathcal{A}_R}[\rho_R^n]_{n=1} \,. \label{eq:entropy_partialn}
\end{align}
The details of this computation can be found in \cite{Balasubramanian:2026ymu}, where we found that the entropy $S(\rho_R)$ naturally decomposes into three pieces:
\begin{align}
	S(\rho_R) = H[\,p_\pi] + \langle \hat{A}\rangle_\rho + S_{\mathrm{bulk}}(\rho_R) \,. \label{eq:SrhoR_gen}
\end{align}
The decomposition of $S(\rho_R)$ into these three terms is the expected structure for a state of an algebra with a center \cite{Casini:2013rba,Harlow:2016vwg,Akers:2024wab,Dong2024,Donnelly:2011hn}, but adapted to the continuous label $\pi$. We will now explain each of these three terms in more detail.

\paragraph{The differential entropy. } The first term in \eqref{eq:SrhoR_gen} is the differential entropy $H[\,p_\pi]$, and is given by
\begin{align}
	H[\,p_\pi] \equiv -\int d\pi \, p_\pi \ln(p_\pi) \,, \qquad p_\pi \equiv \mu(\pi) \, q_\pi \,, \qquad \int d\pi \, p_\pi = 1 \,. \label{eq:Hdef}
\end{align}
This term measures the classical uncertainty about which sector $\pi$ the state $\rho_R$ is in, and can be interpreted as the entropy $S(\mathcal{F}(\rho_R))$ of the state $\mathcal{F}(\rho_R)$ on the algebra $\mathcal{Z}(\mathcal{A}_R)$ \cite{Balasubramanian:2026ymu}. 

The differential entropy is not sign definite when $\widehat{G}$ is continuous, i.e., when $G$ is non-compact. For example, a density $p^\epsilon_\pi$ supported on a set of microcanonical measure $\epsilon$ has $H[\,p^\epsilon_\pi] = \ln \epsilon$, which can be made as negative as we like in the limit $\epsilon \to 0$. Relatedly, $H[\,p_\pi]$ is the only term which depends on the normalization of the Haar measure: under $dg \to \kappa\, dg$,
\begin{align}
	d\mu(\pi) \to \kappa^{-1} d\mu(\pi) \,, \qquad d\pi \to \kappa^{-1} d\pi \,, \qquad H[\,p_\pi] \to H[\,p_\pi] - \ln(\kappa) \,, \label{eq:haarshift}
\end{align}
while $\mu(\pi)$, and with it the area term and the bulk entropy, are invariant. The entropy of a single state is therefore defined only up to a state independent constant, the scheme dependence which was left over in Sec.~\ref{sec:algebras}. Nevertheless, entropy \emph{differences} are the unambiguous, scheme-independent quantities in this model. This is similar to the state independent constant in the entropy of a semiclassical subregion in gravity argued for in  \cite{Witten:2021unn,Chandrasekaran:2022eqq,Chandrasekaran:2022cip,Jensen2023,Penington:2023dql,Kudler-Flam:2023qfl}, although the mechanism here differs from those cases:  $\mathcal{A}_R$ is a direct integral of type I factors, not a type II factor. We will comment on this point again in Sec.~\ref{sec:continuum} when we discuss 3D gravity.

\paragraph{The area contribution. } The second term in \eqref{eq:SrhoR_gen} is the expectation value $\langle \hat{A} \rangle_\rho = \Tr_{\mathcal{A}_R}[\rho_R \hat{A}]$ of the state independent operator
\begin{align}
	\hat{A} = \int d\mu(\pi) \, \ln(\mu(\pi)) \, \Id_{\Ha_R(\pi)} \otimes \Pi_\pi \,, \label{eq:areaoperator}
\end{align}
which we will discuss more in Sec.~\ref{sec:areaop_nomatter}. 

\paragraph{The bulk entropy. } Finally, the last term $S_{\mathrm{bulk}}(\rho_R)$ is defined as
\begin{align}
	S_{\mathrm{bulk}}(\rho_R) &= \int_{\widehat{G}} d\mu(\pi) \, q_\pi S(\rho_\pi) \,, \label{eq:Sbulk_nomatter}\\
	S(\rho_\pi) &= -\tr_{\Ha_R(\pi)}[\rho_\pi \ln(\rho_\pi)] \,.
\end{align}
In other words, $S_{\mathrm{bulk}}(\rho_R)$ is the entanglement entropy of each partial reduced state $\rho_\pi$, averaged over the sectors $\pi$ with respect to the distribution $q_\pi$. We call this term the \emph{bulk entropy}, for reasons which will become more clear in Sec.~\ref{sec:entropy_fixedcut}. The bulk entropy is non-negative, since each $\rho_\pi$ is an honest density matrix on $\Ha_R(\pi)$ with the ordinary trace. This term is not universal: $\Ha_R(\pi)$ depends on the number of marked points in $R$, and $S_{\mathrm{bulk}}$ is generically extensive in that number, with a non-universal coefficient.

\subsubsection{Boundary conditions at the cut }\label{sec:entropy_bcs}

Finally, we explain how the choice of boundary condition at the cut $\gamma$ enters the entropy $S(\rho_R)$. Recall from Sec.~\ref{sec:factorization} that the cut had to be given a topological boundary condition, so that the factorization map (and hence the entropy) is insensitive to the details of the lattice near $\gamma$. For concreteness, we chose the rough boundary condition \cite{Beigi:2010htr,Kitaev:2011dxc,Kong:2014xyz}. Had we chosen a different topological boundary condition, the derivation of the entropy formula would have gone through unchanged, because every step of this derivation used only the topological invariance of the model: the entropy would still decompose into the three terms of \eqref{eq:SrhoR_gen}, with the same interpretation for each term. What would change is the group theoretic data labeling the superselection sectors, which is determined by which line operators can end on the cut, and hence the label set over which the integrals in each of the three terms run. 

For example, the smooth boundary condition, which cuts the lattice along a leg rather than through it (relaxing the magnetic constraint at the plaquettes adjacent to the cut rather than the electric constraint at its endpoints), condenses the pure fluxes instead of the pure charges. In other words, the sectors would be labeled by conjugacy classes $[g]$ of $G$ rather than by representations $\pi$, the Fourier dual data. A general topological boundary condition condenses a mixture of charges and fluxes \cite{Beigi:2010htr,Kitaev:2011dxc,Kong:2014xyz}, and its sectors are labeled by a mixture of the two kinds of data. In each case, the spectrum of the area operator is the logarithm of the ratio of the measures defining the traces on the new algebra $\widetilde{\mathcal{A}}_R$ and its center, evaluated on the new label set. So the spectrum of $\hat{A}$ depends on the choice of boundary condition at the cut, even though its qualitative form does not. We leave a detailed treatment of other topological boundary conditions at the cut to future work. In Sec.~\ref{sec:quantumgroups}, however, we will determine which boundary condition at the cut reproduces the quantum extremal surface formula in 3D gravity. 

\subsubsection{The area operator}\label{sec:areaop_nomatter}

\begin{figure}
	\centering
	\begin{tikzpicture}[scale=1.5]
		\newcommand{\legdisk}[4]{%
			\pgfmathsetmacro{\gap}{asin(#3/#2)}
			\foreach \i in {0,...,\numexpr#1-1\relax}{%
				\pgfmathsetmacro{\ang}{#4 + \i*360/#1}%
				\pgfmathsetmacro{\angnext}{#4 + (\i+1)*360/#1}%
				\draw[thick]
				(\ang+\gap:#2) arc[start angle=\ang+\gap, end angle=\angnext-\gap, radius=#2];
				
				\draw[color=black!50, thick,->-=0.5] (\ang:#2) -- (0,0);
				\filldraw[black!50,thick,fill=white] (\ang:#2) circle (0.1);
				
			}%
			\fill[color=black!50] (0,0) circle (0.05);%
		}
		\def\xx{0.5}
		\legdisk{6}{1.5}{0.1}{55}
		\draw[red,thick] plot[smooth] coordinates {(0,1.5)  (0.25,0) (0,-1.5)};
		\filldraw[red] (0,1.5) circle (0.05);
		\filldraw[red] (0,-1.5) circle (0.05);
		\node[red,anchor=north] at (0,-1.5) {$\hat{A}$};
		\draw[decorate, thick,decoration={brace, amplitude=8pt, mirror}] (1.5+\xx,-1.5) -- (1.5+\xx,1.5) node[midway,xshift=15] {$R$};
		\draw[decorate, thick,decoration={brace, amplitude=8pt}] (-1.5-\xx,-1.5) -- (-1.5-\xx,1.5)  node[midway,xshift=-15] {$\overline{R}$};
		
	\end{tikzpicture}
	\caption{The area operator $\hat{A}$ on the disk. Because $\hat{A}$ is a topological operator (i.e., it is a physical operator that commutes with the gauge constraints) we can think of it as acting on either the $R$ or $\overline{R}$ boundary legs by deforming the line operator slightly to the left. Because $\hat{A}$ can be deformed such that it acts either only on the $R$ legs or not at all, it is central with respect to $\mathcal{A}_R$. }
	\label{fig:areaop}
\end{figure}
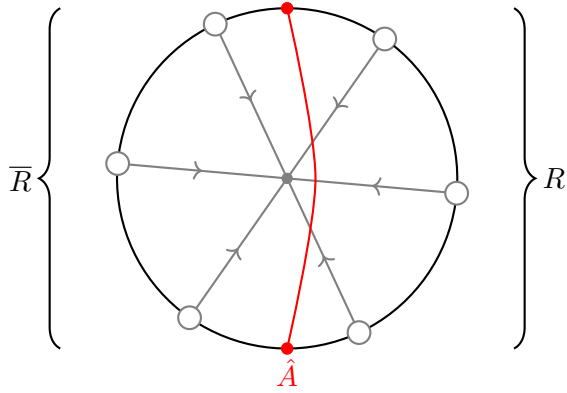

The second term of \eqref{eq:SrhoR_gen} is what plays the role of the area term in the holographic entropy formula, and it has three features which distinguish it from the other two.

First, as discussed above, $\hat{A}$ is built from the density of states $\mu(\pi) = d\mu(\pi)/d\pi$ of \eqref{eq:cfunction}, so its spectrum $\ln(\mu(\pi))$ is completely determined by $G$ and by the topological boundary condition imposed on the cut. It does not depend on the tessellation $\Lambda$, on the state $\dket{\psi}$, or on the number $n$ of marked points on $\partial \Sigma$, so it survives the $n \to \infty$ limit to a conformal boundary condition unchanged. Its expectation value depends on the state only through the distribution $q_\pi$. Furthermore, by the argument below \eqref{eq:cfunction}, $\mu(\pi)$ is independent of the normalization of the Haar measure, so the spectrum of $\hat{A}$ is meaningful even though $S(\rho_R)$ itself is scheme dependent. Note also that $\hat{A} \geq 0$ exactly when $\mu(\pi) \geq 1$ almost everywhere, which is a scheme-independent property of $G$.

Second, $\hat{A}$ is central in $\mathcal{A}_R$, and supported on the cut $\gamma$. In particular, it is of the form \eqref{eq:Crepbasis} with $C(\pi) = \ln(\mu(\pi))$, so it is a gauge invariant line operator along $\gamma$. By the topological invariance of the model, $\hat{A}$ can be deformed to act on the $R$ boundary legs alone or on the $\overline{R}$ boundary legs alone (Fig.~\ref{fig:areaop}). On the disk without matter, $\gamma$ is unique up to homotopy, so there is a single area operator and no minimization over possible cuts $\gamma$; this will change once matter legs are present. 

Third, $\hat{A}$ reduces to a familiar object in condensed matter theory when $G$ is a finite group. In that case, the model we are working with is Kitaev's quantum double model, and $\ln(\mu(\pi))$ agrees, up to a state independent constant of $-\ln|G|$, with the topological entanglement entropy of the sector $\pi$ \cite{Kitaev:2005dm,Levin:2006arx,Balasubramanian:2026ymu}. For general transformable group $G$, $\ln(\mu(\pi))$ is the replacement for $\ln(d_\pi)$, with $\mu(\pi)$ interpreted as the density of states of the sector $\pi$. We will review the connection between the area contribution and the topological entanglement entropy of string nets in Sec.~\ref{sec:doublemodel}, where it is the key to matching the spectrum of $\hat{A}$ to lengths of geodesics in 3D gravity.

\section{Entropy with fixed matter} \label{sec:entropy_matter_norandom}

So far, we have discussed how to factorize tensor networks that do not have any matter legs. 
In that case, the factorization map $V$ of Sec.~\ref{sec:factorization} is defined by cutting open the background manifold $\Sigma$  along a boundary anchored curve $\gamma$ which separates the $R$ and $\overline{R}$ vertices. 
The topological symmetry of the model implies that this definition is unique, because any other cut $\gamma'$ which also separates the boundary subregions is topologically equivalent to $\gamma$.
However, once we include matter legs into the network, there is no longer a unique cut $\gamma$ separating $R$ and $\overline{R}$. Each lollipop factor of Fig.~\ref{fig:reduced} sits at a bulk vertex and plaquette, and so it must be assigned to one side of the cut or to the other. Thus, when we include matter legs into the network, there will be a \emph{family} of factorization maps which split $R$ and $\overline{R}$, one for each topological equivalence class of possible cuts $\gamma$.

In this section, we will define the resulting family of factorization maps and determine the entropy of the corresponding factorized states. We will find that other than the Hilbert space that the reduced state acts on, essentially nothing changes compared to the matter-free case. The assignment of the matter legs enlarges the extended Hilbert space of each subregion, but the cut is still performed on a single corner leg, and the algebras, traces, and operator valued weights of Sec.~\ref{sec:algebras} carry over unmodified. In this section, we will choose the cut $\gamma$ \emph{by hand}. We will defer the question of which cut is preferred, analogously to the minimization over surfaces in the holographic entropy formula, to Sec.~\ref{sec:entropy_matter_random}.

\subsection{The bipartite factorization map} 

\begin{figure}
	\centering
	\begin{tikzpicture}[scale=2,thick]
		\def\Rd{1.5}          
		\def\nn{3}            
		\def\nnm{2}           
		\def\vertcols{1,3}    
		\def\rind{0.16}       
		\def\rbdy{0.05}       
		\def\rblk{0.05}       
		\def\dstem{0.275}      
		\def\rlol{0.1125}       
		\def\lolang{45}       
		\def\brc{0.40}        
		\colorlet{latt}{black}
		\colorlet{matt}{blue}
		\colorlet{cutc}{red}
		\pgfmathsetmacro{\sep}{2*\Rd/(\nn+1)}
		\pgfmathsetmacro{\gap}{asin(\rind/\Rd)}
		
		
		\filldraw[fill opacity=0.15] (0,0) circle (\Rd);
		\foreach \i in {1,...,\nn}{                     
			\pgfmathsetmacro{\X}{-\Rd+\i*\sep}
			\pgfmathsetmacro{\angC}{asin(\X/\Rd)}
			\pgfmathsetmacro{\angD}{180-\angC}
		}

		\foreach \i in {1,...,\nn}{
			\pgfmathsetmacro{\X}{-\Rd+\i*\sep}
			\pgfmathsetmacro{\Ed}{sqrt(\Rd*\Rd-\X*\X)}
			\pgfmathsetmacro{\Vone}{-\Rd+\sep}
			\pgfmathsetmacro{\Vtop}{-\Rd+\nn*\sep}
			\draw[latt,->-=0.5] (-\Ed,\X) -- (\Vone,\X);
			\draw[latt,->-=0.5] (\Vtop,\X) -- (\Ed,\X);
			\foreach \j in {1,...,\nnm}{
				\pgfmathsetmacro{\Ya}{-\Rd+\j*\sep}
				\pgfmathsetmacro{\Yb}{-\Rd+(\j+1)*\sep}
				\draw[latt,->-=0.5] (\X,\Ya) -- (\X,\Yb);
				\draw[latt,->-=0.5] (\Ya,\X) -- (\Yb,\X);
			}
		}
		\foreach \i in \vertcols{
			\pgfmathsetmacro{\X}{-\Rd+\i*\sep}
			\pgfmathsetmacro{\Ed}{sqrt(\Rd*\Rd-\X*\X)}
			\pgfmathsetmacro{\Vone}{-\Rd+\sep}
			\pgfmathsetmacro{\Vtop}{-\Rd+\nn*\sep}
			\draw[latt,->-=0.5] (\X,-\Ed) -- (\X,\Vone);
			\draw[latt,->-=0.5] (\X,\Vtop) -- (\X,\Ed);
		}
		
		\foreach \i in {1,...,\nn}{
			\pgfmathsetmacro{\X}{-\Rd+\i*\sep}
			\foreach \j in {1,...,\nn}{
				\pgfmathsetmacro{\Y}{-\Rd+\j*\sep}
				\pgfmathsetmacro{\cx}{\X+\dstem*cos(\lolang)}
				\pgfmathsetmacro{\cy}{\Y+\dstem*sin(\lolang)}
				\draw[matt,->-=0.6] (\X,\Y) -- (\cx,\cy);
				\fill[matt] (\cx,\cy) circle (0.035);
				\fill[latt] (\X,\Y) circle (\rblk);
			}
		}
		
		\foreach \i in {1,...,\nn}{
			\pgfmathsetmacro{\X}{-\Rd+\i*\sep}
			\pgfmathsetmacro{\angC}{asin(\X/\Rd)}
			\pgfmathsetmacro{\angD}{180-\angC}
			\filldraw[latt,fill=white] (\angC:\Rd) circle (\rbdy);
			\filldraw[latt,fill=white] (\angD:\Rd) circle (\rbdy);
		}
		\foreach \i in \vertcols{
			\pgfmathsetmacro{\X}{-\Rd+\i*\sep}
			\pgfmathsetmacro{\angA}{acos(\X/\Rd)}
			\pgfmathsetmacro{\angB}{-\angA}
			\filldraw[latt,fill=white] (\angA:\Rd) circle (\rbdy);
			\filldraw[latt,fill=white] (\angB:\Rd) circle (\rbdy);
		}
		
		\draw[cutc,very thick] plot[smooth] coordinates
		{(0,1.5) (-0.20,1.22) (-0.34,0.95) (-0.375,0.55) (-0.375,0) (-0.375,-0.55)
			(-0.34,-0.95) (-0.20,-1.22) (0,-1.5)};
		\draw[cutc,very thick,dashed] plot[smooth] coordinates
		{(0,1.5) (0.20,1.22) (0.34,0.95) (0.375,0.55) (0.375,0) (0.375,-0.55)
			(0.34,-0.95) (0.20,-1.22) (0,-1.5)};
		\filldraw[cutc] (0,1.5) circle (0.05);
		\filldraw[cutc] (0,-1.5) circle (0.05);
		\node[cutc] at (-0.42,1.64) {$\gamma_1$};
		\node[cutc] at (0.42,1.64) {$\gamma_2$};
		
		\draw[decorate,decoration={brace,amplitude=8pt,mirror}]
		(\Rd+\brc,-\Rd) -- (\Rd+\brc,\Rd) node[midway,xshift=16] {$R$};
		\draw[decorate,decoration={brace,amplitude=8pt}]
		(-\Rd-\brc,-\Rd) -- (-\Rd-\brc,\Rd) node[midway,xshift=-16] {$\overline{R}$};
	\end{tikzpicture}
	\caption{A lattice $\Lambda$ tessellating the disk $\Sigma$, with a matter leg attached to every bulk vertex. The electric constraint is imposed at every bulk vertex (black), and the magnetic constraint is imposed at every plaquette (gray). As in Fig.~\ref{fig:reduced}, we can use moves 1 and 2 to isolate each matter leg (blue) so that it forms a lollipop factor. The two cuts $\gamma_1$ and $\gamma_2$ split the boundary legs into $R$ and $\overline{R}$, and differ only in which side of the cut we place the matter legs: the central column of matter legs lies in $\mathfrak{l}_R$ for $\gamma_1$ and in $\mathfrak{l}_{\overline{R}}$ for $\gamma_2$.}
	\label{fig:cut_with_matter_legs}
\end{figure}
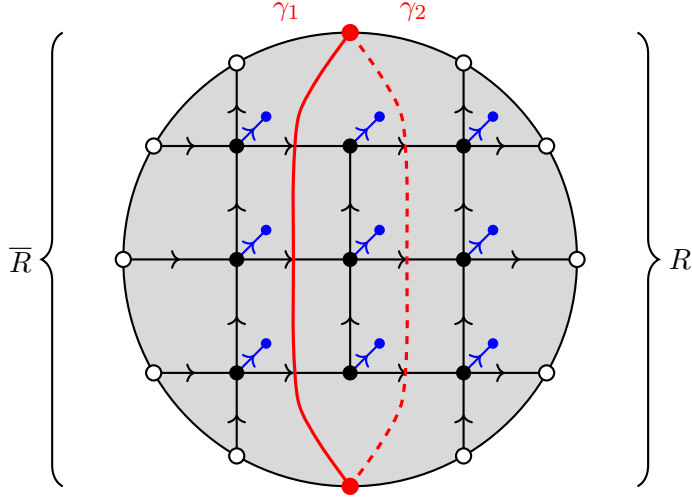

Let $\mathfrak{l}$ denote the set of matter legs of the tensor network. Let $\gamma$ be a boundary anchored curve in $\Sigma$ with $\Sigma \setminus \gamma = \Sigma_R \sqcup \Sigma_{\overline{R}}$, such that the $R$ boundary vertices lie in $\Sigma_R$, the $\overline{R}$ boundary vertices lie in $\Sigma_{\overline{R}}$, and $\gamma$ does not intersect any matter legs. We call any such curve a ``cut'' through the tensor network. This is the same notion of cut that appeared in Sec.~\ref{sec:factorization}, with the additional requirement that $\gamma$ avoid the matter legs, so that each matter leg lies unambiguously on either side of the cut. Such a curve induces a partition of the matter legs
\begin{align}
	\mathfrak{l} = \mathfrak{l}_R \sqcup \mathfrak{l}_{\overline{R}} \,, \label{eq:matterpartition}
\end{align}
where $\mathfrak{l}_R$ is the set of matter legs lying in $\Sigma_R$, and similarly for $\mathfrak{l}_{\overline{R}}$. See Fig.~\ref{fig:cut_with_matter_legs} for two examples of this partition.
Conversely, every partition of the matter legs defines such a curve.
To see this, recall from Sec.~\ref{sec:themodel} that in the reduced lattice of Fig.~\ref{fig:reduced}, every boundary leg and every lollipop attaches to a single bulk vertex. Now apply move 1 (Fig.~\ref{fig:move1}) to that vertex, splitting it into two bulk vertices joined by a new bulk leg, as in the construction of the bowtie lattice of Fig.~\ref{fig:bowtie}.
But in this case, we are free to distribute the lollipop factors between the two new vertices in any way we like, so we can send the $R$ boundary legs and the lollipops in $\mathfrak{l}_R$ to one half of the bowtie, and the $\overline{R}$ boundary legs and the lollipops in $\mathfrak{l}_{\overline{R}}$ to the other. The result is the bowtie lattice of Fig.~\ref{fig:bowtie} decorated with lollipops on either side, which we will call $\Lambda_b[\gamma]$, and the corner leg is defined as before. Gauge invariant operators with support on the corner leg still define boundary anchored line operators, as in the matter-free case, so the partition \eqref{eq:matterpartition} can uniquely (up to homotopy relative to the matter legs) be associated with a cut $\gamma$ that runs through the network. 

With the tessellation $\Lambda_b[\gamma]$ of the disk in hand, the construction  proceeds in the same way as the matter-free case in  Sec.~\ref{sec:factorization}. Let $\vec{m}_R$ denote a basis label for the (kinematic) Hilbert space $\bigotimes_{\ell \in \mathfrak{l}_R} \Hmatt{\ell}$ for the $\mathfrak{l}_R$ matter legs, and similarly for $\vec{m}_{\overline{R}}$. A state in the $\Lambda_b[\gamma]$ presentation has an expansion
\begin{align}
	\dket{\psi} = \int d[\vec{g}_R,\vec{g}_{\overline{R}},h] \sum_{\vec{m}_R, \vec{m}_{\overline{R}}} \psi(\vec{g}_{\overline{R}}, \vec{m}_{\overline{R}}; h ; \vec{m}_R, \vec{g}_R) \, \dket{\vec{g}_{\overline{R}}, \vec{m}_{\overline{R}}; h ; \vec{m}_R,\vec{g}_R} \,,
\end{align}
where $h$ is the group element on the corner leg, as before. We define the factorization map along the cut $\gamma$ defining the partition \eqref{eq:matterpartition} by
\begin{align}
	V[\gamma]\dket{\psi} = \int d[\vec{g}_R,\vec{g}_{\overline{R}}, h, k ] \sum_{\vec{m}_R, \vec{m}_{\overline{R}}} \psi(\vec{g}_{\overline{R}},\vec{m}_{\overline{R}}; kh; \vec{m}_R, \vec{g}_R) \, \dket{\vec{g}_{\overline{R}},\vec{m}_{\overline{R}}, k}_{\overline{R}} \, \dket{h,\vec{m}_R,\vec{g}_R}_R\,. \label{eq:Vgammadef}
\end{align}
Comparing with \eqref{eq:Vgroupbasisfactorized}, the only difference is that the matter labels come along for the ride. Crucially, $V[\gamma]$ only cuts the corner leg, just as $V$ did. 

It may seem surprising that dressing the matter to the gauge field does not force us to introduce additional edge modes at the cut. The reason is that the edge modes of Sec.~\ref{sec:factorization} were introduced to lift the Gauss law at the single bulk vertex that the cut exposes. Because $\gamma$ does not intersect any lollipop factors, the bulk vertex $v$ and bulk plaquette $p$ that dress a given matter leg both lie entirely within $\Sigma_R$ or entirely within $\Sigma_{\overline{R}}$. Furthermore, if a potential cut $\gamma$ does intersect the bulk vertex or plaquette that a matter leg is dressed to, we can always use moves 1 and 2 to separate the matter leg to be fully on either side of $\gamma$ before we factorize the Hilbert space. Thus, the electric constraint $\Pi_A^{(v)}$ and the magnetic constraint $\Pi_B^{(p)}$ which entangle that matter leg with the in-plane legs are imposed inside a single extended Hilbert space, and are never split by the factorization.

From there, $V[\gamma]$ acts by relaxing the Gauss law at the corner leg, which introduces  observer edge modes at the cut. These observer edge modes only depend on the corner leg, and the minimal set of edge modes required to factorize the Hilbert space is insensitive to the state on either side of the cut. In other words, the edge modes introduced by $V[\gamma]$ are the same as those introduced by $V$, and they remain the minimal set that factorization demands, independently of how many matter legs the tensor network contains. Because the observer edge modes are assigned the character function $\ket{\chi_\pi}'$ within each sector, which in particular is bipartite entangled, we will call $V[\gamma]$ the bipartite factorization map, in anticipation of another factorization map relevant for the matter legs we will use in Sec.~\ref{sec:entropy_matter_random} which endows the observer edge modes with a richer entanglement structure.

\subsection{The extended Hilbert space and the reduced state} The extended Hilbert space $\Hext{R}[\gamma]$ of $R \sqcup \mathfrak{l}_R$ is now built from the $R$ boundary legs, the corner leg, and the matter legs in $\mathfrak{l}_R$. Its form under decomposition into superselection sectors is unchanged,
\begin{align}
	\Hext{R}[\gamma] &= \int_{\widehat{G}}^{\oplus} d\mu(\pi) \, \Ha_R(\pi;\gamma) \otimes V_\pi \,, \label{eq:Hextgamma}\\
	\Ha_R(\pi;\gamma) &= \int_{\widehat{G}}^{\oplus} d\mu(\vec{\omega}) \, V_{\vec{\omega}} \otimes \Pi\left[V_{\vec{\omega}}^* \otimes V_\pi^* \otimes \bigotimes_{\ell \in \mathfrak{l}_R} \Hmatt{\ell}\right] \,, \label{eq:HRpigamma}
\end{align}
and similarly for $\overline{R}$. Here, $\Pi$ is the total constraint operator of Sec.~\ref{sec:themodel}, built from the electric and magnetic operators including their action on the matter legs. This notation is shorthand for the co-invariant Hilbert space generated by the quotient of $\Pi$-null states, not by literally acting $\Pi$ on the kinematic Hilbert space in brackets. When $\mathfrak{l}_R$ is empty, this reduces to the matter-free expression of Sec.~\ref{sec:themodel}, as it must. 

Again, note that the matter enters \eqref{eq:Hextgamma} only inside the factor $\Ha_R(\pi;\gamma)$. The observer edge mode factor $V_\pi$, which is the entire content of the observer edge modes, is untouched.
Because of this, the derivation of the reduced state in Sec.~\ref{sec:reducedstate} goes through without modification. The partial trace of the observer edge modes still produces the Plancherel normalized delta function and the projector $\Pi_\pi$ of \eqref{eq:Piproj}, since these came from the character function $\ket{\chi_\pi}'$ on the corner leg, and we find that
\begin{align}
	\rho_R[\gamma] = \Tr_{\Hext{\overline{R}}[\gamma]}\left[V[\gamma] \dket{\psi}\dbra{\psi} V[\gamma]^\dagger\right] = \int d\mu(\pi) \, q^\gamma_\pi \, \rho^\gamma_\pi \otimes \Pi_\pi \,. \label{eq:rhoRgamma}
\end{align}
The coefficients $q^\gamma_\pi$ and the partial density matrices $\rho^\gamma_\pi$ are defined exactly as $q_\pi$ and $\rho_\pi$ in \eqref{eq:rhoRfirst}, with $V \to V[\gamma]$. In other words, $\rho^\gamma_\pi$ is a normalized density matrix on $\Ha_R(\pi;\gamma)$, and $\int d\mu(\pi) \, q^\gamma_\pi = 1$ whenever $\dket{\psi}$ is normalized in the co-invariant inner product. 

We now turn to the algebras, traces, and operator valued weights of Sec.~\ref{sec:algebras}. Let $\mathcal{A}_R[\gamma]$ be the algebra of operators which act non-trivially on the $R$ boundary vertices and on the matter legs in $\mathfrak{l}_R$, trivially on everything else, and completed into a von Neumann algebra using the double commutant. A general element of $\mathcal{A}_R[\gamma]$ can be written as
\begin{align}
	\mathcal{O}_{\mathcal{A}_R[\gamma]} = \int d\mu(\pi) \, \mathcal{O}(\pi) \otimes \Pi_\pi \,,
\end{align}
with $\mathcal{O}(\pi)$ now an operator on $\Ha_R(\pi;\gamma)$. The reduced state \eqref{eq:rhoRgamma} is an element of  $\mathcal{A}_R[\gamma]$. The center of this algebra is unchanged, by the analog of \eqref{eq:fixedareadecompositionintro} with $\Ha_R(\pi)$ replaced by $\Ha_R(\pi;\gamma)$. The reason is that the central operators $\hat{C}$ of \eqref{eq:Crepbasis} act by convolution on the corner leg alone, so $\mathcal{Z}(\mathcal{A}_R[\gamma]) \cong L^\infty(\widehat{G})$ exactly as before.\footnote{Of course, this assumes that each individual matter leg carries an algebra with a trivial center, but this is not important for the discussion at hand.}

Consequently, we can use the same arguments as in  Sec.~\ref{sec:algebras} to fix the traces of $\mathcal{A}_R[\gamma]$ and $\mathcal{Z}(\mathcal{A}_R[\gamma])$ to be the same as they were in the matter-free case: they are \eqref{eq:physicaltrace} and \eqref{eq:centertracemicro}, with the sector-wise trace taken over $\Ha_R(\pi;\gamma)$ rather than $\Ha_R(\pi)$, and likewise the operator valued weights $\mathcal{E}$ and $\mathcal{F}$ are \eqref{eq:ovwdef} and \eqref{eq:AtoZ} with the same replacement. The only change from the matter-free case is which Hilbert space the sector-wise trace is taken over.

\subsection{The entropy along a fixed cut in the presence of matter} \label{sec:entropy_fixedcut}

The area operator associated with the cut $\gamma$ is
\begin{align}
	\hat{A}_\gamma = \int d\mu(\pi) \, \ln(\mu(\pi)) \, \Id_{\Ha_R(\pi;\gamma)} \otimes \Pi_\pi \,, \label{eq:areaoperatorgamma}
\end{align}
which is \eqref{eq:areaoperator} with the sector-wise identity taken on $\Ha_R(\pi;\gamma)$. Thus, the operator $\hat{A}_\gamma$ depends on the cut $\gamma$ only through the homotopy class of the curve it acts along, and it remains state independent. In particular, the spectrum of $\hat{A}_\gamma$ is fixed by the gauge group $G$ and the topological boundary condition at the cut, exactly as in the matter-free case, although of course its expectation value will depend on the state.

Note that $\hat{A}_\gamma$ is central in $\mathcal{A}_R[\gamma]$, as in Sec.~\ref{sec:areaop_nomatter}. Therefore, we can think of $\hat{A}_\gamma$ as an operator that acts only on the corner, and so it is unitarily equivalent to a line operator running along $\gamma$ in the full physical Hilbert space $\Ha_{\mathrm{phys}}(\Sigma)$, as in Fig.~\ref{fig:areaop}. Because it is a topological operator, deforming $\gamma$ without crossing a matter leg does not change it. Thus, $\hat{A}_\gamma$ depends on $\gamma$ only through its homotopy class relative to the matter insertions, which is to say only through the partition $\mathfrak{l} = \mathfrak{l}_R \sqcup \mathfrak{l}_{\overline{R}}$. Two cuts which enclose the same matter legs therefore define the same area operator. However, the cuts which partition the matter legs differently define genuinely different area operators, because the identity operator $\Id_{\Ha_R(\pi;\gamma)} $ does depend on the cut. In particular, following similar arguments to those in \cite{Akers:2024wab}, area operators defined on intersecting curves do not commute. 

Finally, the entropy computation of \cite{Balasubramanian:2026ymu} that we reviewed in Sec.~\ref{sec:entropy_nomatter} may be repeated word for word using the modified ingredients $\rho_R[\gamma]$, $\Tr_{\mathcal{A}_R[\gamma]}$ and $\mathcal{F}$. We find that
\begin{align}
	S(\rho_R[\gamma]) =  H[\,p^\gamma_\pi] + \langle \hat{A}_\gamma \rangle_\rho + S_{\mathrm{bulk}}(\rho_{\mathrm{int}(R \cup \gamma)}) \,, \qquad p^\gamma_\pi = \mu(\pi) \, q^\gamma_\pi \,, \label{eq:Sgamma}
\end{align}
where $\mathrm{int}(R \cup \gamma) = \Sigma_R$ is the region bounded by the $R$ boundary vertices and the cut. Furthermore, the bulk entropy is given by
\begin{align}
	S_{\mathrm{bulk}}(\rho_{\mathrm{int}(R\cup\gamma)}) = \int d\mu(\pi) \, q^\gamma_\pi S(\rho^\gamma_\pi)\,,
\end{align}
which is \eqref{eq:Sbulk_nomatter} with $q_\pi, \rho_\pi \to q^\gamma_\pi, \rho^\gamma_\pi$, and now includes the entropy of the matter legs in $\mathfrak{l}_R$. This is why we call it the bulk entropy: it is the entropy of the state on the bulk region $\Sigma_R = \mathrm{int}(R \cup \gamma)$ carved out by the cut, matter legs included, as opposed to the other two terms, which are carried by the cut itself. The three terms have the same origin and the same interpretation as they did in Sec.~\ref{sec:entropy_nomatter}.

\section{Entropy with random matter}
\label{sec:entropy_matter_random}

In the previous section, we obtained an entropy formula for a boundary subregion $R$ as a function of a fixed cut $\gamma$ through the tensor network, but nothing distinguished one cut $\gamma$ from another. However, from a gravitational perspective, the cut $\gamma$ appearing in the holographic entropy formula is not a free parameter: it is determined by extremizing, and then minimizing, the generalized entropy, as we review in Sec.~\ref{sec:continuum}. Our goal in this section is to explain how such a minimization can arise in our model. We will find that one way to achieve the minimization over cuts that we expect from gravity is to take the state on the matter legs to be  ``random'' in a precise sense we define below.  We defer potential gravitational interpretations of this assumption to the discussion in Sec.~\ref{sec:discussion}.

Traditionally, holographic tensor networks are interpreted as a toy model of the bulk-to-boundary dictionary: we fix a state of the bulk matter, feed it into the network, and read off a state of the boundary. We would like to do the same here. Given a state $\ket{T}$ of the matter legs, we want to contract it into the tensor network and then compute the entropy of a boundary subregion $R$ in the resulting state. If the physical Hilbert space factorized between the matter legs and the in-plane legs, this would be the contraction
\begin{align}
	\ket{\psi[T]} \,``\!=\!"\, \braket{T}{\psi} \,, \label{eq:barecontraction}
\end{align}
which would be a state of the boundary legs alone.

\subsection{The multipartite factorization map} \label{sec:multipartite}

As written, \eqref{eq:barecontraction} is not well-defined, for the same reasons discussed in  Sec.~\ref{sec:factorization}. Taking a partial inner product against $\ket{T}$ presupposes that the Hilbert space splits as a tensor product with the matter legs as one factor. In fact, the physical Hilbert space descending from \eqref{eq:HLambdamatter} does \emph{not} split this way. Recall from Sec.~\ref{sec:themodel} that each matter leg is dressed to a bulk vertex and a bulk plaquette, so that the electric and magnetic constraints act on the matter and the gauge field together. The quotient by null states then mixes the two, and there is no factor of the physical Hilbert space that belongs to the matter alone for $\bra{T}$ to act on.

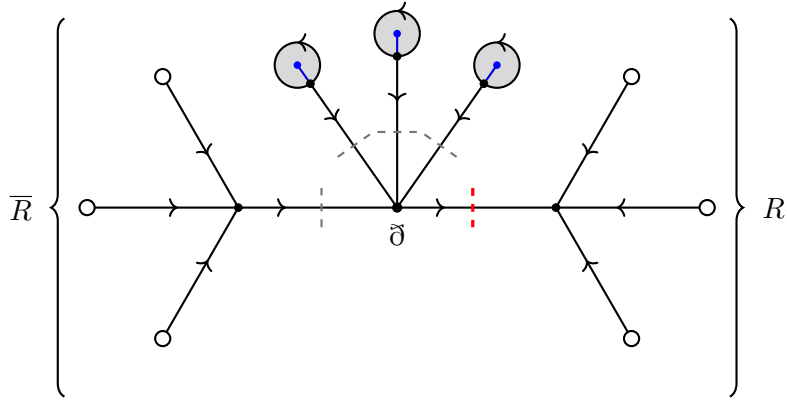
\begin{figure}
	\centering
	\begin{tikzpicture}[scale=2]
		\def\hh{1.05}      
		\def\rr{1.15}      
		\def\rc{0.5}       
		\def\cl{0.13}      
		
		\foreach \xx in {120,180,240}{
			\draw[thick,->-=0.6] ({-\hh+cos(\xx)},{sin(\xx)}) -- (-\hh,0);
			\filldraw[thick,fill=white] ({-\hh+cos(\xx)},{sin(\xx)}) circle (0.05);
		}
		\foreach \xx in {-60,0,60}{
			\draw[thick,->-=0.6] ({\hh+cos(\xx)},{sin(\xx)}) -- (\hh,0);
			\filldraw[thick,fill=white] ({\hh+cos(\xx)},{sin(\xx)}) circle (0.05);
		}
		\draw[thick,->-=0.3] (-\hh,0) -- (0,0);
		\draw[thick,->-=0.3] (0,0)--(\hh,0);
		\foreach \aa in {55,125,90}{
			\draw[thick,->-=0.3] ({cos(\aa)},{sin(\aa)}) -- (0,0);
			\draw[blue,thick] ({cos(\aa)},{sin(\aa)}) -- ({\rr*cos(\aa)},{\rr*sin(\aa)});
			\filldraw ({cos(\aa)},{sin(\aa)}) circle (0.025);
			\filldraw[thick,fill opacity=0.15,->-=0.25] ({\rr*cos(\aa)},{\rr*sin(\aa)}) circle ({\rr-1});
			\fill[blue] ({\rr*cos(\aa)},{\rr*sin(\aa)}) circle (0.025);
		}
		\filldraw (-\hh,0) circle (0.025);
		\filldraw (\hh,0) circle (0.025);
		\filldraw (0,0) circle (0.03);
		\node[anchor=north] at (0,-0.05) {$\eth$};
		\foreach \aa in {180,55,125,90}{
			\draw[black!55,thick,dashed]
			({\rc*cos(\aa)+\cl*cos(\aa+90)},{\rc*sin(\aa)+\cl*sin(\aa+90)}) --
			({\rc*cos(\aa)-\cl*cos(\aa+90)},{\rc*sin(\aa)-\cl*sin(\aa+90)});
		}
		\draw[red,very thick,dashed] ({\rc},{\cl}) -- ({\rc},{-\cl});
		\draw [thick,decorate,decoration={brace,amplitude=5pt}]
		({-\hh-1-0.15},-1.25) -- ({-\hh-1-0.15},1.25) node[midway,xshift=-1.5em]{$\overline{R}$};
		\draw [thick,decorate,decoration={brace,amplitude=5pt,mirror}]
		({\hh+1+0.15},-1.25) -- ({\hh+1+0.15},1.25) node[midway,xshift=1.5em]{$R$};
	\end{tikzpicture}
	\caption{The double bowtie lattice, drawn for a disk with $n=6$ boundary vertices and $m=3$ matter legs. The $R$ boundary legs feed into the right bulk vertex, the $\overline{R}$ boundary legs feed into the left one, and all $m$ lollipops attach to the corner vertex $\eth$ in the middle. As in Fig.~\ref{fig:reduced}, boundary vertices are white, matter legs are blue, and the plaquette of each lollipop is shaded gray. The factorization map $V_{\mathrm{full}}$ cuts all $m+2$ legs meeting at $\eth$, indicated by the dashed marks. The cut drawn in red is the one which defines $\Hext{R}$, and no matter leg lies on its $R$ side.}
	\label{fig:doublebowtie}
\end{figure}

The resolution to this obstruction is the same as before: we first factorize the matter legs from the physical Hilbert space, and then contract the matter into the state on the factorized Hilbert space. To define a factorization map which separates every matter leg, we begin with the reduced lattice of Fig.~\ref{fig:reduced}, in which $n$ boundary legs and $m = |\mathfrak{l}\,|$ lollipops meet at a single bulk vertex. Applying move 1 twice (Fig.~\ref{fig:move1}), we split the bulk vertex into the ``double bowtie lattice'' shown in Fig.~\ref{fig:doublebowtie}. The resulting lattice has three bulk vertices: one into which the $R$ boundary legs feed, one into which the $\overline{R}$ boundary legs feed, and a third carrying the two new legs together with all $m$ bulk legs of the lollipop factors. We call these the $R$ vertex, the $\overline{R}$ vertex, and the corner vertex $\eth$, respectively.

We now factorize the physical Hilbert space by cutting all $m+2$ legs which meet at the corner vertex $\eth$. This defines a map
\begin{align}
	V_{\mathrm{full}} : \Ha_{\mathrm{phys}}(\Sigma) \longrightarrow \Hext{R} \otimes \Hext{\overline{R}} \otimes \bigotimes_{\ell \in \mathfrak{l}} \Hext{\ell} \,, \label{eq:Vfull}
\end{align}
where each factor is built exactly as in Sec.~\ref{sec:factorization}: $\Hext{R}$ is spanned by the $R$ boundary legs together with one corner leg, $\Hext{\ell}$ is spanned by the matter leg $\ell$ together with one corner leg, and every cut supplies its own gauge fixing and observer edge modes. Note that $\Hext{R}$ contains no matter legs at all: it is exactly the same as extended Hilbert space we defined in the matter-free case in Sec.~\ref{sec:factorization} (see also \eqref{eq:Hextdecomp}). 

To explicitly write down the map $V_{\mathrm{full}}$, we need a convenient basis in which to compute its matrix elements. For the moment we set aside questions of convergence and normalizability of $V_{\mathrm{full}}$, returning to them when we discuss the renormalized trace below. Label the $m+2$ legs cut at $\eth$ by an index $i$, running over $R$, $\overline{R}$, and the $m$ lollipops, and let $\ket{\psi_i ; \pi_i, a_i b_i}$ denote a basis for the kinematic Hilbert space of the $i$th piece. Here $\psi_i$ is a basis label for everything in that piece except its corner leg, while $\ket{\pi_i , a_i b_i}$ is a representation basis state \eqref{eq:peterweylnoncompact} for the corner leg itself. We write
\begin{align}
	\dket{\vec{\psi} ; \vec{\pi} , \vec{a} \vec{b}} \equiv \bigotimes_i \ket{\psi_i ; \pi_i , a_i b_i} / \sim
\end{align}
for the corresponding state of the physical Hilbert space, where the tilde denotes the equivalence class modulo null states that defines a co-invariant state, as in \eqref{eq:electricnullstates}. 

If we orient every leg meeting at $\eth$ to point inwards, then by the convention fixed in Sec.~\ref{sec:themodel} the $V_{\pi_i}$ factor of each corner leg sits at its outer end and the $V_{\pi_i}^*$ factor sits at $\eth$. In other words, each $a_i$ index lives at the $R$, $\overline{R}$ or lollipop vertex, and all of the $b_i$ indices live together at the corner vertex. The Gauss law at $\eth$ therefore acts only on the $\vec{b}$ indices, at fixed $\vec{\psi}$, $\vec{\pi}$, and $\vec{a}$, and the states which survive the null quotient there span the subspace
\begin{align}
	\mathcal{I}_{\vec{\pi}} \equiv \Pi_A\left[V_{\vec{\pi}}^*\right] \,. \label{eq:intertwinerspace}
\end{align}
As explained below \eqref{eq:Hphys}, when $G$ is compact, $\mathcal{I}_{\vec{\pi}}$ is the space of intertwiners from $V_{\vec{\pi}}$ to the trivial representation. When $G$ is non-compact, the corresponding states fail to be normalizable in the kinematic inner product, but they are normalizable in the co-invariant inner product, so we may continue to refer to them as intertwiners without ambiguity.

Let $\Gamma_a(\vec{\pi})$ label an orthonormal basis of $\mathcal{I}_{\vec{\pi}}$. Trading the $\vec{b}$ indices for this label gives a basis
\begin{align}
	\dket{\vec{\psi}_{\vec{\pi}} ; \Gamma_a(\vec{\pi})} \label{eq:unfactorizedstatewithmatter}
\end{align}
for the physical Hilbert space in the double bowtie presentation, where $\vec{\psi}_{\vec{\pi}}$ collects the data $(\psi_i, \pi_i, a_i)$ of the individual pieces. This is the more faithful way to label the states in the physical Hilbert space, since it makes manifest that the only data at the corner vertex is a choice of fusion channel from the representations $\vec{\pi}$ to the trivial representation.

As we discussed below \eqref{eq:Vgroupbasis}, the bipartite factorization map $V[\gamma]$ (of which the factorization map $V$ of Sec.~\ref{sec:factorization} is a special case) introduced one auxiliary leg in the state $\ket{e}$ and then applied the Gauss law projector $[\Pi_A]_{\eth}$ at the corner vertex, gluing that leg to the corner leg of the bowtie lattice. In the representation basis \eqref{eq:Vrepresentationbasis} the effect was to insert the unnormalized character state $\ket{\chi_\pi}'$ into each superselection sector. The choice was forced by demanding that the factorization map introduce no defect at the cut: any other wave function produces a relative kink at the corner when the two halves are reglued, which we interpreted as an electric flux through $\eth$.

The language of \eqref{eq:intertwinerspace} explains why this character $\ket{\chi_\pi}'$ appeared in the bipartite factorization map, and also what should replace it. With only two legs at the corner, $\mathcal{I}_{\vec\pi}$ is one dimensional by Schur's lemma, and is spanned by the character function. Once the matter legs are present, there are $m+2$ legs at the corner and $\mathcal{I}_{\vec{\pi}}$ is no longer one dimensional. Therefore, a single character is not enough data to specify the state of the observer edge modes. What is enough, however, is the intertwiner label itself. We define the factorization map to be
\begin{align}
	V_{\mathrm{full}} \dket{\vec{\psi}_{\vec{\pi}} ; \Gamma_a(\vec{\pi})} = \left[\bigotimes_i \dket{\psi_i ; \pi_i}_{\Hext{i}(\pi_i)}\right] \ket{\Gamma_a(\vec{\pi})}' \,, \label{eq:multiparty_factorization}
\end{align}
where $\dket{\psi_i ; \pi_i}_{\Hext{i}(\pi_i)}$ is the state of the $i$th piece in the sector $\pi_i$, carrying its own gauge fixing edge modes, and $\ket{\Gamma(\vec{\pi})}'$ is the state of the $m+2$ observer edge modes created by the cuts. 
Essentially, the corner vertex itself acts as the observer edge mode, and its wave function is multipartite entangled between its $m+2$ tensor factors in each sector.\footnote{That $\ket{\Gamma(\vec{\pi})}' $ is multiparty entangled between the $m+2$ observer edge modes can be demonstrated using \eqref{eq:intertwinerpartialtrace}, which shows that every observer is maximally mixed after tracing out the complementary set. This is not possible if $\ket{\Gamma(\vec{\pi})}' $ only contained bipartite entanglement.} 
The precise multipartite entanglement structure of the observer edge modes $| \Gamma_a(\vec{\pi})\rangle'$ will depend on the representations $\vec{\pi}$. It would be interesting to determine the qualitative kinds of multipartite entanglement achievable by these tensor networks, perhaps using the methods in \cite{Dutta:2019gen,Akers:2021pvd,Gadde:2022cqi,Harper:2020wad,Gadde:2023zzj,Balasubramanian:2024ysu,Gadde:2024taa,Balasubramanian:2025hxg,Iizuka:2025ioc,Balasubramanian:2025kaf,Iizuka:2025caq,Cummings:2025zfe,Balasubramanian:2026chr}.

As with the characters in the bipartite factorization map, the prime indicates that we have not fixed the normalization of the intertwiners $\ket{\Gamma_a(\vec{\pi})}'$, a point to which we return below. Similar to the discussion around \eqref{eq:fakesplitdelta}--\eqref{eq:partialtracecharacters}, we take each $\dket{\psi_i ; \pi_i}_{\Hext{i}(\pi_i)}$ to be an orthonormal basis of $\Hext{i}(\pi_i)$, and let the observer edge mode wave function carry all the delta functions needed to ensure $V_{\mathrm{full}}$ is an isometry after we renormalize the trace. Note that setting $m=0$ recovers the matter-free case, with $\ket{\Gamma(\vec{\pi})}'$ reducing to $\ket{\chi_\pi}'$. Because the factorization map $V_{\mathrm{full}}$ assigns a multipartite entangled state to the observer edge modes, rather than the bipartite state $\ket{\chi_\pi}'$ as $V[\gamma]$ does, we call $V_{\mathrm{full}}$ the multipartite factorization map.

As in Sec.~\ref{sec:Arcenter}, \eqref{eq:multiparty_factorization} is not a unique factorization map. Let $\hat{O}$ be any linear operator from $\mathcal{I}_{\vec{\pi}}$ to itself. Then
\begin{align}
	V^{O}_{\mathrm{full}} \dket{\vec{\psi}_{\vec{\pi}} ; \Gamma(\vec{\pi})} = \left[\bigotimes_i \dket{\psi_i ; \pi_i}_{\Hext{i}(\pi_i)}\right] \hat{O}\ket{\Gamma(\vec{\pi})}'
\end{align}
is also a valid factorization map. Because $\hat{O}$ commutes with the gauge constraints (as one can check explicitly), it is a physical operator on the unfactorized Hilbert space and we can think of it as a line operator: either boundary anchored between $R$ and $\overline{R}$, as in Sec.~\ref{sec:Arcenter}, or a closed loop encircling some subset of the matter legs. When $m=0$ the space $\mathcal{I}_{\vec{\pi}}$ is one dimensional, so $\hat{O}$ reduces to multiplication by a number in each sector, which is exactly the central operator $\hat{C}$ of \eqref{eq:Crepbasis}.\footnote{While this is not relevant for our purposes, we note that unlike the bipartite case, the algebra of operators $\hat{O}$ is not abelian. } The freedom parameterized by $\hat{O}$ is therefore the same freedom we encountered in the matter-free case, now enlarged by the loops which the matter legs make available, as well as the different ways boundary anchored line operators can interleave the matter legs.
Just as before, this freedom is fixed by demanding that the factorization map does not insert any defects into the state before we cut it open. Only the choice $\hat{O} = \Id$ for every sector does so, and \eqref{eq:multiparty_factorization} is the resulting factorization map. We use this factorization map for the rest of this paper. 

\paragraph{The matter-dependent boundary state}

With \eqref{eq:multiparty_factorization} in hand, we can define the contraction we wanted at the start, because the matter legs now sit in factors of an honest tensor product. Let $\dket{T}= \bigotimes_{\ell} \dket{T_\ell}_{\Hext{\ell}}$ be a fixed state of the extended Hilbert space of all the lollipop factors.\footnote{We could also consider superpositions of these product states, i.e., entangled states of the matter legs.} Also, for brevity of notation, we let $\Tr_R$, $\Tr_{\overline{R}}$, and $\Tr_{\mathfrak{l}}$ denote the unique traces on the extended Hilbert spaces $\Hext{R}$, $\Hext{\overline{R}}$, and $\Hext{\mathfrak{l}}$. Then
\begin{align}
	\dket{\psi[T]}_{R\overline{R}}  = \left(\Id_{R\overline{R}} \otimes \dbra{T} \right) V_{\mathrm{full}} \dket{\psi} \label{eq:psiT}
\end{align}
is a vector in $\Hext{R} \otimes \Hext{\overline{R}}$, and it depends on the matter state we contract into the network. This is the state whose bipartite entropy we want to compute between $R$ and $\overline{R}$, and it is the well-defined version of the expression \eqref{eq:barecontraction} we started from. 

What does conditioning on $\dket{T}$ mean physically? A generic $\dket{T_\ell}$ is a superposition over the representations carried by the corner leg of the lollipop $\ell$, and by the discussion above it inserts exactly the kind of kink at the cut between $R$ and $\overline{R}$ which the defect-free condition forbids. This is not a contradiction. The defect-free condition was imposed so that the factorization map itself introduces no structure, and it continues to hold: the map is \eqref{eq:multiparty_factorization}, which by construction inserts no defect. What $\dket{T}$ does afterwards is not part of the factorization map, but a feature of the matter state itself. A defect at a matter cut is the charge of the matter, and conditioning on it is the point of contracting a matter state into the network.  

\subsubsection{The physical trace is unchanged}

Like the bipartite factorization map $V$ of Sec.~\ref{sec:factorization}, the multipartite factorization map $V_{\mathrm{full}}$ is only \emph{projectively} an isometry between the physical and factorized Hilbert spaces, and for the same reason: it relaxes a Gauss law, and the operator which imposes that Gauss law does not square to itself when $G$ is non-compact.

To see this explicitly, consider the case when $G$ is a compact group with volume $\mathrm{Vol}(G)$. Let $g_i$ denote the group element on the $i$th of the $m+2$ legs meeting at the corner vertex $\eth$, with every such leg oriented inwards. We  suppress the boundary and matter labels, which are spectators in what follows. A gauge transformation at $\eth$ acts by a simultaneous right multiplication $g_i \to g_i k^{-1}$ on all $m+2$ legs. Writing $x_i = g_i^{-1} g'_i$ and computing the overlap of two such states, the co-invariant inner product in the double bowtie presentation is
\begin{align}
	\dbraket{\vec{g}}{\vec{g}\,'} = \int_G dk \prod_{i=1}^{m+2} \delta\!\left(g_i^{-1} g'_i k^{-1}\right) = \prod_{i \geq 2} \delta\!\left(x_i x_1^{-1}\right)\,, \label{eq:doublebowtieinner}
\end{align}
while the factorization map is the $m+2$ leg version of \eqref{eq:Vgroupbasis},
\begin{align}
	V_{\mathrm{full}} \dket{\vec{g}} = \int_G dk \bigotimes_{i=1}^{m+2} \ket{g_i k}_i \,. \label{eq:Vfullgroupbasis}
\end{align}
In words, the factorized state is the average over the corner symmetry of the product state, which is the statement that $V_{\mathrm{full}} = [\Pi_A]_{\eth}$ acting on the $m+2$ separated legs, exactly as in \eqref{eq:Vgroupbasis}. Then we can see that
\begin{align}
	\dbra{\vec{g}} V_{\mathrm{full}}^\dagger V_{\mathrm{full}} \dket{\vec{g}\,'}
	= \int d[k,k'] \prod_{i} \delta\!\left(k^{-1} x_i k'\right)
	= \left(\int dk \right)\prod_{i \geq 2} \delta\!\left(x_i x_1^{-1}\right)
	= \mathrm{Vol}(G) \, \dbraket{\vec{g}}{\vec{g}\,'} \,, 
\end{align}
where the $i=1$ delta function fixes $k' = x_1^{-1}k$, and the remaining $m+1$ delta functions are then insensitive to $k$ because $\delta$ is a class function. So
\begin{align}
	V_{\mathrm{full}}^\dagger V_{\mathrm{full}} = \mathrm{Vol}(G)\, \Id_{\mathrm{phys}} \,,\label{eq:Vfullprojective}
\end{align}
which shows that the multipartite factorization map $V_{\mathrm{full}}$ is also a projective isometry, just like the bipartite one $V$. 

As in the matter-free case, the constant in \eqref{eq:Vfullprojective} is state independent, so this divergence is not a property of the state $\dket{\psi}$ we are factorizing, but of the normalization we have implicitly given the observer edge modes wave function in each sector (the intertwiner $\ket{\Gamma(\vec{\pi})}'$). Therefore, the volume divergence of \eqref{eq:Vfullprojective} must come from the norm of the intertwiner state $\ket{\Gamma(\vec{\pi})}'$. This is the direct generalization of the observation below \eqref{eq:character_delta_overlap} to the multiparty setting, where the divergence was traced to the norm of the character state $\ket{\chi_\pi}'$.

To see that this factorization map is renormalized the same way as in Sec.~\ref{sec:algebras}, we compute 
\begin{align}
	\rho_R[\gamma_0] = \Tr_{\overline{R}\,\mathfrak{l}}\left[ V_{\mathrm{full}}\dket{\psi}\dbra{\psi}V_{\mathrm{full}}^\dagger \right] \,, \label{eq:rhoRgamma0}
\end{align}
where $\Hext{\mathfrak{l}} = \bigotimes_{\ell \in \mathfrak{l}} \Hext{\ell}$, the notation $\rho_R[\gamma_0]$ is the same as in \eqref{eq:rhoRgamma}, and $\gamma_0$ is the cut which contains no matter legs at all, i.e., the red cut of Fig.~\ref{fig:doublebowtie}. 
In general, if we do not specify a particular cut $\gamma$ in the argument of $\rho_R[\gamma]$, we mean $\rho_R[\gamma_0]$.
We will now show that the same renormalized trace $\Tr_{\mathcal{A}_R}$ continues to normalize the state. To do so, let $\omega$ denote the representation on the $R$ corner leg and $\vec{\pi}$ the representations on the remaining $m+1$ legs, so that $\ket{\Gamma_a(\omega,\vec{\pi})}'$ is a vector in $V_\omega \otimes V_{\vec{\pi}}$ and the label $a$ runs over an orthonormal basis of $\mathcal{I}_{(\omega,\vec{\pi})}$. 
When we take the partial trace over the $V_{\vec{\pi}}$ observer edge modes, we find that
\begin{align}
	\tr_{V_{\vec{\pi}}}\left[\ket{\Gamma_a(\omega,\vec{\pi})}'\langle\Gamma_b(\eta,\vec{\pi}\,')\vert'\right] = \delta(\omega,\eta) \, \delta(\vec{\pi},\vec{\pi}\,') \, \delta_{ab} \, \Pi_\omega \,, \label{eq:intertwinerpartialtrace}
\end{align}
where $\delta(\vec{\pi},\vec{\pi}\,') = \prod_i \delta(\pi_i,\pi'_i)$, and every delta function over representations is normalized with respect to the Plancherel measure.

There are two important structural features of \eqref{eq:intertwinerpartialtrace}.	The first is that the right hand side is proportional to $\Pi_\omega$, which can be derived using Schur's lemma. More explicitly, the intertwiner $\ket{\Gamma_a(\omega,\vec{\pi})}'$ is by construction invariant under the diagonal action of $G$ on $V_\omega \otimes V_{\vec{\pi}}$, so once the delta functions have set the representations equal, the partial trace on the left hand side is an operator on $V_\omega$ which commutes with the action of $G$ on $V_\omega$.\footnote{This follows by interpreting $U_\omega(g)$, acting on the left of the operator in \eqref{eq:intertwinerpartialtrace}, as the operator $U_\omega(g) \otimes \Id_{\vec{\pi}}$ within the partial trace. Invariance of $\ket{\Gamma_a}$ then lets us trade this action for $U_{\vec{\pi}}(g^{-1})$, and cyclicity of the partial trace lets us place this operator on the right of the ketbra instead. Finally, we use the reverse of the first move to trade $U_{\vec{\pi}}(g^{-1})$ with $U_{\omega}(g)$, so the operator in \eqref{eq:intertwinerpartialtrace} commutes with $U_\omega(g)$ for any $g \in G$.} By Schur's lemma, it is therefore proportional to the identity, which within the algebra $\mathcal{A}_R$ means proportional to $\Pi_\omega$. Furthermore, taking the remaining trace of both sides returns the overlap $\braket{\Gamma_b(\eta,\vec{\pi}\,')}{\Gamma_a(\omega,\vec{\pi})}'$, so the coefficient of $\Pi_\omega$ is simply the norm of the intertwiner state in the primed normalization, exactly as \eqref{eq:character_delta_overlap} expressed the norm of the character state through $\tr(\Pi_\pi)$.

The second important feature of \eqref{eq:intertwinerpartialtrace} is that other than the delta functions, the coefficient of $\Pi_\omega$ is $\delta_{ab}$, with no residual dependence on $\omega$ or on $\vec{\pi}$.\footnote{The delta functions arise because of our interpretation of the factorized basis for $\Hext{i}(\pi_i)$ as orthonormal for each $i$, rather than delta function normalized.} This is the content of the defect-free condition, and replacing $\delta_{ab} \to \mathcal{O}_{ab} \sim (\hat{O}^\dagger \hat{O})_{ab}$ indicates that we have picked a factorization map $V_{\mathrm{full}}^O$ instead of the defect-free factorization map \eqref{eq:multiparty_factorization}.\footnote{Demanding \eqref{eq:intertwinerpartialtrace} holds only fixes that the factorization map is of the form $V_{\mathrm{full}}^{\hat{U}}$, where $\hat{U}$ is a unitary operator within every observer edge mode sector. So the defect-free condition, strictly speaking, is a stronger assumption than is necessary to fix the trace of $\mathcal{A}_R$ (and therefore, after we also fix the trace on $\mathcal{Z}(\mathcal{A}_R)$, the area operator).} This follows from the assumption that the $\Gamma_a$ are orthonormal in the co-invariant inner product, together with the convention $\tr(\Pi_\omega) = 1$ that defines $\Tr_{\mathcal{A}_R}$. 

Recall from \eqref{eq:VCisometry} that requiring that the  factorization map be an isometry determines the trace on $\mathcal{A}_R$.
We have fixed the factorization map in \eqref{eq:multiparty_factorization}, which means we have also fixed a renormalized trace on $\mathcal{A}_R$. The fact that \eqref{eq:intertwinerpartialtrace} contains no additional $\omega$ dependence implies that the physical trace on $\mathcal{A}_R$ is the same as in the matter-free case. If it did have an additional $\omega$ dependence, e.g., of the form $|O(\omega)|^2$, the trace on $\mathcal{A}_R$ would be rescaled to $\Tr^{|O|^{-2}}_{\mathcal{A}_R}$ as in \eqref{eq:CasMeasure}, and by \eqref{eq:areaoperator} the spectrum of the area operator would be shifted by $\ln |O(\omega)|^2$. This would have implied that the spectrum of the area operator depended on the way we factorized out the matter legs. Instead, equation \eqref{eq:intertwinerpartialtrace} says that there is no such dependence, because the renormalized trace on $\mathcal{A}_R$ which makes $V_{\mathrm{full}}$ an isometry is still the physical trace \eqref{eq:physicaltrace} we have been using all along. Furthermore, the trace on the center is the microcanonical one \eqref{eq:centertracemicro}, because the edge mode structure at the corner of $\Hext{R}$ is independent of the number of matter legs we factorized from the Hilbert space. Therefore, by \eqref{eq:cfunction}, the area operator and its spectrum are unchanged. 
This is consistent with the observation of Sec.~\ref{sec:entropy_matter_norandom} that the center $\mathcal{Z}(\mathcal{A}_R) \cong L^\infty(\widehat{G})$ is unchanged by the presence of matter, since the central operators $\hat{C}$ of \eqref{eq:Crepbasis} act by convolution on the $R$ corner leg alone, and the observer edge modes corresponding to this cut do not depend on the matter legs.

Using this partial trace \eqref{eq:intertwinerpartialtrace} over the multipartite observer edge modes, one can use the same arguments as in Sec.~\ref{sec:reducedstate} to show that the reduced state takes the form
\begin{align}
	\rho_R[\gamma_0] &= \int d\mu(\pi) q^{\gamma_0}_\pi \rho^{\gamma_0}_\pi \otimes \Pi_\pi \,, \label{eq:rhoRempty_averaged}\\ 
	\Tr_{\mathcal{A}_R}[\rho_R[\gamma_0]] &= \int d\mu(\pi) q^{\gamma_0}_\pi = 1 \label{eq:rhoRempty_q_averaged}\,. 
\end{align}
For brevity, we do not display the explicit definition of $q^{\gamma_0}_\pi$ or  $\rho^{\gamma_0}_\pi$, but they are defined entirely analogously to $q_\pi$ and $\rho_\pi$ in \eqref{eq:rhoRfirst}, whose explicit expressions can be found in \cite{Balasubramanian:2026ymu}, with additional indices for the matter legs and the possible channels $\Gamma_a$ for each sector, which are the added indices of the wave function \eqref{eq:unfactorizedstatewithmatter} we are factorizing compared to the matter-free case.

\subsubsection{Comparison with the bipartite factorization map }

Note that instead of using the bipartite factorization map in Sec.~\ref{sec:entropy_matter_norandom}, we could also compute the entropy along a fixed cut $\gamma$ through the matter legs using $V_{\mathrm{full}}$ by grouping the factors $\Hext{\ell}$ with $\ell \in \mathfrak{l}_R$, together with $\Hext{R}$, and tracing out the rest of the Hilbert space. The difference between these two procedures is only which lattice we cut open with the factorization map. The multipartite factorization map cuts the $R$ corner leg and the stem of every lollipop in $\mathfrak{l}_R$. In contrast, the bipartite factorization map only cuts open the corner leg of the bowtie lattice $\Lambda_b[\gamma]$. Because we imposed a \emph{topological} boundary condition at the cut, the factorized state should be insensitive to these details near the cut $\gamma$, and the entropy computed using either factorization map should agree.

When $G$ is a finite group, it is indeed the case that these two cuts through the lattice carry unitarily equivalent Hilbert spaces, because a collection of legs terminating on a segment of a topological boundary may be fused into a tree whose single remaining edge meets the boundary \cite{kirillov2011stringnet,Kitaev:2011dxc}. 
Fusing the legs in a different order gives the same answer, so the resulting tree is well-defined. This is a property of the topological boundary condition at the cut, and not only of the fusion rules of $\mathrm{Rep}(G)$, because the legs end on the cut rather than in the bulk.\footnote{The move which fuses two neighboring legs on the boundary is the boundary $F$ move of \cite{Kitaev:2011dxc}, and what makes the result independent of the order of fusion is the mixed pentagon identity (a consistency condition for this associativity) which it satisfies, rather than the pentagon identity of $\mathrm{Rep}(G)$ alone.} The internal edges of the tree are not new degrees of freedom, because the electric constraint at each of its vertices fixes them. The single edge which remains anchored to the cut carries a representation $\omega \in \widehat{G}$, together with a choice of channel whenever the original legs on the cut can fuse into $\omega$ in more than one way. 

This choice of channel is exactly the data carried by the intertwiner label $\Gamma_a$ of \eqref{eq:multiparty_factorization}. However, the bipartite factorization map has no such label: with only two legs meeting at the corner vertex, Schur's lemma leaves a single channel, as we noted below \eqref{eq:intertwinerspace}, so the cut of $\Lambda_b[\gamma]$ carries no channel label at all. The resolution to this apparent mismatch is that the recoupling of the multipartite map to the tree-like fusion structure splits the channel $a$ into the representation $\omega$ on the corner leg of $\Lambda_b[\gamma]$, together with a choice of channel on each side of the cut. Those two remaining choices are part of the intertwiner in \eqref{eq:HRpigamma}: in the bipartite presentation, they are bulk data of $R$ and of $\overline{R}$ rather than data living on the cut. They therefore give the same reduced state \eqref{eq:rhoRgamma}, and the same entropy \eqref{eq:Sgamma}, so the two factorization maps are equivalent.

We emphasize that this argument uses that $\gamma$ is an interval. Legs ending on a closed boundary circle cannot be fused into one in the same way, since a leg may be slid all the way around the circle, which would force the labels to commute. The boundary values of a circle then form the Drinfeld center $Z(\mathrm{Rep}(G)) = D(G)$, rather than $\mathrm{Rep}(G)$ itself \cite{kirillov2011stringnet}. This is the same fact which gives the closed ribbon operators of Appendix~\ref{app:ribbons} a flux in addition to a charge, while a boundary anchored cut sees only the representation label $\pi$. For a boundary anchored cut, then, the choice between $V[\gamma]$ and $V_{\mathrm{full}}$ is a choice of presentation of the cut itself, and nothing more. 

We add, however, that these results are established rigorously for categories with finitely many simple objects, whereas for a general transformable group $G$, the fusion of the legs is a direct integral over $\widehat{G}$ rather than a finite sum. When $G$ is finite, the identification of the two presentations carries a normalization by the total quantum dimension, which diverges for continuous and/or non-compact groups. However, this is the same obstruction which required us to use the co-invariant inner product to define the physical Hilbert space, rather than the divergent invariant inner product, so we expect that these results can be adapted to our setting in some form. We leave a more careful treatment of this to future work.

\subsection{Averaging over the matter legs} \label{sec:ensembledef}

The entropy \eqref{eq:Sgamma} was computed along a cut $\gamma$ we chose by hand. No assumptions about the matter legs were required to derive this formula.
In contrast, in gravity, the cut through spacetime that accounts for the boundary entanglement entropy is determined by a minimization principle \cite{Ryu_2006,Faulkner:2013ana,Hubeny:2007xt,Engelhardt:2014gca}, as we review in Sec.~\ref{sec:gen_entropy}. In this section we will show that our tensor networks reproduce this phenomenon if we average over a random ensemble of matter states.  This is similar to how averages over random tensor networks reproduce entanglement properties of gravity such as the Ryu--Takayanagi formula \cite{Hayden_2016}. We will defer discussion of the possible gravitational meaning of the ensemble of matter states in our tensor networks to  Sec.~\ref{sec:discussion}.

\subsubsection{The Gaussian random ensemble is the renormalized Haar ensemble} \label{sec:gaussianvshaar}

For a $d$ dimensional Hilbert space,  work on random tensor networks \cite{Hayden_2016} suggests that we should average over states $\dket{T}_\ell = U\dket{T_0}_\ell$ with $U \in \mathrm{U}(d)$ drawn from the Haar measure for $\mathrm{U}(d)$.  We can write the average as
\begin{align}
	\int dU \, U^{\otimes n} \dketbra{T_0}{T_0}^{\otimes n} (U^\dagger)^{\otimes n} = \frac{1}{N(d)} \sum_{\sigma \in P_n} \sigma \,, \label{eq:haarnormalization}
\end{align}
where $P_n$ is the symmetric group on $n$ elements,\footnote{We write the symmetric group as $P_n$ rather than the more common $S_n$, to avoid a clash with the R\'enyi entropies $S_n$ that appear below.} $\sigma$ acts by permuting the $n$ copies, and $N(d)$ is a degree $n$ polynomial in $d$ fixed by matching the traces of the two sides.
However, this ensemble is not immediately available to us  because the extended matter Hilbert spaces $\Hext{\ell}$ appearing in \eqref{eq:Vfull} are generically infinite dimensional when $G$ is non-compact. Thus, the normalization constant $N(d)$ is infinite, and this ensemble of states does not exist for the matter leg Hilbert space $\Hext{\ell}$.

Furthermore, even if we tried to regulate this ensemble by truncating each matter leg Hilbert space to a $d$ dimensional subspace, we would find that the cutoff is in tension with gauge invariance. To see this, note that a subspace of $\Hext{\ell}$ which is invariant under the corner symmetry must organize as a direct sum (or direct integral) of representation spaces $V_\pi$. When $G$ is compact, each $V_\pi$ is finite dimensional, so a cutoff on the representation label is both finite dimensional and gauge invariant, and no tension arises. When $G$ is non-compact, every $V_\pi$ is infinite dimensional, so there is no finite dimensional subspace invariant under the corner symmetry at all. A hard dimension cutoff on the matter legs therefore necessarily breaks the corner symmetry. Gauge invariance is only restored in the $d\to\infty$ limit, so it is more convenient to work directly in infinite dimensions.

To explain the resolution of this tension, let us first set up the problem at finite dimension. Truncate each matter leg to a subspace $\Ha^{(d)}_\ell$ of dimension $d$ (as explained above, this step will break gauge invariance if we try to glue back the extended Hilbert spaces), and let $\dket{T_0} \in \Ha^{(d)}_\ell$ be a fixed unit vector. Then using \eqref{eq:haarnormalization} on each matter leg,
\begin{align}
	\mathbb{E}_{\mathrm{Haar}}\left[\left(\dket{T}\dbra{T}\right)^{\otimes n}\right] 
	&= \frac{1}{N(d)^m}\bigotimes_{\ell \in \mathfrak{l}} \left(\sum_{\sigma_\ell \in P_n} \sigma_\ell\right) \\
	&=  \frac{1}{N(d)^m} \sum_{\vec{\sigma} \in (P_n)^{\otimes m}} \vec{\sigma}  \,, \label{eq:haaraverage}
\end{align}
where $\vec{\sigma}$ permutes the $n$ replicas of each lollipop independently. The normalization is fixed by matching traces of both sides of \eqref{eq:haaraverage}, and it diverges as $d \to \infty$ for every $n \geq 1$.

The Gaussian random tensor ensemble of \cite{Cheng:2022ori} is defined by instead fixing an orthonormal basis $\dket{j}_\ell$ of each $\Hext{\ell}$ and expanding
\begin{align}
	\dket{T_\ell} = \sum_j \frac{1}{\sqrt{2}}\left(x^\ell_j + i y^\ell_j\right) \dket{j}_\ell \,, \label{eq:gaussianexpansion}
\end{align}
with the wave function coefficients $x^\ell_j$ and $y^\ell_j$ treated as independent, real, Gaussian random variables of mean zero and unit variance. Notice that $\dket{T_\ell}$ is not normalized, and indeed, a typical draw will have $\dbraket{T}{T} \sim d$. We will deal with this normalization below. This is the same ensemble used in \cite{Akers:2024wab} to study typical matter states of topological tensor networks with finite gauge groups. In particular, the states on distinct lollipop factors are drawn independently, so the Wick contractions defining the $n$th moment pair up replicas within each lollipop separately, and
\begin{align}
	\mathbb{E}_{\mathrm{Gaussian}}\left[\left(\dket{T}\dbra{T}\right)^{\otimes n}\right] = \bigotimes_{\ell \in \mathfrak{l}} \left(\sum_{\sigma_\ell \in P_n} \sigma_\ell\right) = \sum_{\vec{\sigma} \in (P_n)^{\otimes m}} \vec{\sigma} \,, \label{eq:gaussianmoments}
\end{align}
where $\vec{\sigma} = (\sigma_\ell)_{\ell \in \mathfrak{l}}$ assigns an independent permutation of the $n$ replicas to each of the $m$ lollipop factors, and acts on $(\Hext{\mathfrak{l}})^{\otimes n}$ by permuting the replicas of each lollipop separately.\footnote{Recall that $\Hext{\mathfrak{l}} = \bigotimes_{\ell \in \mathfrak{l}} \Hext{\ell}$. }

Comparing \eqref{eq:haarnormalization} and \eqref{eq:gaussianmoments}, the two ensembles have the same moments up to the constant factor $N(d)^{-m}$. This means that for any functional $f_k: \Hext{\mathfrak{l}} \to \C$ which satisfies $f_k(\kappa\dket{T}) = \kappa^k f_k(\dket{T})$ for any nonzero constant $\kappa$, the two ensemble averages will be proportional to each other. 
In particular, if a functional $f_0$ of the matter leg is independent of $\kappa$, then the two ensemble averages will agree with each other, regardless of the dimension $d$. With this in mind, suppose that we did take a hard dimension cutoff on the matter legs, so that both ensembles are well-defined. For a fixed $\dket{T}$, define the reduced state of $R$ in the presence of that matter state as
\begin{align}
	\rho_R[T] = \Tr_{\overline{R} \,\mathfrak{l}}\left[\left(\Id_{\overline{R}} \otimes \dket{T}\dbra{T}\right) V_{\mathrm{full}}\dket{\psi}\dbra{\psi}V_{\mathrm{full}}^\dagger \right] \,, \label{eq:rhoRT}
\end{align}
This is the same object as $\Tr_{\overline{R}}[\dket{\psi[T]}\dbra{\psi[T]}]$, with $\dket{\psi[T]}$ the contracted state \eqref{eq:psiT}, written so that the average we are about to take acts in an obvious way. 
Note that $\rho_R[\kappa\, T] = |\kappa|^2 \rho_R[T]$. Then, define the normalized reduced state
\begin{align}
	\overline{\rho}_R[T] = 	\rho_R[T] / \Tr_{\mathcal{A}_R} [\rho_R[T]] \,. \label{eq:reducedstate_matter_normalized}
\end{align}
Notice that because both the numerator and denominator are degree two in the matter leg state, this implies that $\overline{\rho}_R[\kappa\, T] = \overline{\rho}_R[T]$ for any non-zero constant $\kappa$. Notice that this $\kappa$ independence subsumes the need to normalize the state $\dket{T}$ in \eqref{eq:gaussianexpansion}.
Thus, \emph{any} functional $f$ of the normalized reduced state satisfies
\begin{align}
	\mathbb{E}_{\mathrm{Haar}}[f(\overline{\rho}_R)] = \mathbb{E}_{\mathrm{Gaussian}}[f(\overline{\rho}_R)] \,. \label{eq:gaussian_haar}
\end{align}
In particular, the Haar and Gaussian ensemble averages of $\Tr_{\mathcal{A}_R}[\overline{\rho}_R[T]^n]$ will agree, because we have normalized the reduced state $\overline{\rho}_R[T]$. This equality holds for any cutoff $d$: therefore, we can freely take the limit $d\to\infty$.

What distinguishes the Haar and Gaussian ensembles is which one has non-zero moments in the limit $d \to \infty$ for functionals which depend on the normalization $\kappa$ of the matter states.  Other than the number of random variables that defines the ensemble, the Gaussian moments \eqref{eq:gaussianmoments} are independent of the dimension $d$ of the matter Hilbert spaces. In contrast, the moments of the Haar ensemble \eqref{eq:haarnormalization} vanish in the limit $d\to\infty$, so the Haar ensemble in infinite dimensions does not exist.\footnote{Relatedly, the Haar measure $dU$ which defines the Haar ensemble does not exist in infinite dimensions, because the group $\mathrm{U}(\infty)$ is not a locally compact group.} We may therefore regard the Gaussian ensemble as a version of the Haar ensemble which has been renormalized so that the $d\to\infty$ limit exists. 

Let us examine in more detail why the Haar ensemble fails to exist in higher dimensions, while the Gaussian ensemble continues to be well-defined. In $d$ dimensions, a fixed unit vector $\dket{j}$ and a Haar random unit vector $\dket{T}$ have an approximate squared overlap 
\begin{align}
	\mathbb{E}_{\mathrm{Haar}}[|\dbraket{j}{T}|^2] = \frac{1}{d} \label{eq:Haaroverlap} \,.
\end{align}
As we send $d\to\infty$, this overlap vanishes, so a Haar random vector is almost surely orthogonal to any fixed vector. However, the Haar average explicitly conditions on the state $\dket{T}$ being unit normalized, and the normalization of $\dket{T}$ is not physical from the boundary perspective: it is the contracted state $\dket{\psi[T]}$ of \eqref{eq:psiT} which must be normalized in order to compute observables, and \eqref{eq:Haaroverlap} implies that $\mathbb{E}_{\mathrm{Haar}}[\dbraket{\psi[T]}{\psi[T]}] = \mathcal{O}(d^{-1})$ as well. Thus, although $\dket{T}$ is normalizable in the $d\to\infty$ limit of the Haar ensemble, $\dket{\psi[T]}$ is not. On the other hand, the Gaussian ensemble is precisely defined by the overlap equation $\mathbb{E}_{\mathrm{Gaussian}}[|\dbraket{j}{T}|^2] = 1$, which follows from \eqref{eq:gaussianexpansion}. 
This is achieved by scaling $\dbraket{T}{T} = d$ with the dimension of the Hilbert space, and it implies that $\mathbb{E}_{\mathrm{Gaussian}}[\dbraket{\psi[T]}{\psi[T]}] = \mathcal{O}(1)$.
Furthermore, this average overlap is independent of the choice of fixed unit vector $\dket{j}$, because for any unitary $U$ this ensemble satisfies 
\begin{align}
	\mathbb{E}_{\mathrm{Gaussian}}[|\dbra{j} U \dket{T}|^2] = \mathbb{E}_{\mathrm{Gaussian}}[|\dbraket{j}{T}|^2] = 1  \,. \label{eq:gaussian_unitary_equiv}
\end{align}
So in some sense, we can think of the Gaussian ensemble over $\dket{T}$ as a kind of Haar average over $\dket{\psi[T]}$. Because we are interested in computing the entropy of $\dket{\psi[T]}$, and because it is the ensemble which preserves the corner symmetry of the extended Hilbert space of the matter legs, we will work exclusively with the Gaussian ensemble from now on, and drop the subscript on $\mathbb{E}$.

Because the trace over the matter legs is linear, the average of \eqref{eq:rhoRT} over an ensemble of matter states depends on that ensemble only through the first moment,
\begin{align}
	\mathbb{E}\left[\rho_R\right] = \Tr_{\overline{R}\,\mathfrak{l}}\left[\left(\Id_{\overline{R}} \otimes \mathbb{E}\!\left[\dket{T}\dbra{T}\right]\right) V_{\mathrm{full}}\dket{\psi}\dbra{\psi}V_{\mathrm{full}}^\dagger \right] \,. \label{eq:rhoRbar}
\end{align}
At $n=1$, \eqref{eq:gaussianmoments} is simply
\begin{align}
	\mathbb{E}\left[\dket{T}\dbra{T}\right] = \Id_{\Hext{\mathfrak{l}}} \,. \label{eq:firstmoment}
\end{align}
Inserting \eqref{eq:firstmoment} into \eqref{eq:rhoRbar}, the projection onto the matter state disappears and $\mathbb{E}[\rho_R]$ is obtained from $V_{\mathrm{full}}\dket{\psi}$ by tracing out everything except $\Hext{R}$. This is exactly the reduced state $\rho_R[\gamma_0]$ we computed in \eqref{eq:rhoRempty_averaged}. Taking the trace and using \eqref{eq:rhoRempty_q_averaged},
\begin{align} 
	\Tr_{\mathcal{A}_R}\left[\mathbb{E}[\rho_R]\right] = \int d\mu(\pi)\, q^{\gamma_0}_\pi = 1 \label{eq:avgTrrho_1}\,.
\end{align}

We would like to use \eqref{eq:avgTrrho_1} to conclude that $\mathbb{E}\left[\Tr[\rho_R]\right] = 1$ as well. This works perfectly well for operators $\rho_R[T]$ which are already in the algebra $\mathcal{A}_R$, as the state $\rho_R[T]$ can be expanded into a countable sum over the random variables that define the ensemble.
However, we need to be careful about \emph{which} trace sits inside the expectation value for a generic draw $\rho_R[T]$. The factorization map $V_{\mathrm{full}}$ produces a state invariant under the diagonal action of the corner symmetry of each extended Hilbert space in its image. Denote that corner symmetry action as $u(g) = U_R(g) \otimes U_{\overline{R}}(g) \otimes u_{\mathfrak{l}}(g)$. Thus, $V_{\mathrm{full}} \dket{\psi}= u(g) V_{\mathrm{full}} \dket{\psi}$.
Plugging this into \eqref{eq:rhoRT}, this implies that
\begin{align}
	U_R(g) \rho_R[T] = \, \rho_R\big[u_{\mathfrak{l}}(g)^{-1}T\big] \, U_R(g)  \,. \label{eq:equivariance}
\end{align}
Thus, $\rho_R[T]$ commutes with the corner symmetry $U_R(g)$ of the extended Hilbert space $\Hext{R}$ if and only if the draw $\dket{T}$ is invariant under the corner symmetry. A generic draw from the Gaussian ensemble is not invariant under this symmetry, which is the precise version of the observation we made below \eqref{eq:psiT}: a generic $\dket{T}$ inserts a defect at each matter cut.
Now recall that $\mathcal{A}_R$ was defined in \eqref{eq:ARcommutant} by the operators in the extended Hilbert space which commute with the corner symmetry. This means that for a typical draw $\dket{T}$, $\rho_R[T]$ will not be a state for $\mathcal{A}_R$, so we cannot immediately use the renormalized trace $\Tr_{\mathcal{A}_R}$ to take its trace. 

However, because $\mathbb{E}[\mathcal{O}] \in \mathcal{A}_R$ for any random operator $\mathcal{O}[T] \in \mathcal{B}_R$, we can \emph{define} a weight on the space of random operators, denoted formally as $\widetilde{\mathbb{E}} \circ \Tr_{\mathcal{B}_R}$, by
\begin{align}
    \widetilde{\mathbb{E}}[\Tr_{\mathcal{B}_R}[\mathcal{O}]] := \Tr_{\mathcal{A}_R}[\mathbb{E}[\mathcal{O}]]\,. \label{eq:Etildedef}
\end{align}
In other words, our rule for computing the average trace of an operator is to first average the operator, and then take the trace. This order of operations ensures that the ensemble average computes finite quantities. This definition of ``ensemble average outside of the trace'' effectively implements the renormalization of the trace we explained in Sec.~\ref{sec:algebras}. 

This step is logically independent of the need to use the Gaussian ensemble, because $\mathbb{E}$ still averages over an infinite dimensional vector space. For example, note that when $G$ is a compact group, $\widetilde{\mathbb{E}} = (\mathrm{Vol}(G))^{-1} \mathbb{E}$, despite the infinite dimensionality of $L^2(G)$ when $G$ is continuous. When $G$ is non-compact, we simply take \eqref{eq:Etildedef} as the definition of $\widetilde{\mathbb{E}}$. More generally, if there were a non-perturbative effect which imposed a gap of width $\Delta \sim e^{- S_0}$ between the continuum of representations $\pi$, then (as explained in \cite{Balasubramanian:2026ymu}) $\Tr_{\mathcal{A}_R} = e^{S_0}\Tr_{\mathcal{B}_R}|_{\mathcal{A}_R}$ in that case. Then, we could simply define $\widetilde{\mathbb{E}} = e^{- S_0} \mathbb{E}$. So the reason we need to define $\widetilde{\mathbb{E}}$ in this way is because of the continuum of representations, not because of the infinite dimensionality of the matter Hilbert space.

Crucially, because we define $\widetilde{\mathbb{E}}$ in this way, $\widetilde{\mathbb{E}}[\Tr_{\mathcal{B}_R}[\rho_R^2]] = \Tr_{\mathcal{A}_R}[\mathbb{E}[\rho_R^2]]$, which in general does not equal $\Tr_{\mathcal{A}_R}[\mathbb{E}[\rho_R]^2]$. In other words, this definition of $\widetilde{\mathbb{E}}$ still allows us to compute the impact of fluctuations of the state $\rho_R$ on the average entropy, rather than the entropy of the averaged state $\mathbb{E}[\rho_R]$. For notational simplicity, we now drop the tilde on $\widetilde{\mathbb{E}}$, and abuse notation by also writing $\mathbb{E}$, so that 
\begin{align}
	\widetilde{\mathbb{E}} \circ \Tr_{\mathcal{B}_R} := \mathbb{E} \circ \Tr_{\mathcal{A}_R} := \Tr_{\mathcal{A}_R} \circ\, \mathbb{E}\,.
\end{align}
When $G$ is compact (or there is a non-perturbatively small gap in the continuum of representations), this is an exact statement. When $G$ is a non-compact group, the left-hand side is defined by regulating the renormalized trace $\Tr_{\mathcal{A}_R}$ (or whatever renormalized trace we need to commute past $\mathbb{E}$) using this $S_0$, commuting the trace with the ensemble average, and taking the limit where the regulator is removed.

Finally, it is interesting to note that because we can view the ensemble average as defining a map $\mathbb{E}: \mathcal{B}_R \to \mathcal{A}_R$ it plays a similar role to the operator valued weight $\mathcal{E}$ which we used to renormalize the trace in Sec~\ref{sec:algebras}. It is tempting to speculate that these two operations are therefore related. In other words, perhaps it is the gauge symmetry defining the tensor network which leads to the requirement of randomness in the matter for a consistent interpretation. Some work in two dimensional de Sitter space suggests this identification \cite{Kolchmeyer:2024fly} by linking the group averaging needed to solve the constraints to chaotic dynamics near horizons. We will comment on the possible connections between this observation and the definition of finite subregions in gravity in Sec.~\ref{sec:finite_regions}.

\subsubsection{Measure concentration} \label{sec:concentration}

By itself, \eqref{eq:avgTrrho_1} does not imply that $\Tr_{\mathcal{A}_R}[\rho_R[T]]$ is close to one, or even finite, for a given draw $\dket{T}$ from the Gaussian ensemble. Since we need to divide by this quantity to define the normalized state $\overline{\rho}_R[T]$ in \eqref{eq:reducedstate_matter_normalized}, which we used to justify the Gaussian ensemble through \eqref{eq:gaussian_haar}, we must establish that it is a well-behaved random variable before we discuss the entropy of a random state. 

To draw conclusions about the normalizability of $\rho_R[T]$ for a typical draw, we will show that the higher moments of $ \Tr_{\mathcal{A}_R}\big[\rho_R[T]\big]$ are finite for every $n$. We write
\begin{align}
	\mathcal{N}[T] \equiv \Tr_{\mathcal{A}_R}\big[\rho_R[T]\big] = \dnorm{\psi[T]} \,.
	\label{eq:VarN}
\end{align}
Thinking of $\mathcal{N}$ as a random variable, we need to show that the higher moments $\mathbb{E}[\,\mathcal{N}^n]$ are both strictly positive and bounded above for any $n$ to conclude that a typical draw produces a normalizable state.\footnote{Strictly speaking, bounding the second moment is sufficient for this purpose, but we will use the boundedness of the higher moments when we compute the entropy.}
To do so, we define the ``matter density matrix''
\begin{align}
	\rho_{\mathfrak{l}} \equiv \Tr_{R\overline{R}}\left[V_{\mathrm{full}}\dket{\psi}\dbra{\psi}V_{\mathrm{full}}^\dagger\right] \,, \label{eq:Mdef}
\end{align}
where the trace here is the renormalized one that ensures $\Tr_{\mathfrak{l}}[\rho_{\mathfrak{l}}] = \dnorm{\psi} = 1$. With this definition, we have 
\begin{align}
	\mathcal{N}[T] = \dbra{T} \rho_{\mathfrak{l}} \dket{T} \,. \label{eq:Nquadratic}
\end{align}
The operator $\rho_{\mathfrak{l}}$ is the reduced state of the bulk matter legs in the network. This form for $\mathcal{N}$ will be convenient when computing its moments.

We can now use $\rho_{\mathfrak{l}}$ to compute the moments of $\mathcal{N}$. Note that $\rho_{\mathfrak{l}}$ is a positive operator, and \eqref{eq:avgTrrho_1} says exactly that $\Tr_{\mathfrak{l}}[\rho_{\mathfrak{l}}] = 1$. A positive operator of unit trace is always trace class, so $\rho_{\mathfrak{l}}$ is a genuine density matrix on $\Hext{\mathfrak{l}}$ even though the extended Hilbert space is infinite dimensional. Using $\rho_{\mathfrak{l}}$, we can now bound all the moments of $\mathcal{N}$. Using \eqref{eq:gaussianmoments}, we can see that
\begin{align}
	\mathbb{E}\left[\,\mathcal{N}^n\right] &= \mathbb{E}\left[\Tr_{\mathfrak{l}^{\otimes n}} \!\left[\rho_{\mathfrak{l}}^{\otimes n} (\dket{T}\dbra{T})^{\otimes n} \right]\right]\\
	&= \sum_{\vec{\sigma} \in (P_n)^{\otimes m}} \Tr_{\mathfrak{l}^{\otimes n}}\!\left[\rho_{\mathfrak{l}}^{\otimes n} \, \vec{\sigma}\,\right] \,. \label{eq:Nmoments}
\end{align}
There are exactly $(n!)^m$ terms in the sum of \eqref{eq:Nmoments}, which in particular is finite for every $n$. Furthermore, because $\rho_{\mathfrak{l}}$ is a positive operator of unit trace, its eigenvalues are always between $0$ and $1$, and each term in the sum can be decomposed into sums of products of the eigenvalues of $\rho_{\mathfrak{l}}$. Therefore, we can bound the $n$th moment of $\mathcal{N}$ by
\begin{align}
	1 \leq \mathbb{E}\left[\,\mathcal{N}^n\right] \leq (n!)^m \label{eq:Nbounds}
\end{align}
for every $n$. The lower bound comes from noticing that the identity permutation always contributes exactly 1, and because every term in the sum is positive, there are no cancellations which could spoil this lower bound. Because of the bounds \eqref{eq:Nbounds} satisfied by the moments of $\mathcal{N}$, it follows \cite{Vladimirova_2020} that
\begin{align}
	\mathrm{Prob}(|\mathcal{N}| > t) \leq \alpha e^{-\beta t^{1/m}}\,,
\end{align}
for some $\alpha,\beta > 0$. Thus, the tails of the distribution over possible values of  $|\mathcal{N}[T]|$ decay faster than any polynomial for any finite number of matter legs. Furthermore, the lower bounds on the moments imply that $\mathcal{N}$ is almost surely positive. A typical draw thus produces a normalizable, though not necessarily normalized, state of $\mathcal{A}_R$, and we may divide by $\mathcal{N}[T]$ inside expectation values. 

The case $n=2$ is special and is worth writing out in more detail, because $P_2$ has only two elements. Every configuration $\vec{\sigma}$ therefore assigns to each lollipop either the identity or the transposition. We can therefore think of each assignment $\vec{\sigma}$ as defining a bipartition of the matter legs $\mathfrak{l} \to \mathfrak{l}_{\vec{\sigma} = e} \sqcup \mathfrak{l}_{\vec{\sigma} = \tau}$, which by the discussion in Sec.~\ref{sec:entropy_matter_norandom} defines a cut $\gamma$ through the original network. Let $\rho_{\mathfrak{l}}[\gamma]$ denote the reduced density matrix on the $\mathfrak{l}_{\vec{\sigma} = \tau}$ matter legs. Because this cut does not contain any boundary legs, we can think of it as a closed loop which encircles the legs $\mathfrak{l}_{\vec{\sigma} = \tau}$, and indicates that the matter legs which are assigned $\mathfrak{l}_{\vec{\sigma} = e}$ should be traced out of the factorized state, along with all the boundary legs. We refer to this collection $\mathfrak{l}_{\vec{\sigma} = \tau} = \mathrm{int}(\gamma)$ as the ``interior'' of $\gamma$, and write $|\mathrm{int}(\gamma)|$ for the number of legs in the interior of the cut. Therefore, we can trade the sum over permutations $\vec{\sigma}$ for an equivalent sum over cuts $\gamma$:
\begin{align}
	\mathbb{E}\left[\,\mathcal{N}^2\right] &= \sum_{\vec{\sigma} \in (P_2)^{\otimes m}} \Tr_{\mathfrak{l}^{\otimes 2}}\!\left[\rho_{\mathfrak{l}}^{\otimes 2} \, \vec{\sigma}\,\right] = \sum_{\gamma} \Tr_{\mathrm{int}(\gamma)}[\rho_{\mathfrak{l}}[\gamma]^2] \,. \label{eq:permstocuts}
\end{align}
With this notation, it is useful to define the R\'enyi entropies
\begin{align}
	S_n(\rho) = \frac{1}{1-n} \ln \Tr\left[\rho^n\right]  \label{eq:renyidef}
\end{align}
to simplify the expression for $\mathrm{Var}(\mathcal{N})$. We can then use \eqref{eq:permstocuts} to write the second moment of $\mathcal{N}$ as
\begin{align}
	\mathbb{E}\left[\mathcal{N}^2\right] = \sum_{\gamma} \tr\!\left[\rho_{\mathfrak{l}}[\gamma]^2\right] = \sum_{\gamma} e^{-S_2(\rho_{\mathfrak{l}}[\gamma])} \,, \qquad \rho_{\mathfrak{l}}[\gamma] \equiv \Tr_{\mathfrak{l}_{\vec{\sigma} = e}}[\rho_{\mathfrak{l}}] \,, \label{eq:N2cuts}
\end{align}
where the sum runs over all cuts and $S_2$ is the second R\'enyi entropy. The cut $\gamma_0$ that encloses no matter leg simply computes the norm of the global state $\dket{\psi}$, i.e., it computes the first moment squared. Therefore, subtracting this contribution out of \eqref{eq:N2cuts}, we can see that
\begin{align}
	\mathrm{Var}(\mathcal{N}) = \sum_{\gamma \neq \gamma_0} e^{-S_2(\rho_{\mathfrak{l}}[\gamma])} \,. \label{eq:Nvariance}
\end{align}
This is an exact identity for the variance of $\mathcal{N}$, and it already exhibits the sum over cuts which will eventually organize the entropy calculation. The distribution of $\mathcal{N}$ over possible draws concentrates about its mean precisely when \eqref{eq:Nvariance} is small, which is to say when every nonempty collection of matter legs carries a large bulk R\'enyi entropy in the state $\rho_{\mathfrak{l}}$. We will refer to a state satisfying this condition as \emph{semiclassical}. For simplicity, we will further assume the second R\'enyi entropy of every collection of matter legs is extensive in the number of legs, i.e.,  $S_2(\rho_{\mathfrak{l}}[\gamma]) \gtrsim s_2 \cdot |\mathrm{int}(\gamma)|$ for some state dependent constant $s_2$ (essentially a R\'enyi entropy density for the matter). This assumption is not essential for $\mathrm{Var}(\mathcal{N}) $ to be small. But when it does hold, $\mathrm{Var}(\mathcal{N})$ will be small whenever
\begin{align}
	s_2 \gtrsim \ln(m)\,, \,\label{eq:semiclassicalcriterion}
\end{align}
where $m$ is the total number of matter legs.\footnote{Furthermore, note that the only scheme dependence in our renormalization procedure is a possible shift in each $S_2$ by the state independent constant $\ln(\kappa)$ of \eqref{eq:haarshift}. This corresponds to an overall shift of the lower bound by that same constant, so we should interpret the semiclassical condition as a requirement on the entropy density measured relative to the vacuum, which is a difference of entropies and therefore scheme independent.} This is the large bond dimension condition familiar from the semiclassical limit of random tensor networks \cite{Hayden_2016}, restated intrinsically: the effective bond dimension $\exp(s_2)$ of each matter leg must exceed the total number of matter legs. 

This suggests that the semiclassical observables of the matter legs may be well-approximated by tensor networks whose bond dimensions of the matter legs are $\approx e^{s_2}$. However, note that this effective bond dimension will depend on the state $\dket{\psi}$ in the physical Hilbert space. It is therefore interesting to think of the gauge invariant tensor networks we constructed as a model of a tensor network with a ``state dependent'' bond dimension. It would be interesting to make this more precise, and perhaps also connect this more closely to Cauchy slice holography, where similar ideas have appeared \cite{Soni:2024aop}. Finally, when \eqref{eq:semiclassicalcriterion} holds, division by $\mathcal{N}[T]$ is close to dividing by one. This means we can approximate $\overline{\rho}_R[T] \approx \rho_R[T]$ for each draw $T$, which will simplify the entropy calculation we will now perform.

\subsection{The average entropy of random matter} \label{sec:averageentropy}

We want to compute the average of the entanglement entropy of the boundary subregion $R$,
\begin{align}
	\mathbb{E}\left[S(\overline{\rho}_R)\right] \,, \qquad \overline{\rho}_R[T] = \rho_R[T] \,/\, \mathcal{N}[T] \,, \label{eq:avgentropydef}
\end{align}
over the Gaussian ensemble of matter states defined in Sec.~\ref{sec:ensembledef}.  As we discussed above, this ensemble provides a renormalized version of the Haar average over the matter Hilbert space. Physically, \eqref{eq:avgentropydef} is the average entanglement entropy over the entire ensemble of matter states, but the concentration result of Sec.~\ref{sec:concentration} implies that for global states $\dket{\psi_{\mathrm{sc}}}$ which satisfy the semiclassical condition \eqref{eq:semiclassicalcriterion}, the entropy of a single draw is close to its ensemble average: 
\begin{align}
	S(\overline{\rho}_R[T]) \approx \mathbb{E}[S(\overline{\rho}_R)]\,. \label{eq:semiclassical_is_average}
\end{align}
So \eqref{eq:avgentropydef} is not only a statement about the ensemble: it is also the approximate entanglement entropy of the class of semiclassical states that we discussed below \eqref{eq:Nvariance}.

\subsubsection{The replica trick}
We compute \eqref{eq:avgentropydef} using the replica trick \eqref{eq:entropy_partialn}. Because $\overline{\rho}_R[T]$ is normalized for every draw, the replica trick holds draw by draw, and we may exchange the ensemble average with the derivative,
\begin{align}
	\mathbb{E}\left[S(\overline{\rho}_R)\right] = -\,\partial_n \, \mathbb{E}\!\left[\Tr_{\mathcal{A}_R}\!\big[\,\overline{\rho}_R^{\;n}\big]\right]_{n=1} \,. \label{eq:avgreplica}
\end{align}
The average in \eqref{eq:avgreplica} still involves the ratio $\Tr_{\mathcal{A}_R}[\rho_R^n]/\mathcal{N}^n$, which is not polynomial in the matter state $\dket{T}$ because of the $\mathcal{N}$ in the denominator, and so cannot be evaluated with \eqref{eq:gaussianmoments} directly. However, if the state is semiclassical, i.e., if \eqref{eq:semiclassicalcriterion} holds, $\mathrm{Var}(\mathcal{N})$ is small and $\mathcal{N}[T] \approx \mathbb{E}[\mathcal{N}] = 1$ almost surely by \eqref{eq:avgTrrho_1}. Thus, we can approximate
\begin{align}
	\mathbb{E}\!\left[\Tr_{\mathcal{A}_R}\!\big[\,\overline{\rho}_R^{\;n}\big]\right] \approx \frac{\mathbb{E}\!\left[\Tr_{\mathcal{A}_R}\!\big[\rho_R^{\,n}\big]\right]}{\mathbb{E}\left[\mathcal{N}\right]^{n}} = \mathbb{E}\!\left[\Tr_{\mathcal{A}_R}\!\big[\rho_R^{\,n}\big]\right] \,. \label{eq:quenchedtoannealed}
\end{align}
For brevity of notation, we  write
\begin{align}
	\rho_{\mathrm{full}} \equiv V_{\mathrm{full}}\dket{\psi}\dbra{\psi}V_{\mathrm{full}}^\dagger
\end{align}
for the factorized (pure) state on $\Hext{R} \otimes \Hext{\overline{R}} \otimes \Hext{\mathfrak{l}}$. In this notation, $\rho_R[T] = \Tr_{\overline{R}\,\mathfrak{l}}[\, \rho_{\mathrm{full}}  \,(\Id_{\overline{R}} \otimes \dket{T}\dbra{T})]$ as in \eqref{eq:rhoRT}, and $\rho_{\mathfrak{l}} = \Tr_{R\overline{R}}[\rho_{\mathrm{full}}]$ as in \eqref{eq:Mdef}. 

We can rewrite $\rho_R[T]$ as follows. Let $\tau_n \in P_n$ denote the cyclic permutation of the $n$ replicas. We can then write the $n$th power of a reduced density matrix as a cyclic permutation acting on $n$ replicas of the full factorized state,
\begin{align}
	\Tr_{\mathcal{A}_R}\!\big[\rho_R^n\big] = \Tr\!\left[\rho_{\mathrm{full}} ^{\otimes n}\left(\tau^{(n)}_{R} \otimes \Id_{\overline{R}}^{\otimes n} \otimes \big(\dket{T}\dbra{T}\big)^{\otimes n}\right)\right] \,,
\end{align}
where $\tau_R^{(n)}$ cyclically permutes the $n$ replicas of the extended Hilbert space $\Hext{R}$.\footnote{The unlabeled $\Tr$ denotes the trace over the completely factorized Hilbert space $\Hext{R} \otimes \Hext{\overline{R}} \otimes \Hext{\mathfrak{l}}$.} Averaging with \eqref{eq:gaussianmoments} replaces the matter state by the free sum over a permutation of each lollipop,
\begin{align}
	\mathbb{E}\!\left[\Tr_{\mathcal{A}_R}\!\big[\rho_R^{\,n}\big]\right] = \sum_{\vec{\sigma} \in (P_n)^{\otimes m}} \Tr\!\left[\rho_{\mathrm{full}} ^{\otimes n}\left(\tau^{(n)}_{R} \otimes\Id_{\overline{R}}^{\otimes n} \otimes \vec{\sigma}_{\,\mathfrak{l}}\right)\right] \,. \label{eq:replicapermsum}
\end{align}
Here, $\tau^{(n)}_{R}$ is as above and $\vec{\sigma}_{\,\mathfrak{l}} = (\sigma_{\ell_1},...,\sigma_{\ell_m})$ acts as a permutation $\sigma_{\ell_i}$ on the $n$ replicas of the $i$th matter leg.
This is the familiar structure of random tensor networks \cite{Hayden_2016}, written in the form appropriate to our model: the boundary legs carry \emph{fixed} permutations, $\tau_n$ on $R$ and $e$ on $\overline{R}$, inherited from the replica trace and from the partial trace respectively, while each matter leg is summed over freely. It differs from the moments \eqref{eq:Nmoments} of the norm only in the boundary condition at $R$, which was the trivial permutation there, and is the cyclic permutation $\tau_n$ here. This change is what converts the closed loops of Sec.~\ref{sec:concentration} into curves anchored on the endpoints of $R$, as we will now see.

\subsubsection{Replica symmetry}

At $n=2$, the sum \eqref{eq:replicapermsum} can be simplified without further assumptions, because $P_2 = \{e,\tau_2\}$ has only two elements. Exactly as below \eqref{eq:Nmoments}, each configuration $\vec{\sigma}$ is a bipartition of the matter legs and therefore a cut $\gamma$ through the network in the sense of Sec.~\ref{sec:entropy_matter_norandom}. What differs is which side of the cut the boundary legs sit on: because the $R$ legs now carry $\tau_2$, the matter legs assigned $\tau_2$ join $R$. Therefore, in this case, we can identify the matter leg bipartition $\mathfrak{l}_R$ defining the bipartite factorization map $V[\gamma]$ in Sec.~\ref{sec:entropy_matter_norandom} with the matter leg bi-partition $\mathfrak{l}_{\vec{\sigma}=\tau_2}$ of Sec.~\ref{sec:concentration}. The cuts appearing at $n=2$ in the calculation of \eqref{eq:replicapermsum} are  exactly the family of cuts $\gamma$ over which the fixed cut entropy \eqref{eq:Sgamma} of $S(\rho_R[\gamma])$ was defined. We denote the family of all cuts $\gamma$ which are boundary anchored between $R$ and $\overline{R}$ as $\gamma \sim R$, i.e., the cuts $\gamma$ which are homologous to $R$.
We then have the exact statement 
\begin{align}
	\mathbb{E}\!\left[\Tr_{\mathcal{A}_R}\!\big[\rho_R^{\,2}\big]\right] = \sum_{\gamma \sim R} \Tr_{\mathcal{A}_R[\gamma]}\!\left[\rho_R[\gamma]^2\right] = \sum_{\gamma \sim R} e^{-S_2(\rho_R[\gamma])} \,. \label{eq:secondmomentcuts}
\end{align}
No approximation beyond \eqref{eq:quenchedtoannealed} was needed to reach \eqref{eq:secondmomentcuts}.

For $n > 2$, the group of permutations $P_n$ contains elements other than $e$ and $\tau_n$, and a configuration which assigns one of them to some matter leg defines an $n$-way split of the matter legs instead of a bipartition. We proceed by assuming \emph{replica symmetry}: that the sum \eqref{eq:replicapermsum} is dominated by configurations in which every $\sigma_\ell \in \{e , \tau_n\}$. This assumption is made in much of the random tensor network literature \cite{Hayden_2016,Akers:2024wab}, and is explicitly assumed in derivations of the generalized entropy formula in gravity \cite{Lewkowycz_2013,Dong:2016fnf,Dong:2016hjy}. For convenience, we will simply restrict to states where it is a good approximation in this paper, rather than justifying this assumption further.\footnote{Corrections from relaxing this assumption have been studied in random tensor networks \cite{Vasseur:2018gfy} and in gravity \cite{Akers:2020pmf}, so one could imagine including the corrections beyond the replica symmetric saddle in the tensor model and comparing explicitly to gravity.} With the assumption of replica symmetry and semi-classicality, and using \eqref{eq:replicapermsum}, we have therefore demonstrated that
\begin{align}
	\mathbb{E}\!\left[\Tr_{\mathcal{A}_R}\!\big[\,\overline{\rho}_R^{\;n}\big]\right] \approx \sum_{\gamma \sim R} \Tr_{\mathcal{A}_R[\gamma]}\!\left[\rho_R[\gamma]^n\right] = \sum_{\gamma \sim R} e^{-(n-1) S_n(\rho_R[\gamma])} \,. \label{eq:replicacutsum}
\end{align}
Every ingredient on the right hand side was computed in Sec.~\ref{sec:entropy_matter_norandom}: $\rho_R[\gamma]$ is the reduced state \eqref{eq:rhoRgamma} obtained by factorizing along the fixed cut $\gamma$, and the trace is the renormalized trace of $\mathcal{A}_R[\gamma]$. Averaging over the matter states has done nothing except replace the single cut we chose by hand in Sec.~\ref{sec:entropy_matter_norandom} with a sum over all of them, each weighted by its own R\'enyi entropy.

\subsubsection{The saddle point approximation}
The sum \eqref{eq:replicacutsum} runs over the $2^m$ bipartitions of the matter legs, and is therefore a finite sum of exponentials. Suppose that a single cut $\gamma^{(n)}_*$ minimizes $S_n(\rho_R[\gamma])$, and that it does so by a margin
\begin{align}
	\Delta_n \equiv \min_{\gamma \neq \gamma^{(n)}_*} S_n(\rho_R[\gamma]) \; - \; S_n\!\left(\rho_R\!\left[\gamma^{(n)}_*\right]\right) \,. \label{eq:cutgap}
\end{align}
To reduce the $2^m$ terms in the sum \eqref{eq:replicacutsum}, we will make the \emph{saddle point approximation}, namely, that this gap $\Delta_n$ is sufficiently large
\begin{align}
	\Delta_n \gg m / (n-1) \,. \label{eq:saddlegap}
\end{align}
In this regime, a single term dominates \eqref{eq:replicacutsum}, and we can approximate
\begin{align}
	\mathbb{E}\!\left[\Tr_{\mathcal{A}_R}\!\big[\,\overline{\rho}_R^{\;n}\big]\right] \approx \exp\left(-(n-1)\min_{\gamma \sim R} S_n(\rho_R[\gamma])\right) \,. \label{eq:saddle}
\end{align}
Finally, differentiating at $n=1$ with \eqref{eq:avgreplica}, we find that
\begin{align}
	\mathbb{E}\left[S(\overline{\rho}_R)\right] \approx \min_{\gamma \sim R} S(\rho_R[\gamma]) \,. \label{eq:minoverS}
\end{align}

Note that \eqref{eq:saddlegap} is a separate assumption from the semiclassical condition \eqref{eq:semiclassicalcriterion}. The semiclassical condition controls the \emph{total} weight carried by the non-minimal cuts, which is what makes $\mathrm{Var}(\mathcal{N})$ small in \eqref{eq:Nvariance}; it says nothing about whether one cut dominates the rest, and it is perfectly consistent with many cuts being nearly degenerate. Dominance of a single cut instead requires the gap \eqref{eq:saddlegap}, which is a statement about the spectrum of cut entropies rather than about their overall size. The $n\to1$ limit deserves particular care, because the suppression factor $e^{-(n-1)\Delta_n}$ degenerates there for any fixed $\Delta_n$. As is usual in the replica trick, \eqref{eq:saddle} should be established at every integer $n>1$, where the suppression is valid, and then analytically continued to $n=1$; the step from \eqref{eq:saddle} to \eqref{eq:minoverS} is the assumption that this continuation commutes with the minimization over cuts. When it does not, several cuts contribute at once and the sharp minimization is replaced by a smoothed version of it \cite{Akers:2020pmf}. We will leave a detailed analysis of the saddle point approximation to the future.

\paragraph{The final result.}
Finally, inserting the fixed cut entropy \eqref{eq:Sgamma} into \eqref{eq:minoverS}, we obtain our  result:
\begin{align}
	\mathbb{E}\left[S(\overline{\rho}_R)\right] \approx \min_{\gamma \sim R}\left[H[\,p^\gamma_\pi] + \langle \hat{A}_\gamma \rangle_\rho + S_{\mathrm{bulk}}\!\left(\rho_{\mathrm{int}(R \cup \gamma)}\right)\right] \,, \qquad p^\gamma_\pi = \mu(\pi) \, q^\gamma_\pi \,. \label{eq:mainresult}
\end{align}
In words, the entropy of a boundary subregion in a typical semiclassical state is the fixed cut entropy of Sec.~\ref{sec:entropy_fixedcut}, evaluated on the cut which minimizes it. There are three interesting features of \eqref{eq:mainresult}:
\begin{enumerate}
	\item The minimization is over the entire \emph{generalized} entropy, and not just over the area. The spectrum of $\hat{A}_\gamma$ is fixed by the gauge group and the topological boundary condition at the cut, as we saw in Sec.~\ref{sec:entropy_fixedcut}, and the differential entropy $H[\,p^\gamma_\pi]$ is not sign definite when $G$ is non-compact, as we saw in Sec.~\ref{sec:entropy_nomatter}. All three terms therefore compete, and the minimizing cut is the analog of a quantum extremal surface rather than of a Ryu--Takayanagi surface. 
	\item The minimizing cut is state dependent: it enters \eqref{eq:mainresult}  through $q^\gamma_\pi$ and $\rho^\gamma_\pi$, which depend on the state $\dket{\psi}$ we factorized. 
	\item  The minimization is insensitive to the scheme dependence of $S(\rho_R)$. Under a rescaling $dg \to \kappa \, dg$ of the Haar measure, the area term and the bulk term are invariant while $H[\,p^\gamma_\pi] \to H[\,p^\gamma_\pi] - \ln(\kappa)$ for \emph{every} cut, so the generalized entropies of all cuts shift together and the minimizing cut $\gamma_*$ is unchanged. The value of \eqref{eq:mainresult} thus carries the state independent ambiguity discussed in Sec.~\ref{sec:entropy_nomatter}, but the surface it selects does not depend on this ambiguity. We will consider the gravitational interpretation of all of this in Sec.~\ref{sec:continuum}.
\end{enumerate}

\section{Three-dimensional gravity} \label{sec:continuum}

In this section we will consider applications of the methods developed above to 3D gravity.

\subsection{The generalized entropy in gravity} \label{sec:gen_entropy}

Consider a two dimensional holographic CFT, and let $\ket{\psi}$ be a global state of this CFT with a semiclassical bulk dual that is well-described by the low energy limit of 3D gravity. In AdS/CFT, the generalized entropy \cite{Ryu_2006,Hubeny:2007xt,Faulkner:2013ana,Engelhardt:2014gca} of a reduced state $\sigma_R \sim \tr_{\overline{R}}[\ketbra{\psi}]$ is defined by
\begin{align}
	S(\sigma_R) \;=\; \underset{\gamma \,\sim\, R}{\mathrm{min\;ext}} \left[ \frac{A_\gamma}{4G_N} + S_{\mathrm{bulk}}\bigl(\rho_{\mathrm{int}(R \cup\gamma)}\bigr) \right] \,. \label{eq:QESformula}
\end{align}
Here, $A_\gamma$ is the area of a boundary anchored curve $\gamma$ which is homologous to $R$, and $S_{\mathrm{bulk}}\bigl(\rho_{\mathrm{int}(\gamma \cup R)}\bigr) $ is the von Neumann entropy of all matter fields in the bulk subregion $\mathrm{int}(\gamma \cup R)$. The $\mathrm{min\;ext}$ then indicates that we are instructed to first extremize the sum of these two terms over all possible curves $\gamma$, and then minimize over all possible extremal curves. 

The generalized entropy formula \eqref{eq:QESformula} is a good approximation to the entanglement entropy shared between a boundary subregion $R$ and its complement when the gravitational path integral which computes the R\'enyi entropy $S_n(\sigma_R)$ is dominated by a single, replica symmetric saddle point for $n \gtrsim 1$ \cite{Lewkowycz_2013,Dong:2016hjy}.

Above, one of our main results was \eqref{eq:mainresult}, which computes the entanglement entropy of a semiclassical state $\rho_R$ (as defined by \eqref{eq:semiclassicalcriterion}) whose R\'enyi entropies are dominated by a single, replica symmetric saddle point for $n \gtrsim 1$. We reproduce \eqref{eq:mainresult} here for convenience:
\begin{align}
	\mathbb{E}\left[S(\rho_R)\right] \approx \min_{\gamma \sim R}\left[H[\,p^\gamma_\pi] + \langle \hat{A}_\gamma \rangle_\rho + S_{\mathrm{bulk}}\!\left(\rho_{\mathrm{int}(R \cup \gamma)}\right)\right]  \,. \label{eq:mainresult2}
\end{align}

Let us compare the structure of \eqref{eq:QESformula} and \eqref{eq:mainresult2} more closely. First,  the minimization over cuts in the tensor network model runs over a finite set of curves which interleave the matter legs, rather than over a continuous family of curves as in the generalized entropy formula. Second, there is no extremization step in the tensor network: therefore, for \eqref{eq:mainresult2} to match \eqref{eq:QESformula}, the extremization over curves must somehow be implicit even when there are no matter legs (so that the minimization over cuts is trivial). We will provide evidence that this is the case in Sec.~\ref{sec:examples}.

Third, our formula includes the differential entropy $H[\,p^\gamma_\pi]$ of \eqref{eq:Hdef}, which does not appear in \eqref{eq:QESformula} as it is usually written. Finally, and most importantly, our $\hat{A}_\gamma$ does not have to be the area of anything: it is the operator \eqref{eq:areaoperator}, whose spectrum $\ln \mu(\pi)$ is fixed by the gauge group $G$ alone. In gravity the area term is derived from the gravitational path integral; to genuinely match the gravitational answer, we must match the spectrum $\ln(\mu(\pi))$ to the lengths of geodesic curves in the bulk. After discussing the reduction of the phase space of $\SL(2,\R) \times \SL(2,\R)$ Chern--Simons theory to impose the invertibility of the metric in Sec.~\ref{sec:CTV}, we will argue in Sec.~\ref{sec:examples} that the spectrum of the area operator actually should match the length of a geodesic in the homotopy class of $\gamma$, which would agree with the gravitational answer for the entropy of a connected interval on the boundary of the disk.

\subsection{Canonical quantization of 3D gravity} \label{sec:canonicalquant}

Consider pure gravity in $2+1$-dimensional, Lorentzian, globally hyperbolic, anti-de Sitter (AdS) spacetimes. Let $\Sigma$ be a Cauchy slice of such a spacetime $\mathcal{M}$.\footnote{Of course, AdS spacetimes are not globally hyperbolic, but for a given Cauchy slice $\Sigma$ we can just focus on the domain of dependence $D[\Sigma]$ without loss of generality. Then the construction below should be interpreted as conditional on the boundary conditions for the Cauchy slice $\Sigma$, i.e., the intersection of the Cauchy slice $\Sigma$ and the conformal boundary.}
The phase space of this theory is labeled by the induced metric $h_{ij}$  and  extrinsic curvature $K_{ij}$ of the Cauchy slice, subject to the constraint equations of gravity. Because of  unitarity, we can define the phase space of the theory on any Cauchy slice, but if we choose it to be the unique maximum volume slice of $\mathcal{M}$ \cite{Krasnov:2005dm,Moncrief:1989dx,Scarinci:2011np,Bonsante2010:xyz,Couch:2018phr,Witten:2022xxp},\footnote{The maximum volume slice is unique after fixing the intersection of the Cauchy slice $\Sigma$ and the conformal boundary.} which is a diffeomorphism invariant specification of $\Sigma$, then the constraint equations simplify.

Specifically, let us decompose the phase space data as
\begin{align}
	h_{ij} = e^{\phi} \hat{h}_{ij} \,,&& K_{ij} = \hat{K}_{ij} + \frac{1}{2} K h_{ij} \,.
\end{align}
Here, $\hat{h}_{ij}$ is a conformal equivalence class for the induced metric, $\phi$ is a Liouville field, and $K$, $\hat{K}_{ij}$ are the trace and traceless part of the extrinsic curvature, respectively. On the maximum volume slice, the extrinsic curvature is traceless, so we can set $K = 0$. Furthermore, because $K$ is canonically conjugate to the conformal factor $\phi$ of the induced metric, the phase space on the maximum volume slice only depends on the reduced phase space data $(\hat{h}_{ij},\hat{K}_{ij})$. This is equivalent to the larger set of data $(h_{ij},K_{ij})$ because the equations of motion uniquely determine the Liouville field $\phi$ in terms of just the reduced phase space data \cite{Krasnov:2005dm}, which is where the Lorentzian AdS assumption comes in.

Conformal equivalence classes $\hat{h}_{ij}$ of two dimensional metrics are parameterized by a point in the Teichm\"uller space $\mathcal{T}_{g,n}$, where $g$ is the genus of $\Sigma$ and $n$ is the number of boundary components. The phase space $(\hat{h}_{ij},\hat{K}_{ij})$ can therefore be identified with the cotangent space $T^*\mathcal{T}_{g,n}$,\footnote{Actually, we have to also gauge the phase space by the large diffeomorphisms of the spatial slice as well, so the configuration space of metrics is actually the moduli space $M_{g,n} = \mathcal{T}_{g,n} / \mathrm{Map}(\Sigma)$, where $\mathrm{Map}(\Sigma)$ is the mapping class group of $\Sigma$. We defer further discussion of this point to the discussion in Sec.~\ref{sec:discussion}.} and the extrinsic curvature $\hat{K}_{ij}$ can be identified with the fibers of this phase space \cite{Krasnov:2005dm}.

The tensor network Hilbert spaces reviewed in Sec.~\ref{sec:themodel} are analogs of quantizations of this phase space. Indeed, Teichm\"uller space of a surface $\Sigma$ with negative Euler characteristic can be identified with flat $\PSL(2,\R)$ connections on $\Sigma$ \cite{Goldman1988Components,Witten:1988hc}, subject to a constraint on the connection that we review below. So when we use our tensor networks to solve the constraint equations and construct a flat connection on the spatial surface $\Sigma$, the wavefunctions that we construct are analogs of wavefunctions over Teichm\"uller space \cite{Kim:2015qoa}. Therefore, we can picture the tensor networks of this paper as living on the maximum volume Cauchy slice of a 3D Lorentzian AdS spacetime.

Suppose we instead quantize $3$d gravity using the first order formalism of vielbeins $e^a_\mu$ and spin connections $\omega^{a}_\mu = \frac{1}{2}\epsilon^{abc} \omega_{ab\mu}$, as opposed to induced metrics and extrinsic curvatures. Then when the cosmological constant is negative,\footnote{We will work in units where the AdS radius is set to unity.} it is convenient to first make the field redefinition 
\begin{align}
	(A_\pm)^a_{\mu} = \omega^{a}_\mu \pm e^a_{\mu} \,.
\end{align}
With this redefinition the Einstein--Hilbert action reduces to two copies of Chern--Simons theory \cite{Witten:1988hc}, and the fields $A_\pm$ play the role of Chern--Simons gauge fields, with gauge group $\SL(2,\R)$. The level $k$ of the Chern--Simons theory and Newton's constant $G_N$ are related by $k = (4 G_N)^{-1}$ to leading order in $G_N$ \cite{Witten:1988hc}. So in AdS spacetimes, 3D gravity and Chern--Simons theory are equivalent at the level of the action.

However, $3$d gravity and Chern--Simons are not the same quantum theory. For example, $A_\pm = 0$ is an on-shell solution in Chern--Simons theory, which corresponds to a completely degenerate spacetime metric. In contrast, the field space of gravity is restricted to invertible spacetime metrics, so at best gravity should be thought of as a subtheory of Chern--Simons theory. Remarkably, it is possible to make this notion precise using Virasoro TQFT (VTQFT) \cite{Collier_2023,Collier:2024mgv} or Conformal Turaev--Viro (CTV) theory \cite{Hartman:2025cyj,Hartman:2025ula}, where one quantizes only the subspace of the Chern--Simons phase space corresponding to invertible metrics.\footnote{See also \cite{Mertens:2022ujr,Mertens:2025ydx,Jafferis:2025vyp,Jafferis:2025jle,Jafferis:2025yxt} } The key fact is that the moduli space of Chern--Simons connections has many connected components, and for a fixed orientation of the spacetime, invertible metrics correspond to a single connected component. Operationally, this deforms the representation theoretic data that defines the TQFTs, where instead of $\SL(2,\R)$, one must consider the so-called modular double of the quantum group $\mathcal{U}_q(\sl(2,\R))$ \cite{Faddeev:1999fe,Ponsot:2000mt}. So to construct a tensor network for VTQFT/CTV (as opposed to Chern--Simons theory), one should likely associate a copy of the modular double of $\mathcal{U}_q(\sl(2,\R))$ to the legs of the tensor network. Finally, in addition to the invertibility of the metric, 3D gravity requires (at least) a gauging of the mapping class group and a sum over topologies. 

In our setting, we are quantizing the full Chern--Simons theory because we allow arbitrary group elements on the legs of the tensor network. Equivalently, we allow arbitrary irreducible representations on the legs. This is also where the implicit $G_N \to 0$ limit comes into our model. For any finite level $k$, not every representation of $\SL(2,\R)$ is allowed \cite{Maldacena:2000hw}, and the fusion rules between the representations are modified by the finite level. But in the limit $k\to\infty$, every representation is allowed again, and their fusion rules are the same as those of $\SL(2,\R)$. So because we are allowing \emph{every} representation of $\SL(2,\R)$, we are implicitly working in the $k\to\infty$ (equivalently, $G_N\to 0$) limit.

\subsection{Imposing invertibility of the metric}
\label{sec:CTV}

In Sec.~\ref{sec:canonicalquant} we reviewed one of the main reasons that 3D gravity and Chern--Simons theory are not the same quantum theory: to all orders in perturbation theory, gravity can be thought of as a subtheory of Chern--Simons theory, obtained by restricting to the connected component of the Chern--Simons phase space on which the spacetime metric is invertible. The tensor networks of this paper quantize the full Chern--Simons theory, because we allow every unitary irreducible representation of $G$ to appear on every leg. In this section, we will investigate what would change in the entropy formula \eqref{eq:mainresult2} if we imposed invertibility of the metric, and in particular what would happen to the spectrum of the area operator $\hat{A}$.

There is an obstruction to answering this question directly. Every step in the derivation of $\hat{A}$ referred to the gauge group $G$: the physical Hilbert space was decomposed over $\widehat{G}$ in Sec.~\ref{sec:factorization}, the traces on $\mathcal{A}_R$ and on its center were fixed by the Plancherel and microcanonical measures in Sec.~\ref{sec:algebras}, and the spectrum $\ln(\mu(\pi))$ of \eqref{eq:areaoperator} came out as the measure-theoretic derivative relating the two. Imposing invertibility replaces the input data of the model altogether, and once it is replaced none of those intermediate steps survive verbatim. We therefore cannot simply follow $\mu(\pi)$ through the restriction and see what it becomes.

Our strategy is to first restate the result in a form which does not mention $G$ at all. In Sec.~\ref{sec:doublemodel} we do this by comparing with the case of a finite gauge group, where the same contribution to the entropy is a familiar object in condensed matter theory, and where its standard expression is written in terms of the anyons of the theory rather than in terms of the group. In Sec.~\ref{sec:CTVtheory} we identify the theory that should be substituted for Chern--Simons theory, namely conformal Turaev--Viro theory, and explain in what sense our construction is and is not a quantization of it. In Sec.~\ref{sec:quantumgroups} we collect the representation theoretic data that the substitution requires. We will then investigate the consequences of this conjectured substitution in Sec.~\ref{sec:examples} using this data.

\subsubsection{The quantum double model and Turaev--Viro theory} \label{sec:doublemodel}

The area operator $\hat{A}$ of \eqref{eq:areaoperator} has a counterpart in condensed matter theory. As we showed in \cite{Balasubramanian:2026ymu}, its eigenvalue in a given sector is the topological entanglement entropy of that sector \cite{Kitaev:2005dm,Levin:2006arx}, once the state independent constant of Sec.~\ref{sec:entropy_nomatter} is accounted for. We review that identification here, for it lets us express our results in terms of the anyons of the theory alone, with no reference to the gauge group we built the theory from. It is this group-free form which Sec.~\ref{sec:CTVtheory} requires to discuss the generalization of our results to 3D gravity.

Our topological tensor networks are, as noted in Sec.~\ref{sec:themodel}, a generalization of string nets \cite{Levin_2005,kirillov2011stringnet}. In a string net, one decorates the legs of a graph $\Lambda$ with the objects of a tensor category and its vertices with intertwiners between them, and one declares two decorated graphs to represent the same state whenever some fixed collection of local moves relates them; this is what makes the Hilbert space a function of the surface $\Sigma$ rather than of $\Lambda$. For us, the objects are the representations in $\mathrm{Rep}(G)$, the intertwiners are the multipartite observer edge modes $|\Gamma_a(\vec{\pi})\rangle'$ of Sec.~\ref{sec:multipartite}, and the moves are moves 1 and 2 of Sec.~\ref{sec:themodel}. The one respect in which we depart from the standard definition is finiteness: an ordinary string net takes as input a fusion category with a finite set of simple objects (in our case, the irreducible representations of $G$), and this condition only holds for $\mathrm{Rep}(G)$ when $G$ is a finite group. In that case, our construction reduces to Kitaev's quantum double model \cite{Kitaev:1997wr}, and its physical Hilbert space is the Hilbert space of Dijkgraaf--Witten theory \cite{Dijkgraaf:1989pz,Hu:2012wx}, or equivalently of the Turaev--Viro theory with input category $\mathrm{Rep}(G)$ \cite{TuraevViro1992,BarrettWestbury1996,kirillov2011stringnet,Buerschaper_2009}. For any other transformable $G$, the TQFT equivalent to our model has not been identified, to the best of our knowledge.

The entanglement entropy of a finite string net across a curve $\gamma$ which separates $\Sigma$ into two pieces is \cite{Kitaev:2005dm,Levin:2006arx,Dong:2008ft,Bonderson:2017osr}
\begin{align}
	S \;=\; \underbrace{- \sum_\xi p_\xi \ln (p_\xi)}_{H[\,p_\xi]} \;+\; \underbrace{\sum_\xi p_\xi \ln \bigl(\mathcal{S}_1^{\,\xi}\bigr)}_{\langle \hat{A}\rangle + \mathrm{constant}} \;+\; \sum_\xi p_\xi \, S(\rho_\xi) \,, \label{eq:stringnetentropy}
\end{align}
where $p_\xi$ is the probability that the state carries the anyon $\xi$ through $\gamma$, and $\mathcal{S}_1^{\,\xi}$ is the matrix element of the modular $S$ transformation between the anyon $\xi$ and the vacuum line, which is the quantum dimension of $\xi$ (denoted $d_\xi$) divided by the total quantum dimension of the theory,
\begin{align}
	\mathcal{S}_1^{\,\xi} = \frac{d_\xi}{\mathcal{D}} \,, \qquad \mathcal{D}^2 = \sum_\xi d_\xi^2 \,. \label{eq:S1adef}
\end{align}
When only the trivial anyon contributes, the middle term of \eqref{eq:stringnetentropy} reduces to $-\ln(\mathcal{D})$, the topological entanglement entropy of Kitaev and Preskill and of Levin and Wen \cite{Kitaev:2005dm,Levin:2006arx}. In \cite{Balasubramanian:2026ymu}, we showed that \eqref{eq:stringnetentropy} and our formula \eqref{eq:SrhoR_gen} agree term by term when $G$ is finite: for the double model, $\mathcal{D} = |G|$, and the pure charge $\xi = (e,\pi)$ has $d_\xi = d_\pi$, so that $\ln(\mathcal{S}_1^{\,(e,\pi)}) = \ln(\mu(\pi)) - \ln|G|$ (with $\mathrm{Vol}(G) = 1$, as in Sec.~\ref{sec:factorization}). Thus, $\mathcal{S}_1^{\,\xi}$ plays the role of the density of states $\mu(\pi)$, and the two formulas agree up to the state independent constant $-\ln|G| = -\ln \mathcal{D}$. For a general transformable group, this constant diverges, and is in fact the divergence which the renormalized trace of Sec.~\ref{sec:algebras} removes: because it is state independent, it is absorbed into the scheme dependent constant \eqref{eq:haarshift}, and it drops out of the entropy differences which are physical in our model.

Note that our formula \eqref{eq:SrhoR_gen} runs over $\widehat{G}$ alone, a single representation label and no flux, whereas \eqref{eq:stringnetentropy} runs over every anyon which can end on $\gamma$. The two agree because we imposed the rough boundary condition of Sec.~\ref{sec:factorization} at the cut $\gamma$ in this paper, and the rough boundary condition is the one on which only the pure charges $(e,\pi)$ condense, so a boundary anchored cut on the disk can only see anyons of this form \cite{Bravyi:1998sy,Beigi:2010htr,Kitaev:2011dxc,Balasubramanian:2026ymu}.\footnote{Such a maximal set of anyons condensing on a topological boundary is called a Lagrangian algebra, and its anyons are sometimes called Lagrangian anyons \cite{Kapustin:2010hk,Kong:2014xyz,KONG2021115607}. The name refers to a Lagrangian subalgebra of the boundary category, not to a local action density.} As we discussed in Sec.~\ref{sec:entropy_nomatter}, a different topological boundary condition at the cut changes only which anyons the cut can see, and not the form of any of our results.

The gauge group therefore entered our derivation in two places: it fixed the anyon content of the theory, with the quantum dimensions \eqref{eq:S1adef}, and, through the boundary condition at the cut, it fixed which anyons a boundary anchored cut can see. The invariant content of the area operator is then that its eigenvalue in the sector labeled by an anyon $\xi$ is
\begin{align}
	\langle \hat{A} \rangle_\xi \sim \ln \bigl( \mathcal{S}_1^{\,\xi} \bigr) \,, \label{eq:areainvariant}
\end{align}
where $\xi$ runs over the anyons the cut in question can support, and $\sim$ denotes equality up to the state independent constant above. This suggests two generalizations of \eqref{eq:areainvariant}.

First, if a cut $\gamma$ that we use to factorize the Hilbert space is not boundary anchored, we expect all the anyons of the theory to contribute to the spectrum of the area operator. For example, the sectors shared between the two boundary components of a cylinder are labeled by the entire anyon spectrum of the theory \cite{Zhang:2011jd,Delcamp:2016eya,kirillov2011stringnet}. When $G$ is a finite group, the anyons of the quantum double model are labeled by $([g],\lambda)$, where $[g]$ is a conjugacy class of $G$, and $\lambda$ is an irreducible representation of the centralizer $C_G(g)$ of $g$. Equivalently, these pairs are the simple objects of the quantum double $D[G]$, the symmetry structure described by the combination of electric and magnetic operators in Sec.~\ref{sec:themodel}. When we replace a finite group $G$ with a compact quantum group $G_k$ at level $k$, the anyons are instead parameterized by two chiral labels of the doubled $G_k \times G_{-k}$ Chern--Simons theory, as described in Sec.~\ref{sec:themodel}. Returning to the finite group case, the quantum dimension of $([g],\lambda)$ is $|[g]|\,d_\lambda$, so that
\begin{align}
	\ln \bigl( \mathcal{S}_1^{\,([g],\lambda)} \bigr) = \ln \left( \frac{|[g]| \, d_\lambda}{|G|} \right) \,. \label{eq:finiteGquantumdim}
\end{align}
Here, $|[g]|$ is the number of elements in the conjugacy class, and $d_\lambda$ is the dimension of the $\lambda$ representation.

In Appendix~\ref{app:ribbons}, we construct the closed ribbon operators which correspond to anyons of an arbitrary transformable group, and show that they carry exactly this pair of labels. More precisely, the label space of these anyons is given by the diagonal quotient \eqref{eq:diagonalquotient}; the pure charges of the main text are a special case of these anyons with trivial flux $t=e$.\footnote{It would be interesting to prove that this class of anyons is the complete set when $G$ is a general transformable group, analogous to the finite $G$ case.} We expect \eqref{eq:areainvariant}, with $\xi$ running over all of \eqref{eq:diagonalquotient}, to give the area spectrum along a closed cut, up to the same constant $-\ln|G|$.

Second, we propose, following \cite{Donnelly_2016,Strominger:1997eq,Wong:2022eiu,Lin:2017uzr,Jafferis:2019wkd,Balasubramanian:2023dpj,Balasubramanian:2025rcr,Hartman:2025cyj,Mertens:2022ujr,Mertens:2025ydx,Mertens:2022aou,Chua:2023ios} (see especially \cite{Carlip:1994gy,McGough:2013gka,Mertens:2022ujr}), that the formula \eqref{eq:areainvariant} continues to hold when the input $\mathrm{Rep}(G)$ to our tensor network model is no longer given by the representation theory of a classical group, and when the topological boundary conditions at the cut are generalized beyond the rough boundary condition. We explore this generalization now.

\subsection{Conformal Turaev--Viro theory} \label{sec:CTVtheory}

As we explained in Sec.~\ref{sec:canonicalquant}, the phase space of 3D gravity on a fixed background topology can be thought of as the restriction of the $\SL(2,\R) \times \SL(2,\R)$ Chern--Simons phase space to invertible metrics, the so called Teichm\"uller component \cite{Witten:1988hc, Goldman1988Components, Scarinci:2011np}.\footnote{This is true locally. The global phase space of 3D gravity needs an additional quotient by the large diffeomorphisms of the spatial surface, also called the mapping class group \cite{Witten:1988hc, Moncrief:1989dx, Carlip:1998uc, Maloney:2007ud}. We defer this to Sec.~\ref{sec:discussion}.} At the level of the path integral, there are two constructions which make this restriction precise.\footnote{See also \cite{Mertens:2022ujr,Chen:2024unp}, as well as \cite{Jafferis:2025vyp,Jafferis:2025yxt,Jafferis:2025jle} for another possible path integral construction which may match pure 3D gravity, especially after the sum over the mapping class group and non-trivial topologies are accounted for.} The first is Virasoro TQFT (VTQFT) \cite{Collier_2023,Collier:2024mgv}, which builds a three-dimensional topological theory whose states on hyperbolic surfaces are Virasoro conformal blocks, and whose partition functions on hyperbolic three manifolds reproduce the 3D gravity path integral on a fixed topology. The second is Conformal Turaev--Viro theory \cite{Hartman:2025cyj,Hartman:2025ula}, which is defined by a state sum construction that produces the same physics as two copies of VTQFT,\footnote{More precisely, the two theories are equivalent up to a transformation in their boundary conditions.} in the same way that ordinary Turaev--Viro theory produces a doubled Chern--Simons theory from a triangulation \cite{TuraevViro1992, Turaev:1994xb, kirillov2010corners,Turaev:2010pp, Roberts1995}. In the language of the previous subsection, Virasoro TQFT is a chiral half of 3D gravity on a fixed topology, and the full gravitational path integral is reproduced by the doubled theory $|\mathrm{VTQFT}|^2$, with conformal Turaev--Viro theory its state sum presentation.

What is new about these TQFTs compared to traditional Turaev--Viro theory is exactly the feature which we overcame in Sec.~\ref{sec:themodel}. Ordinary Turaev--Viro theory takes a fusion category with finitely many simple objects as input, and therefore has finitely many anyons. Conformal Turaev--Viro theory instead has an unbounded continuum of them, labeled by pairs of Liouville momenta $(P_+,P_-) \in \R_+^2$, and so it cannot be described by such a category at all. We met the same phenomenon in our own model: we showed in \cite{Balasubramanian:2026ymu} that the Hilbert space our networks assign to the torus is infinite dimensional, which means that the associated TQFT has infinitely many anyon types, and cannot come from a finite fusion category either. The two constructions break the same finiteness assumption, and they break it in the same way. This is the structural reason to expect that the machinery of Sec.~\ref{sec:factorization} through Sec.~\ref{sec:entropy_matter_random}, which did not assume a finite list of anyons, may apply to conformal Turaev--Viro theory as well.

That said, note that our construction is not a quantization of conformal Turaev--Viro theory. The symmetry structure of Virasoro TQFT and of conformal Turaev--Viro theory is the modular double of the quantum group $\mathcal{U}_q(\sl(2,\R))$ \cite{Faddeev:1999fe,Ponsot:2000mt,Meusburger:2008bs,Mertens:2022aou}, not the representation category $\mathrm{Rep}(G)$ for a group $G$.\footnote{This structure can, however, be thought of as a deformation of $\mathrm{Rep}(\SL(2,\R))$. More precisely, the modular double is constructed using two copies of this deformed representation theoretic structure \cite{Faddeev:1999fe}.} As we explained in Sec.~\ref{sec:canonicalquant}, allowing every representation of $\SL(2,\R)$ on the legs is the same as working at infinite Chern--Simons level, and the quantum group deformation to the modular double is what both restricts the metric to be invertible and moves the theory to finite level.

However, $\mathrm{Rep}(\SL(2,\R))$ and $\mathcal{U}_q(\sl(2,\R))$ are closely related, and the relation is of the kind that a string net construction can potentially generalize. A string net needs an input category together with its $6j$ symbols, and those of the modular double are known \cite{Ponsot:2000mt,Teschner:2012em,Teschner:2003em}. It should therefore be possible to build a string net whose legs carry the modular double of $\mathcal{U}_q(\sl(2,\R))$ in place of $\mathrm{Rep}(\SL(2,\R))$, and such a string net should be a canonical quantization of conformal Turaev--Viro theory in the same sense that the $\mathrm{Rep}(G_k)$ string net is a canonical quantization of the corresponding Turaev--Viro theory.\footnote{See \cite{Chen:2024unp} for a possibly related model based on BCFTs, which is defined on the boundary of Euclidean manifolds rather than Cauchy slices of Lorentzian spacetimes.} We do not carry out that construction explicitly in this paper, but we plan to return to it in future work \cite{Cummings:2026xyz}.

The reason this matters for us is the spectrum of the area operator \eqref{eq:areainvariant}. Because the area spectrum can be stated without reference to $G$, the entropy formula \eqref{eq:mainresult2} is a statement about string nets and not about $\SL(2,\R)$ specifically. If \eqref{eq:areainvariant} survives the change of input category, with only the label set and the function $\mathcal{S}_1^{\,\xi}$ replaced, then the resulting string net should satisfy an entropy formula similar to \eqref{eq:mainresult2}. Since conformal Turaev--Viro theory agrees with 3D gravity on a fixed topology, an entropy computed in the hypothetical modular double string net would then be an entropy computed in 3D gravity on a fixed topology. Genuinely matching gravity also requires gauging the mapping class group of $\Sigma$ and summing over topologies, neither of which we address, and we defer both to Sec.~\ref{sec:discussion}.

\subsection{The role of the quantum group} \label{sec:quantumgroups}

We close this section by collecting the data that the substitution would require. In the examples of Sec.~\ref{sec:examples} we will assume that the naive replacement of $\mathrm{Rep}(\SL(2,\R))$ by the modular double of $\mathcal{U}_q(\sl(2,\R))$ is valid, in the sense that the formulas of the previous sections continue to hold with the following three dictionary entries, and we will investigate the consequences.

First, we discuss the labels of the anyons. The unitary irreducible representations $\pi \in \widehat{G}$ appearing on the internal legs are replaced by the non-degenerate representations of the modular double, labeled by a Liouville momentum $P \in \R_+$ \cite{Ponsot:1999uf, Ponsot:2000mt, Faddeev:1999fe, Bytsko:2002br,Mertens:2022ujr}. On the $\SL(2,\R)$ side these are the images of the principal series, which are the representations of zero $U(1)$ charge, and they are the representations which survive the restriction to invertible metrics after the quantum deformation \cite{Teschner:2003em, Collier_2023, Jackson_2015}.

Second, we need to know what happens to the spectrum of the area operator. As in Sec.~\ref{sec:doublemodel}, the anyons of the doubled theory come in pairs, one label from each chiral half, so an anyon of $\mathrm{CTV}=|\mathrm{VTQFT}|^2$ is a pair $(P_+,P_-)$. Under time reversal the two chiral halves are exchanged and $P_+ \leftrightarrow P_-$, so a state with $P_+ = P_-$ is time reversal invariant. The modular $\mathcal{S}$ matrix of CTV is known, and its matrix elements correspondingly factorize as
\begin{align}
	\mathcal{S}_1^{\,(P_+,P_-)} = \rho_0(P_+) \, \rho_0(P_-) \,, \label{eq:S1VTQFT}
\end{align}
where $\rho_0(P) = S_1^{\,P}$ is the vacuum row of the modular $S$ transformation of a single chiral half. More explicitly, $\rho_0(P)$ is given by the Cardy density
\begin{align}
	\rho_0(P) = 4\sqrt{2} \, \sinh(2\pi b P) \, \sinh(2\pi b^{-1} P) \,, \label{eq:cardydensity}
\end{align}
where the deformation parameter of $\mathcal{U}_q(\sl(2,\R))$ is written as $q = e^{i \pi b^2}$. This function is the density of states of the boundary Liouville theory. Note that this is the same statement as \eqref{eq:S1adef}: the vacuum row of the modular $S$ matrix is a density of states because a modular transformation exchanges two channels in which the same partition function can be expanded, and reading the vacuum character in the crossed channel returns the density of states of the theory. In fact, the same argument can be used to justify the appearance of $\mathcal{S}_1^\xi$ for general string nets \cite{Kitaev:2005dm} as well. This justifies our terminology of the Plancherel measure as a ``density of states'': for the CFT dual of the bulk theory, the Plancherel measure is precisely the density of states of the (extended) primaries of the theory.

Third, we discuss the central charge, which is how $G_N$ enters. The deformation parameter $b$ is related to the central charge of the associated Liouville theory by \cite{Zamolodchikov:1995aa, Teschner:2001rv}
\begin{align}
	c = 1 + 6 \left( b + b^{-1} \right)^2 \,, \label{eq:c_liouville}
\end{align}
so we may take $b \in (0,1]$ without loss of generality. The semiclassical limit $c \to \infty$ is then the limit $b \to 0$. On the gravity side, Brown and Henneaux showed that the asymptotic symmetry algebra of 3D gravity is a Virasoro algebra with central charge\footnote{This formula holds to leading order in $G_N$, which is the only order at which we will use it.} \cite{Brown1986}
\begin{align}
	c = \frac{3 L_{\mathrm{AdS}}}{2 G_N} \,. \label{eq:brownhenneaux}
\end{align}
We will work in units where $L_{\mathrm{AdS}} = 1$, so that $c = 3/(2 G_N)$. Expanding \eqref{eq:c_liouville} at small $b$ gives $c = 6b^{-2} + 13 + \mathcal{O}(b^2)$, so comparing the two, we see that
\begin{align}
	b^2 = 4 G_N \label{eq:bsquared}
\end{align}
to leading order in $G_N$. 

Putting everything together, the assumption we will carry into Sec.~\ref{sec:examples} is the following. A string net built from the modular double of $\mathcal{U}_q(\sl(2,\R))$ should have an area operator $\hat{A}$, whose eigenvalues are given by \eqref{eq:areainvariant} with $\mathcal{S}_1^{\,\xi}$ as in \eqref{eq:S1VTQFT}, so that its spectrum is
\begin{align}
	\langle \hat{A} \rangle_{(P_+,P_-)} = \ln \bigl( \rho_0(P_+) \, \rho_0(P_-) \bigr) \,, \label{eq:areaspectrum_VTQFT}
\end{align}
with $(P_+,P_-)$ running over the anyons that the boundary conditions at the cut can support. To see how the boundary conditions of the main text generalize to this setting, we make the following observations. Recall from Sec.~\ref{sec:factorization} that the rough boundary condition at the cut is a Dirichlet condition for the gauge field at $\gamma$. For the finite level $G_k \times G_{-k}$ string nets of Sec.~\ref{sec:themodel}, the topological boundary condition which imposes Dirichlet conditions on both gauge fields condenses the diagonal anyons $(j,\overline{j})$, with conjugate labels in each chiral half \cite{Kitaev:2011dxc,Kong:2014xyz}, and in Virasoro TQFT the corresponding sectors are $(P,P)$. Note that this is a time reversal invariant boundary condition, since time reversal exchanges the two chiral halves of the theory. We therefore expect a boundary anchored cut with the Dirichlet topological boundary condition to see only the time reversal invariant sectors $P_+ = P_- = P$, in which case the spectrum is $2\ln(\rho_0(P))$. For a closed cut, we assume that the spectrum can run over all pairs of anyons, just as in the finite group case.

\section{Examples} \label{sec:examples}

In this section, we assume that the substitution of Sec.~\ref{sec:quantumgroups} is valid, so that a string net built from the modular double of $\mathcal{U}_q(\sl(2,\R))$ has an area operator with the spectrum \eqref{eq:areaspectrum_VTQFT}, and we work out what that spectrum is in the semiclassical limit $G_N \to 0$ for two examples. The first is a boundary anchored subregion of the disk with time symmetric boundary conditions, which restrict the sectors (labeled by the Liouville momenta $P_\pm$ explained in Sec.~\ref{sec:quantumgroups}) to $P_+ = P_-$. The answer we obtain should be compared to the Ryu--Takayanagi formula \cite{Ryu_2006}. The second is the entropy shared between the two asymptotic boundaries of a rotating BTZ black hole, where the cut is closed, both labels $P_\pm$ are free, and the answer should be compared to the Hubeny--Rangamani--Takayanagi formula \cite{Hubeny:2007xt} for the entropy of a subregion in a non-static spacetime. In both cases we will find that the eigenvalues of $\hat{A}$ are the length of a bulk geodesic divided by $4G_N$. The second example will also let us address the question raised in Sec.~\ref{sec:gen_entropy}, of how the extremization over curves in \eqref{eq:QESformula} can be implicit in \eqref{eq:mainresult2}.

\subsection{Time symmetric boundary anchored regions} \label{sec:timesymmetric}

Consider the disk with time symmetric conditions, and a boundary anchored cut $\gamma$ separating $R$ from $\overline{R}$. By the discussion at the end of Sec.~\ref{sec:quantumgroups}, the only sectors which such a cut can see are the time reversal invariant ones, where the two Liouville momenta defining the superselection sectors (analogous to the representations above) satisfy $P_+ = P_- \equiv P$. As we will see below, we can think of $P$ as measuring the lengths of geodesics in the spacetime. The spectrum of the area operator is then $2 \ln (\rho_0(P))$, with $\rho_0$ the Cardy density \eqref{eq:cardydensity}.

Before taking the limit, recall what the restriction $P_+ = P_-$ means geometrically. From Sec.~\ref{sec:canonicalquant},  the two Chern--Simons connections $A_\pm$ each define a flat $\SL(2,\R)$ connection on $\Sigma$, and a flat $\SL(2,\R)$ connection in the Teichm\"uller component is the same thing as a hyperbolic metric on $\Sigma$. A Lorentzian AdS$_3$ spacetime with Cauchy slice $\Sigma$ therefore comes with \emph{two} hyperbolic metrics on $\Sigma$, one from each of $A_\pm$, which are sometimes called the left and right metrics of the spacetime \cite{Krasnov:2005dm,Mess2007,Scarinci:2011np,Benedetti:2005vy}. The Liouville momentum $P_\pm$ of the sector through $\gamma$ measures the holonomy of $A_\pm$ around $\gamma$, and so it measures the length of $\gamma$ in the left and right hyperbolic metrics, respectively. The two metrics coincide precisely when the extrinsic curvature $K_{ij}$ of $\Sigma$ vanishes identically \cite{Krasnov:2005dm}, i.e., when $\Sigma$ is at a moment of time symmetry. This is the tensor network version of the fact that the Ryu--Takayanagi formula applies to time symmetric slices. Since other topological boundary conditions at the cut let other anyons terminate on it (Sec.~\ref{sec:entropy_nomatter} and \cite{Balasubramanian:2026ymu}), we expect that some of them allow sectors with $P_+ \neq P_-$ to terminate on the cut, and these would be the boundary conditions relevant to boundary anchored regions of spacetimes without a moment of time symmetry. We do not pursue this here, and instead access the sectors with $P_+ \neq P_-$ through a closed cut in Sec.~\ref{sec:btz}.

We now take the semiclassical limit. By \eqref{eq:c_liouville}, $c \to \infty$ is the limit $b \to 0$, and by \eqref{eq:bsquared} this is the limit $G_N \to 0$. However, we cannot simply set $b \to 0$ at fixed $P$. The scaling dimension of the state in the sector $(P,P)$ is
\begin{align}
	\Delta = h_+ + h_- = \frac{c-1}{12} + 2P^2 \,, \qquad h_\pm = \frac{(b + b^{-1})^2}{4} + P_\pm^2 \,, \label{eq:liouvilledimension}
\end{align}
so a state whose energy above the vacuum is of order $c$, which is the regime in which the bulk geometry is semiclassical, has $P = \mathcal{O}(b^{-1})$. We therefore define
\begin{align}
	P = \frac{\ell}{4 \pi b} \,, \label{eq:Pofell}
\end{align}
and hold $\ell$ fixed as $b \to 0$. The normalization of $\ell$, including the factor of $4\pi$, is a convention at this point; we will see in Sec.~\ref{sec:btz} that it is the convention in which $\ell$ is a geodesic length. Inserting \eqref{eq:Pofell} into the Cardy density \eqref{eq:cardydensity}, the two hyperbolic sines become $\sinh(\ell/2)$ and $\sinh(\ell/(2b^2))$, and the second is exponentially large. Thus, we find that the eigenvalues of the area operator will be given by
\begin{align}
	2 \ln (\rho_0(P)) = \frac{\ell}{b^2} + \mathcal{O}(b^0)\,. \label{eq:areaexpansion}
\end{align}
Using $b^2 = 4G_N$ from \eqref{eq:bsquared}, the leading term is
\begin{align}
	\langle \hat{A} \rangle_{(P,P)} = \frac{\ell}{4 G_N} + \mathcal{O}(G_N^0) \,. \label{eq:areasemiclassical}
\end{align}

Equation \eqref{eq:areasemiclassical} has the form of the area term in \eqref{eq:QESformula}, with $\ell$ playing the role of the area $A_\gamma$ of the curve. To see that $\ell$ is indeed the length of a geodesic, recall that $P$ measures the holonomy of $A_+$ (equivalently, of $A_-$, since $P_+ = P_-$) along $\gamma$. The holonomy of a flat $\SL(2,\R)$ connection in the Teichm\"uller component along a curve $\gamma$ is a hyperbolic element of $\SL(2,\R)$, whose trace is $2\cosh(\ell_\gamma/2)$, where $\ell_\gamma$ is the length of the unique geodesic in the homotopy class of $\gamma$ in the hyperbolic metric determined by the connection.\footnote{For a boundary anchored curve on a surface with boundary, the relevant notion is the geodesic in the homotopy class of $\gamma$ relative to the boundary, whose length is regularized in the usual way \cite{Ryu_2006,Hubeny:2007xt}. The regularization is supplied by the boundary condition on $\partial\Sigma$, which does not affect the sector label $P$ \cite{Balasubramanian:2026ymu}.} The identification of the Liouville momentum with this length, in precisely the normalization \eqref{eq:Pofell}, is standard in the quantization of Teichm\"uller space \cite{Verlinde:1989ua,Teschner:2003em,Hadasz:2005gk}. 
We will verify this directly in Sec.~\ref{sec:btz}, where the same normalization \eqref{eq:Pofell} reproduces the horizon length of the BTZ black hole, and where the relevant geodesic is closed, so no regularization is needed. Taking this identification for granted for the moment, \eqref{eq:areasemiclassical} says that the area operator of the $\mathcal{U}_q(\sl(2,\R))$ string net has eigenvalues $\ell/4G_N$, with $\ell$ the length of the geodesic in the homotopy class of the cut $\gamma$.

We can now assemble the entropy of a boundary anchored region in a time symmetric state. Without matter legs there is a single cut up to homotopy, and \eqref{eq:SrhoR_gen} gives
\begin{align}
	S(\rho_R) = \langle \hat{A} \rangle_\rho + H[\,p_\pi] + S_{\mathrm{bulk}}(\rho_R) \approx \frac{\langle \ell \rangle_\rho}{4G_N} + \cdots \,,
\end{align}
where $\langle \ell \rangle_\rho$ is the expectation value of the geodesic length in the state, and the remaining terms are subleading in $G_N$ for a semiclassical state. This is the Ryu--Takayanagi formula \cite{Ryu_2006} for the entropy of $R$, which in 3D gravity on a time symmetric slice is the length of the boundary anchored geodesic homologous to $R$ divided by $4G_N$. 

With matter legs, \eqref{eq:mainresult2} instructs us to minimize over the cuts $\gamma \sim R$ which interleave the matter legs, and each such cut carries its own geodesic length $\ell_\gamma$ and its own bulk entropy. Since the state is time symmetric, every geodesic on $\Sigma$ is also extremal in the spacetime, and the minimization over homotopy classes of cuts is therefore the minimization over extremal curves in \eqref{eq:QESformula}. So for time symmetric states, the tensor network reproduces the quantum extremal surface formula, in the form which is valid for time symmetric states.  

The extremization step of \eqref{eq:QESformula} is invisible in this example, because on a time symmetric slice it is automatic, since $\ell$ is always a geodesic length. The next example will make the requirement of spacetime extremality of the curves measured by the area operator more visible, which we explain in detail in Sec.~\ref{sec:offmaxvol}. 

\subsection{Rotating black holes} \label{sec:btz}

The phase space of 3D gravity contains many states with no moment of time symmetry, and the simplest of these are the rotating BTZ black holes \cite{Banados:1992wn, Banados:1992gq}, with inner and outer horizon radii $r_-$ and $r_+$, respectively. In coordinates, the metric of a rotating BTZ black hole is given by
\begin{align}
	ds^2 = -f(r)^2 dt^2 + f(r)^{-2} dr^2 + r^2 \left(d\phi - \frac{r_+ r_- dt}{r^2}\right)^2 \,, \qquad f(r)^2 = \frac{(r^2 - r_+^2)(r^2 - r_-^2)}{r^2}  \,. \label{eq:btzmetric}
\end{align}
The mass and angular momentum of the black hole are
\begin{align}
	M = \frac{r_+^2 + r_-^2}{8 G_N} \,, \qquad J = \frac{r_+ r_-}{4 G_N} \,. \label{eq:btzMJ}
\end{align}
The outer horizon at $r = r_+$ is a circle of proper length $2\pi r_+$, so that the Bekenstein--Hawking entropy of a rotating BTZ black hole is $2\pi r_+ / 4 G_N$. When $J \neq 0$ the metric is stationary but not static, and so it does not admit a moment of time symmetry. In the language of Sec.~\ref{sec:timesymmetric}, the nonzero angular momentum implies that the left and right hyperbolic metrics (defined by the Chern--Simons connections $A_\pm$) of the spacetime are different.

We now consider the tensor network state which prepares the two sided black hole. The Cauchy slice that this tensor network tessellates is the cylinder $\Sigma_{0,2}$, which we studied in Appendix~\ref{app:ribbons},  with one asymptotic boundary at each end. We will now compute the entropy shared between the two boundaries in the tensor network model, under the assumptions of Sec.~\ref{sec:quantumgroups}. The cut $\gamma$ separating the two boundaries is the closed curve wrapping the cylinder, so by the first generalization of Sec.~\ref{sec:doublemodel} its sectors run over all of the anyons of the theory. Together with the substitution of Sec.~\ref{sec:quantumgroups}, this gives the spectrum \eqref{eq:areaspectrum_VTQFT} for the area operator, with both labels $P_\pm$ free. Assuming that the state is peaked around a particular pair of Liouville momenta, the area contribution to the entropy is given by
\begin{align}
	\langle \hat{A} \rangle_{(P_+,P_-)} = \ln(\rho_0(P_+)) + \ln (\rho_0(P_-)) \,. \label{eq:btzareaspectrum}
\end{align}
We now investigate the consequences of \eqref{eq:btzareaspectrum}.\footnote{See \cite{Cvetic:1997vp} for some interesting work about interpreting each of these terms as a ``chiral entropy'' of the BTZ black hole.}  Note though that \eqref{eq:btzareaspectrum} rests on the assumptions stated in Sec.~\ref{sec:quantumgroups}, and that the computation below is a consistency check of these assumptions rather than a derivation.

The state we have in mind is one whose distribution $q_{P_\pm}$ over the sectors $(P_+, P_-)$ is concentrated near a single pair, so that $\langle \hat{A} \rangle_\rho$ is well approximated by the eigenvalue \eqref{eq:btzareaspectrum} at that pair. The two labels are fixed by the mass and spin of the black hole as follows. The sector $(P_+,P_-)$ carries chiral scaling dimensions $h_\pm$ given by \eqref{eq:liouvilledimension}, and the asymptotic Virasoro charges of the BTZ metric \eqref{eq:btzmetric} are therefore given by \cite{Strominger:1997eq, Banados:1992gq, Brown1986}
\begin{align}
	h_\pm - \frac{c}{24} = \frac{M \pm J}{2} \,, \label{eq:btzcharges}
\end{align}
in units where $L_{\mathrm{AdS}} = 1$. Combining \eqref{eq:liouvilledimension}, \eqref{eq:btzMJ} and \eqref{eq:btzcharges}, and dropping the $\mathcal{O}(G_N^0)$ difference between $c/24$ and $(c-1)/24$, we can see that
\begin{align}
	P_\pm^2 = \frac{M \pm J}{2} = \frac{(r_+ \pm r_-)^2}{16 G_N} \qquad \Longrightarrow \qquad P_\pm = \frac{r_+ \pm r_-}{2\sqrt{4G_N}} \,, \label{eq:btzmomenta}
\end{align}
In terms of the variables $\ell_\pm = 4\pi b P_\pm$ of \eqref{eq:Pofell}, this reads
\begin{align}
	\ell_\pm = 2\pi \left(r_+ \pm r_-\right) \,. \label{eq:btzlengths}
\end{align}
Now we expand \eqref{eq:btzareaspectrum} as in \eqref{eq:areaexpansion}, with $\ell_\pm$ held fixed as $b \to 0$, and keep the leading term:
\begin{align}
	\langle \hat{A} \rangle_{(P_+,P_-)} \approx \frac{\ell_+ + \ell_-}{2 b^2}  = \frac{2\pi r_+}{4 G_N}  \,. \label{eq:btzarea}
\end{align}
This leading term is the Bekenstein--Hawking entropy of the rotating black hole, for any mass and spin. Since the differential entropy $H[\,p_\pi]$ and the bulk entropy are subleading in $G_N$ for a semiclassical state, the entropy \eqref{eq:mainresult2} shared between the two boundaries is $2\pi r_+/4G_N$ to leading order. In particular, we have reproduced the entropy of a family of states which are not time symmetric, as promised in the introduction.

Note that \eqref{eq:btzlengths} fixes the normalization convention \eqref{eq:Pofell} that we left open in Sec.~\ref{sec:timesymmetric}. Setting $r_- = 0$, the non-rotating black hole has $\ell_+ = \ell_- = 2\pi r_+$, which is exactly the length of the horizon: the closed geodesic of the slice in the homotopy class of $\gamma$. This is an internal check that $\ell = 4\pi b P$ is a geodesic length, including the factor of $4\pi$, which used only the dictionary of Sec.~\ref{sec:quantumgroups} and the BTZ solution. The same normalization then applies to the boundary anchored case, where $P_+ = P_-$ and the sector label is the same object. 

\paragraph{Fixed area states} Finally, the two examples of this section give a concrete meaning to the direct integral over sectors that has organized this paper since \eqref{eq:fixedareadecompositionintro}. In the dictionary of Sec.~\ref{sec:quantumgroups}, the sector label of a cut is a pair of Liouville momenta $(P_+,P_-)$, and we have just seen that $P_\pm = \ell_\pm/4\pi b$ are the lengths of the closed geodesic through the cut in the left and right hyperbolic metrics, with the eigenvalue of $\hat{A}$ the length of the corresponding extremal curve in the spacetime divided by $4 G_N$. The superselection sectors of the tensor network are therefore labeled by the areas of extremal curves, and a state supported in a single sector is a fixed area state in the sense of \cite{Dong:2018seb,Akers:2018fow,Dong:2022ilf}: within each sector the observer edge modes are maximally mixed, so the R\'enyi spectrum is flat, and the entropy formula \eqref{eq:mainresult2} is literally an integral over fixed area states weighted by $q_\pi\, d\mu(\pi)$. In this sense, the decomposition \eqref{eq:fixedareadecompositionintro} is the tensor network version of the fixed area state decomposition of the gravitational Hilbert space, with the Plancherel measure supplying the density of fixed area states.

\subsection{Extremal surfaces off the maximum volume slice} \label{sec:offmaxvol}

Note that the two lengths $\ell_\pm$ in \eqref{eq:btzlengths} have a geometric meaning of their own. As explained in Sec.~\ref{sec:timesymmetric}, $P_\pm$ measures the holonomy of $A_\pm$ around $\gamma$, so $\ell_\pm$ is the length of the closed geodesic around the cylinder in the left and right hyperbolic metrics of the spacetime \cite{Krasnov:2005dm,Mess2007, Benedetti:2005vy, Scarinci:2011np}. For the rotating black hole these are two hyperbolic cylinders with minimal lengths $2\pi(r_+ + r_-)$ and $2\pi(r_+ - r_-)$ around their non-contractible cycles, and the length of the outer horizon of the black hole is their average,
\begin{align}
	2\pi r_+ = \frac{\ell_+ + \ell_-}{2} \,. \label{eq:averagelength}
\end{align}
This is a general fact about locally AdS$_3$ spacetimes: a pair of hyperbolic holonomies $(g_+, g_-)$ around a cycle determines a unique, closed, spacelike geodesic in the spacetime which is invariant under  $(g_+, g_-)$. Furthermore, its length is the average of the two hyperbolic lengths, while their difference $(\ell_+ - \ell_-)/2$ measures the boost, or time shift, along it \cite{Carlip:1994gc,Mess2007, Benedetti:2005vy}. So the area operator along a closed cut does not measure a length in either hyperbolic metric separately, nor a length on the maximal slice, where the tensor network naturally lives. It measures the length of the closed geodesic determined by $(g_+, g_-)$ in the spacetime, which for the black hole is the horizon.

This last point is the answer to the question raised in Sec.~\ref{sec:gen_entropy}. The generalized entropy formula \eqref{eq:QESformula} instructs us to extremize over curves in the spacetime before minimizing, while \eqref{eq:mainresult2} contains no extremization. We can now see why the extremization is implicit in \eqref{eq:mainresult2}. The eigenvalues of $\hat{A}$ are functions of the holonomy of the flat connections $A_\pm$ around $\gamma$, and these holonomies are constants of motion: they are unchanged by any evolution of the state generated by operators supported on the boundary legs. In the tensor network, this is the statement that a closed ribbon operator wrapping the cylinder can be deformed away from either boundary, so it commutes with any operator acting on the legs of one boundary alone, and hence with a Hamiltonian of the form $H_L + H_R$. On the other hand, the maximal slice on which the tensor network lives does move with $H_L + H_R$. At $t_L = t_R = 0$, the maximal volume slice of \eqref{eq:btzmetric} passes through the bifurcation circle of the horizon, but under evolution to $t_L = t_R = t$, the maximal slice anchored at the new boundary times pulls away from it \cite{Hartman:2013qma, Stanford:2014jda, Couch:2018phr}, and the minimal length closed curve on the new maximum volume slice is no longer extremal in the spacetime. 
But the expectation value of $\hat{A}$ does not change under this evolution, because the reduced state on either boundary transforms unitarily under the flow of the boundary Hamiltonian $H_L + H_R$, so the entropy remains invariant too. This is precisely the behavior of the HRT surface \cite{Hubeny:2007xt}, whose area is independent of the boundary times for the two sided black hole, and it is not the behavior of the minimal curve on the maximal slice. 

In this sense, the area operator knows about a surface that the slice it is defined on does not intersect. The extremization in \eqref{eq:QESformula} is implicit in \eqref{eq:mainresult2} because the sector labels are holonomies, and holonomies see the whole domain of dependence of the slice rather than the slice itself. It would be very interesting to understand this more deeply, for example by studying the fusion of the horizon sector $(P_+,P_-)$ with the sectors of boundary anchored cuts, which would probe how the HRT surface of a boundary subregion of the black hole sweeps through the spacetime. We leave this for future work.

\section{Finite regions in gravity} \label{sec:finite_regions}

So far, every entropy we have computed has been the entropy of a boundary subregion $R$. In this section we use the same machinery to propose a definition of the entropy of a compact region in the bulk in the tensor model, and compare the result to the proposals for such entropies in the gravity literature.

Compact bulk subregions are difficult to define in any diffeomorphism invariant theory, and TQFTs make the difficulty especially clear. Suppose we draw a closed curve $\gamma$ on the spatial surface $\Sigma$ and declare its interior to be ``the subregion''. Any other curve $\gamma'$ related to $\gamma$ by a diffeomorphism of $\Sigma$ defines a gauge equivalent subregion, and when the bulk is topological every closed curve which is homotopic to $\gamma$ is of this kind. So a curve drawn by hand carries no gauge invariant information beyond its homotopy class, and on the disk there is only one such class. We can see this explicitly in the tensor networks without matter legs: the only physical operators are boundary anchored line operators, and the bulk vertices, edges and plaquettes of $\Lambda$ can be removed entirely by moves 1 and 2 of Sec.~\ref{sec:themodel}. A region defined by a collection of bulk plaquettes therefore has no meaning in the physical Hilbert space.

However, the situation is different when we include matter legs in the tensor network. Recall from Sec.~\ref{sec:themodel} that each matter leg $\ell \in \mathfrak{l}$ is dressed to a bulk vertex and a bulk plaquette, forming a lollipop factor, and that moves 1 and 2 can only be performed away from the lollipops. In other words, the lollipops are the only bulk features of the lattice which survive the quotient by the graphical moves. The matter legs thus give us a gauge invariant way to ``track'' a collection of plaquettes of $\Lambda$, which are codimension zero in $\Sigma$, through the physical Hilbert space. This suggests that we can \emph{define} a compact bulk subregion in this model to be a subset $a \subset \mathfrak{l}$ of the lollipop factors, and its Hilbert space to be the extended Hilbert space of those legs. This is essentially the definition of a subregion in a 3D TQFT proposed in \cite{Delcamp:2016eya}, and it is structurally the same as the definition of a finite region in gravity advocated for in \cite{Balasubramanian:2023dpj,Ciambelli:2021nmv}: the bulk degrees of freedom of the region, together with the minimal set of edge modes required to make the region gauge invariant on its own.\footnote{Furthermore, recent investigations of compact regions in gravity \cite{Bousso:2026btw,Bousso:2026txw} suggest the possible importance of using the conformal equivalence class of the spatial metric and the trace of the extrinsic curvature as the primary variables to study a compact region. It is interesting to note that, as explained in Sec.~\ref{sec:canonicalquant}, these are the same variables used by our tensor networks, suggesting a possible connection.} We will adopt this as a working definition and investigate its consequences.\footnote{See also a recent proposal \cite{Wei:2026syv}, which argues that finite subregions in gravity may be pure states. In our tensor model, however, finite regions described by finite subsets of matter legs will always be mixed.}

\subsection{Boundary dressed compact regions}

The first definition uses the multipartite factorization map $V_{\mathrm{full}}$ of \eqref{eq:multiparty_factorization}, which we already have in hand. Let $a \subset \mathfrak{l}$ be a subset of the matter legs, and let $\overline{a} = \mathfrak{l} \setminus a$ denote the remaining matter legs. Recall from \eqref{eq:Vfull} that $V_{\mathrm{full}}$ embeds the physical Hilbert space into $\Hext{R} \otimes \Hext{\overline{R}} \otimes \bigotimes_{\ell \in \mathfrak{l}} \Hext{\ell}$ by cutting every leg which meets the corner vertex $\eth$ of the double bowtie lattice (Fig.~\ref{fig:doublebowtie}). The Hilbert space of the region $a$ is then the tensor factor
\begin{align}
	\Hext{a} \equiv \bigotimes_{\ell \in a} \Hext{\ell} \,,
\end{align}
and we define the reduced state of the region $a$ to be
\begin{align}
	\rho_a = \Tr_{R \overline{R}\, \overline{a}}\left[V_{\mathrm{full}} \dketbra{\psi}{\psi} V_{\mathrm{full}}^\dagger\right] \,, \label{eq:rhoadef}
\end{align}
where the trace is over the extended Hilbert spaces of both boundary subregions and of the complementary matter legs. The trace which makes $\rho_a$ unit normalized is analogous to the renormalized trace defined in Sec.~\ref{sec:entropy_matter_random}.

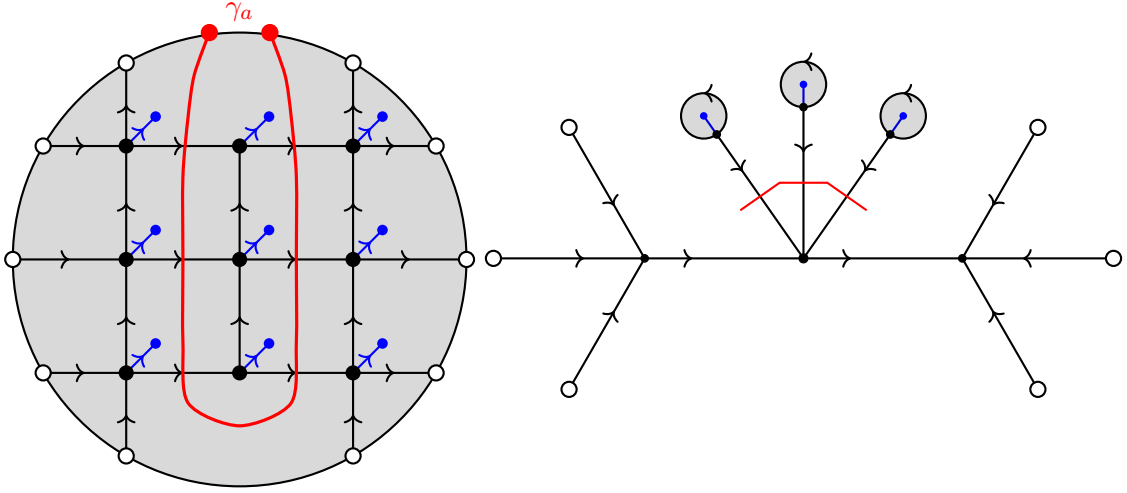
\begin{figure}
	\centering
    \begin{tikzpicture}[scale=2,thick]
		\def\Rd{1.5}          
		\def\nn{3}            
		\def\nnm{2}           
		\def\vertcols{1,3}    
		\def\rind{0.16}       
		\def\rbdy{0.05}       
		\def\rblk{0.05}       
		\def\dstem{0.275}      
		\def\rlol{0.1125}       
		\def\lolang{45}       
		\def\brc{0.40}        
		\colorlet{latt}{black}
		\colorlet{matt}{blue}
		\colorlet{cutc}{red}
		\pgfmathsetmacro{\sep}{2*\Rd/(\nn+1)}
		\pgfmathsetmacro{\gap}{asin(\rind/\Rd)}
		
		
		\filldraw[fill opacity=0.15] (0,0) circle (\Rd);
		\foreach \i in {1,...,\nn}{                     
			\pgfmathsetmacro{\X}{-\Rd+\i*\sep}
			\pgfmathsetmacro{\angC}{asin(\X/\Rd)}
			\pgfmathsetmacro{\angD}{180-\angC}
		}

		\foreach \i in {1,...,\nn}{
			\pgfmathsetmacro{\X}{-\Rd+\i*\sep}
			\pgfmathsetmacro{\Ed}{sqrt(\Rd*\Rd-\X*\X)}
			\pgfmathsetmacro{\Vone}{-\Rd+\sep}
			\pgfmathsetmacro{\Vtop}{-\Rd+\nn*\sep}
			\draw[latt,->-=0.5] (-\Ed,\X) -- (\Vone,\X);
			\draw[latt,->-=0.5] (\Vtop,\X) -- (\Ed,\X);
			\foreach \j in {1,...,\nnm}{
				\pgfmathsetmacro{\Ya}{-\Rd+\j*\sep}
				\pgfmathsetmacro{\Yb}{-\Rd+(\j+1)*\sep}
				\draw[latt,->-=0.5] (\X,\Ya) -- (\X,\Yb);
				\draw[latt,->-=0.5] (\Ya,\X) -- (\Yb,\X);
			}
		}
		\foreach \i in \vertcols{
			\pgfmathsetmacro{\X}{-\Rd+\i*\sep}
			\pgfmathsetmacro{\Ed}{sqrt(\Rd*\Rd-\X*\X)}
			\pgfmathsetmacro{\Vone}{-\Rd+\sep}
			\pgfmathsetmacro{\Vtop}{-\Rd+\nn*\sep}
			\draw[latt,->-=0.5] (\X,-\Ed) -- (\X,\Vone);
			\draw[latt,->-=0.5] (\X,\Vtop) -- (\X,\Ed);
		}
		
		\foreach \i in {1,...,\nn}{
			\pgfmathsetmacro{\X}{-\Rd+\i*\sep}
			\foreach \j in {1,...,\nn}{
				\pgfmathsetmacro{\Y}{-\Rd+\j*\sep}
				\pgfmathsetmacro{\cx}{\X+\dstem*cos(\lolang)}
				\pgfmathsetmacro{\cy}{\Y+\dstem*sin(\lolang)}
				\draw[matt,->-=0.6] (\X,\Y) -- (\cx,\cy);
				\fill[matt] (\cx,\cy) circle (0.035);
				\fill[latt] (\X,\Y) circle (\rblk);
			}
		}
		
		\foreach \i in {1,...,\nn}{
			\pgfmathsetmacro{\X}{-\Rd+\i*\sep}
			\pgfmathsetmacro{\angC}{asin(\X/\Rd)}
			\pgfmathsetmacro{\angD}{180-\angC}
			\filldraw[latt,fill=white] (\angC:\Rd) circle (\rbdy);
			\filldraw[latt,fill=white] (\angD:\Rd) circle (\rbdy);
		}
		\foreach \i in \vertcols{
			\pgfmathsetmacro{\X}{-\Rd+\i*\sep}
			\pgfmathsetmacro{\angA}{acos(\X/\Rd)}
			\pgfmathsetmacro{\angB}{-\angA}
			\filldraw[latt,fill=white] (\angA:\Rd) circle (\rbdy);
			\filldraw[latt,fill=white] (\angB:\Rd) circle (\rbdy);
		}
		
		\draw[cutc,very thick] plot[smooth] coordinates
		{(-0.2,1.5) (-0.30,1.22) (-0.34,0.95) (-0.375,0.55) (-0.375,0) (-0.375,-0.55)
			(-0.34,-0.95)  
			(0,-1.1)
			(0.34,-0.95)
			(0.375,-0.55)
			(0.375,0) (0.375,0.55) (0.34,0.95) (0.30,1.22) (0.2,1.5)} ;
		\filldraw[cutc] (0.2,1.5) circle (0.05);
		\filldraw[cutc] (-0.2,1.5) circle (0.05);
		\node[cutc] at (0,1.64) {$\gamma_a$};
	
	\end{tikzpicture}
	\begin{tikzpicture}[scale=2]
		\def\hh{1.05}      
		\def\rr{1.15}      
		\def\rc{0.5}       
		\def\cl{0.16}      
		
		\foreach \xx in {120,180,240}{
			\draw[thick,->-=0.6] ({-\hh+cos(\xx)},{sin(\xx)}) -- (-\hh,0);
			\filldraw[thick,fill=white] ({-\hh+cos(\xx)},{sin(\xx)}) circle (0.05);
		}
		\foreach \xx in {-60,0,60}{
			\draw[thick,->-=0.6] ({\hh+cos(\xx)},{sin(\xx)}) -- (\hh,0);
			\filldraw[thick,fill=white] ({\hh+cos(\xx)},{sin(\xx)}) circle (0.05);
		}
		\draw[thick,->-=0.3] (-\hh,0) -- (0,0);
		\draw[thick,->-=0.3] (0,0)--(\hh,0);
		\foreach \aa in {55,125,90}{
			\draw[thick,->-=0.3] ({cos(\aa)},{sin(\aa)}) -- (0,0);
			\draw[blue,thick] ({cos(\aa)},{sin(\aa)}) -- ({\rr*cos(\aa)},{\rr*sin(\aa)});
			\filldraw ({cos(\aa)},{sin(\aa)}) circle (0.025);
			\filldraw[thick,fill opacity=0.15,->-=0.25] ({\rr*cos(\aa)},{\rr*sin(\aa)}) circle ({\rr-1});
			\fill[blue] ({\rr*cos(\aa)},{\rr*sin(\aa)}) circle (0.025);
		}
		\filldraw (-\hh,0) circle (0.025);
		\filldraw (\hh,0) circle (0.025);
		\filldraw (0,0) circle (0.03);
		\foreach \aa in {55,125,90}{
			\draw[red,thick]
			({\rc*cos(\aa)+\cl*cos(\aa+90)},{\rc*sin(\aa)+\cl*sin(\aa+90)}) --
			({\rc*cos(\aa)-\cl*cos(\aa+90)},{\rc*sin(\aa)-\cl*sin(\aa+90)});
		}
		\node at (0,-1.45) { };
	\end{tikzpicture}
	\caption{Left: A boundary anchored cut $\gamma_a$. 
		If we imagine taking a large number of boundary points, and always demanding that $\gamma_a$ enclose no such points, then the boundary region that $\gamma_a$ is anchored to can be interpreted as infinitesimally sized. The corresponding ``tube'' in the bulk connecting the matter legs to the boundary is what allows the bulk region to remain dressed to the boundary. While this ``tube'' is essential for the gravitational dressing which forces the matter legs have the same edge mode structure as a boundary anchored region, we do not expect this tube to contribute significantly to the entropy of the region, because of its infinitesimal size.
		Right: The same cut on the double bowtie lattice.}
	\label{fig:compactregionboundarycut}
\end{figure}

It is worth being more explicit about which cut through the network \eqref{eq:rhoadef} corresponds to, which we draw in Fig.~\ref{fig:compactregionboundarycut}. After using moves 1 and 2 to present the state in the reduced lattice, the legs cut by the factorization map are the stems joining the lollipops in $a$ to the corner vertex $\eth$. 
The curve on $\Sigma$ which crosses exactly these stems, and no other leg, is a boundary anchored curve encircling only the lollipops of $a$ and no boundary legs, which we denote $\gamma_a$. In the notation of Sec.~\ref{sec:concentration}, we can say  $a = \mathrm{int}(\gamma_a)$. 
Note that because $\mathrm{int}(\gamma_a)$ contains no boundary vertices, $a$ describes non-boundary degrees of freedom. However, $\gamma_a$ is still a boundary anchored curve: it enters $\Sigma$ between two adjacent boundary vertices, encircles the lollipops of $a$, and returns to the boundary through the same interval. Therefore, the edge modes $\gamma_a$ introduces are the same as for any other boundary anchored curve, one representation $\pi \in \widehat{G}$ per stem. This means we can still use the factorization map $V_{\mathrm{full}}$ along it to factorize the Hilbert space. In the language of Sec.~\ref{sec:doublemodel}, the sectors which $\gamma_a$ can see are the pure charges $(e,\pi)$, the anyons which condense on the rough boundary condition we impose at the cut. 

Two features of $\gamma_a$ deserve more explanation. First, the interval of $\partial\Sigma$ between the two endpoints of $\gamma_a$ contains no marked points. Thus, in the limit $n \to \infty$ of Sec.~\ref{sec:themodel} in which the open boundary condition becomes a conformal one, this interval is of the order of the lattice spacing, and the boundary degrees of freedom it contains are negligible. From the bulk point of view, the strip between $\gamma_a$ and this interval is a thin tube connecting $a$ to $\partial\Sigma$, with the rough boundary condition on both of its walls. Since the pure charges are the anyons which can end on a rough boundary, a charged line inside the tube can simply terminate on its walls. In other words, the charge $\pi$ flowing out of $a$ through its stems is absorbed by the cut, which is what labels the sectors of $\rho_a$. The line operators which can run along the tube without ending are instead the pure fluxes, which a rough boundary does not absorb. These are the central operators $\hat{C}$ of Sec.~\ref{sec:Arcenter}, which act on the stems through class functions and whose eigenvalues measure the charge $\pi$ of the region, so the tube carries exactly the operators which measure the total charge of $a$.

Second, the choice of boundary interval to anchor this cut to is arbitrary: the reduced state \eqref{eq:rhoadef} is defined by tracing out $R$, $\overline{R}$ and $\overline{a}$, and makes no reference to where $\gamma_a$ meets $\partial\Sigma$, so anchoring the tube in any other interval which avoids the marked points gives the same $\rho_a$. Conversely, a closed ribbon operator\footnote{See Appendix~\ref{app:ribbons} for the definition of a closed ribbon operator. At least for finite gauge groups, they exhaust the gauge invariant, line-like operators which can form closed loops on the surface, and are in one-to-one correspondence with the anyons of the TQFT associated with the tensor network.} with nontrivial flux cannot end on the rough cut, so it does not label a sector of the boundary dressed factorization. Such operators do exist along a closed loop $\widetilde{\gamma}_a$ encircling $a$, but $V_{\mathrm{full}}$ treats them as operators within each sector $\pi$ rather than as edge modes. We return to these additional line operators in Sec.~\ref{sec:compactlysupported}.

In this sense, we can think of the region $a$ as being ``dressed'' to the boundary, despite the fact that its extended Hilbert space contains no boundary degrees of freedom. In the continuum, the picture to keep in mind is a ``thin tube'' stretching from the boundary to the region $a$, along which only the pure fluxes which measure the charge of $a$ are allowed to propagate all the way to the boundary. We will call bulk regions defined in this way ``boundary dressed''. We will now compute the entanglement entropy of a boundary dressed bulk region in two different models for the matter legs, and compare the results.

\subsubsection{Fixed exterior} 

Suppose first that we do not average over any of the matter legs. Then computing the entropy of $\rho_a$ is completely analogous to the computation along a fixed cut in Sec.~\ref{sec:entropy_fixedcut}. The reduced state takes the form \eqref{eq:rhoRgamma} with $\Ha_R(\pi;\gamma)$ replaced by the sector-wise Hilbert space of the legs in $a$, the algebra of operators on $a$ has a center generated by the operators $\hat{C}$ acting on the stems, and the traces on that algebra and on its center are fixed by the same two principles as before, because the observer edge modes of $V_{\mathrm{full}}$ are insensitive to which legs we group together, as we showed around \eqref{eq:intertwinerpartialtrace}. Repeating the steps of Sec.~\ref{sec:entropy_nomatter}, we find
\begin{align}
	S(\rho_a) = H[\,p^{\gamma_a}_\pi] + \langle \hat{A}_{\gamma_a} \rangle_\rho + S_{\mathrm{bulk}}(\rho_a) \,, \label{eq:Sa_fixed}
\end{align}
which is \eqref{eq:Sgamma} with $\rho_R[\gamma] \to \rho_a$. In particular, the area operator $\hat{A}_{\gamma_a}$ is the \emph{same} operator as in the case of a boundary subregion, now supported on the cut $\gamma_a$, and its spectrum is once again $\ln(\mu(\pi))$. Equation \eqref{eq:Sa_fixed} is the entropy of the region $a$ as it is, with the area term evaluated on the boundary of the region itself. This is the structure of the entropy of a finite region proposed in \cite{Jensen2023}, where the region is fixed by hand and its generalized entropy includes the area of its own boundary.

\subsubsection{Random exterior} Alternatively, we can contract the matter legs in $\overline{a}$ into a Gaussian random state $\dket{T_{\overline{a}}}$ drawn from the ensemble of Sec.~\ref{sec:ensembledef}, and leave the legs in $a$ alone. We can think of this as a model of a situation in which we ``know'' the state of the bulk inside the region $a$, but are ignorant of the state of the bulk outside of it. A maximum ignorance principle would then instruct us to average over all bulk states in $\overline{a}$ which are compatible with a fixed state on $a$. This prepares the $\dket{T_{\overline{a}}}$-dependent state
\begin{align}
	\rho_a[T] = \Tr_{R \overline{R}\, \overline{a}}\left[\left(\Id_{R \overline{R} a} \otimes \dketbra{T_{\overline{a}}}{T_{\overline{a}}}\right) V_{\mathrm{full}} \dketbra{\psi}{\psi} V_{\mathrm{full}}^\dagger\right] \,, \label{eq:rhoaT}
\end{align}
which is the analog of \eqref{eq:rhoRT} for the region $a$. Its ensemble average is exactly \eqref{eq:rhoadef}, by \eqref{eq:firstmoment}, and the concentration argument of Sec.~\ref{sec:concentration} goes through with $\rho_{\mathfrak{l}}$ replaced by the reduced state of the legs in $\overline{a}$.

The replica computation of Sec.~\ref{sec:averageentropy} also goes through, with one change in the boundary conditions of the permutation sum. In \eqref{eq:replicapermsum}, the $R$ legs carried the cyclic permutation $\tau_n$ and the $\overline{R}$ legs carried the identity, and every matter leg was summed over. Here, the legs in $a$ carry $\tau_n$ because it is their R\'enyi entropy we are computing. Additionally, both boundary subregions $R$ and $\overline{R}$ carry the identity permutation. Only the legs in $\overline{a}$ are summed over after averaging over the state of the matter legs there. Under the same assumptions of replica symmetry and a dominant saddle point, each configuration of the sum is again a bipartition of the matter legs, and hence a cut $\gamma$ in the sense of Sec.~\ref{sec:entropy_matter_norandom}. The difference from the boundary anchored case is entirely in which cuts are allowed to appear. Because the legs in $a$ are fixed to $\tau_n$, every cut must contain $a$ in its interior. Because neither boundary subregion is ``twisted'', every cut separates the $\tau_n$ domain from a domain containing all of the boundary legs, so the interior of the cut will contain no boundary points. We therefore find
\begin{align}
	\mathbb{E}\left[S(\rho_a)\right] \approx \min_{\gamma \supset a}\left[H[\,p^{\gamma}_\pi] + \langle \hat{A}_{\gamma} \rangle_\rho + S_{\mathrm{bulk}}\!\left(\rho_{\mathrm{int}(\gamma)}\right)\right] \,, \label{eq:finiteregionresult}
\end{align}
where the minimization runs over the closed cuts $\gamma$ which enclose $a$ and do not enclose any boundary legs, and $\rho_{\mathrm{int}(\gamma)}$ is the reduced state of all the matter legs enclosed by $\gamma$. The condition $\gamma \supset a$ plays the role of the homology condition $\gamma \sim R$ in \eqref{eq:mainresult}: the region whose entropy we compute must be contained in the region bounded by the minimizing cut, and the minimizing cut may enlarge the region by absorbing legs from $\overline{a}$ but may never reach the boundary.

Equation \eqref{eq:finiteregionresult} says that the entropy of a compact region in a typical semiclassical state is the minimum of the generalized entropy over all regions of the network which contain it. This is the structure of the generalized entanglement wedge proposed by Bousso and Penington \cite{Bousso:2022hlz,Bousso:2023sya}, in which the entropy of a bulk region $a$ is computed by the generalized entropy of the smallest region containing $a$, rather than by the generalized entropy of $a$ itself.\footnote{See also \cite{Kaya:2025vof,Bozanic:2026ios,Sahu:2025upe} for other work on understanding the Bousso--Penington proposal within tensor networks. } It is also the same structure as the schematic derivation of the entropy of a finite region in \cite{Balasubramanian:2023dpj}: one first uses edge modes to define the state of the region, and then computes its entropy.

The two paragraphs above give a single mechanism which interpolates between the two proposals in the literature, at least in this topological model. If the exterior of the region is fixed, the entropy is that of the region itself, as in \cite{Jensen2023}, while if the exterior is random, the boundary of the region is free to fluctuate outward until it finds the minimal generalized entropy, as in \cite{Bousso:2022hlz,Bousso:2023sya}. 

In this model, it is the average over the matter legs outside the region which produces the outward deformations of the boundary of $a$. We noted in Sec.~\ref{sec:ensembledef} that this average is not obviously independent of the gauge structure of the network. This is because the ensemble average $\mathbb{E}$ over random matter and the renormalizing operator valued weight $\mathcal{E}$ implementing the group averaging both serve as maps from $\mathcal{B}_R \to \mathcal{A}_R$. A similar identification between chaotic behavior of matter and group averaging has been suggested in two dimensional de Sitter space \cite{Kolchmeyer:2024fly}. If the average is in this sense supplied by the group averaging over the edge modes, then it is the gauge symmetry itself which causes the boundary of the region to fluctuate, which is the mechanism for outward fluctuations of a finite region suggested in \cite{Balasubramanian:2023dpj}. It would be very interesting to determine whether the Gaussian average is in fact superfluous for non-compact gauge groups. We leave this for future work.

\subsection{What is the area of a compact region?} There is one feature of \eqref{eq:Sa_fixed} and \eqref{eq:finiteregionresult} which we should be explicit about. A compact region in this model is specified by its \emph{content}, namely the matter legs it contains, and not by a curve drawn on $\Sigma$. Correspondingly, the area operator $\hat{A}_{\gamma_a}$ measures the representation flowing out of $a$ through its stems, which by the Gauss law at the corner vertex $\eth$ is the total charge of the matter inside under the (quantum deformed) diffeomorphism group. In other words, the ``area'' of a compact region is not independent geometric data: it is fixed by the state of the matter within the region, in the same way that the holonomy around a closed curve in a flat connection is fixed by the flux it encircles.

This has a consequence for the gravitational interpretation of Sec.~\ref{sec:examples}. There, the sectors relevant to a boundary anchored cut were the principal series, where the Liouville momentum $P$ is real and the geodesic length $\ell$ of \eqref{eq:Pofell} is the length of a boundary anchored geodesic. However, an arbitrary closed curve on a hyperbolic surface need not have a geodesic in its homotopy class. For example, the holonomy around a conical defect, i.e., a particle below the black hole threshold, is elliptic rather than hyperbolic, and no closed geodesic encircles it. In the dictionary of Sec.~\ref{sec:quantumgroups}, these are sectors with imaginary $P$, for which $\rho_0(P)$ is oscillatory rather than exponentially large, and \eqref{eq:areaexpansion} has no leading term.
For such a matter configuration, the fact that the entropy is $\mathcal{O}(G_N^0)$ can be interpreted as a ``boundary area'' which is Planckian in size, so we probably should not trust the effective field theory defining the tensor network in the first place in this case.
On the other hand, a region whose content is above the black hole threshold has hyperbolic holonomy, a closed geodesic in the homotopy class of $\gamma_a$, and an area eigenvalue $\ell/4G_N$ exactly as in Sec.~\ref{sec:btz}, where the closed geodesic was the horizon. It would be very interesting to determine what the analog of a ``subset of matter legs'' is directly in the continuum, and in particular whether the content-defined regions of this model correspond to the regions of \cite{Bousso:2022hlz,Bousso:2023sya,Balasubramanian:2023dpj} or some coarse graining of them.

\subsection{Compactly supported regions} \label{sec:compactlysupported}

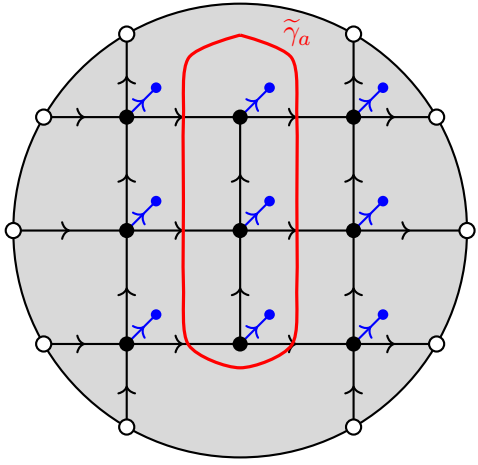
\begin{figure}
	\centering
    \begin{tikzpicture}[scale=2,thick]
    \def\Rd{1.5}          
    \def\nn{3}            
    \def\nnm{2}           
    \def\vertcols{1,3}    
    \def\rind{0.16}       
    \def\rbdy{0.05}       
    \def\rblk{0.05}       
    \def\dstem{0.275}      
    \def\rlol{0.1125}       
    \def\lolang{45}       
    \def\brc{0.40}        
    \colorlet{latt}{black}
    \colorlet{matt}{blue}
    \colorlet{cutc}{red}
    \pgfmathsetmacro{\sep}{2*\Rd/(\nn+1)}
    \pgfmathsetmacro{\gap}{asin(\rind/\Rd)}
    
    
    \filldraw[fill opacity=0.15] (0,0) circle (\Rd);
    \foreach \i in {1,...,\nn}{                     
        \pgfmathsetmacro{\X}{-\Rd+\i*\sep}
        \pgfmathsetmacro{\angC}{asin(\X/\Rd)}
        \pgfmathsetmacro{\angD}{180-\angC}
    }

    \foreach \i in {1,...,\nn}{
        \pgfmathsetmacro{\X}{-\Rd+\i*\sep}
        \pgfmathsetmacro{\Ed}{sqrt(\Rd*\Rd-\X*\X)}
        \pgfmathsetmacro{\Vone}{-\Rd+\sep}
        \pgfmathsetmacro{\Vtop}{-\Rd+\nn*\sep}
        \draw[latt,->-=0.5] (-\Ed,\X) -- (\Vone,\X);
        \draw[latt,->-=0.5] (\Vtop,\X) -- (\Ed,\X);
        \foreach \j in {1,...,\nnm}{
            \pgfmathsetmacro{\Ya}{-\Rd+\j*\sep}
            \pgfmathsetmacro{\Yb}{-\Rd+(\j+1)*\sep}
            \draw[latt,->-=0.5] (\X,\Ya) -- (\X,\Yb);
            \draw[latt,->-=0.5] (\Ya,\X) -- (\Yb,\X);
        }
    }
    \foreach \i in \vertcols{
        \pgfmathsetmacro{\X}{-\Rd+\i*\sep}
        \pgfmathsetmacro{\Ed}{sqrt(\Rd*\Rd-\X*\X)}
        \pgfmathsetmacro{\Vone}{-\Rd+\sep}
        \pgfmathsetmacro{\Vtop}{-\Rd+\nn*\sep}
        \draw[latt,->-=0.5] (\X,-\Ed) -- (\X,\Vone);
        \draw[latt,->-=0.5] (\X,\Vtop) -- (\X,\Ed);
    }
    
    \foreach \i in {1,...,\nn}{
        \pgfmathsetmacro{\X}{-\Rd+\i*\sep}
        \foreach \j in {1,...,\nn}{
            \pgfmathsetmacro{\Y}{-\Rd+\j*\sep}
            \pgfmathsetmacro{\cx}{\X+\dstem*cos(\lolang)}
            \pgfmathsetmacro{\cy}{\Y+\dstem*sin(\lolang)}
            \draw[matt,->-=0.6] (\X,\Y) -- (\cx,\cy);
            \fill[matt] (\cx,\cy) circle (0.035);
            \fill[latt] (\X,\Y) circle (\rblk);
        }
    }
    
    \foreach \i in {1,...,\nn}{
        \pgfmathsetmacro{\X}{-\Rd+\i*\sep}
        \pgfmathsetmacro{\angC}{asin(\X/\Rd)}
        \pgfmathsetmacro{\angD}{180-\angC}
        \filldraw[latt,fill=white] (\angC:\Rd) circle (\rbdy);
        \filldraw[latt,fill=white] (\angD:\Rd) circle (\rbdy);
    }
    \foreach \i in \vertcols{
        \pgfmathsetmacro{\X}{-\Rd+\i*\sep}
        \pgfmathsetmacro{\angA}{acos(\X/\Rd)}
        \pgfmathsetmacro{\angB}{-\angA}
        \filldraw[latt,fill=white] (\angA:\Rd) circle (\rbdy);
        \filldraw[latt,fill=white] (\angB:\Rd) circle (\rbdy);
    }
    
    \draw[cutc,very thick,yshift=.5em] plot[smooth] coordinates
    {(0,1.1) 
        (-0.34,0.95) (-0.375,0.55) 
        (-0.375,0) 
        (-0.375,-0.55) (-0.34,-0.95) 
        (0,-1.1) 
        (0.34,-0.95) (0.375,-0.55) 
        (0.375,0) 
        (0.375,0.55) (0.34,0.95) 
        (0,1.1) } ;
    \node[cutc,anchor=south] at (.38,1.15) {$\widetilde{\gamma}_a$};

    \end{tikzpicture}
	\caption{Closed cut $\widetilde{\gamma}_a$. This definition of the bulk subregion $a$ is still gauge invariant because of the complete set of edge modes which live on a closed cut.}
	\label{fig:compactregionbulkcut}
\end{figure}

The boundary dressed definition has a limitation which we have already met in Sec.~\ref{sec:btz}. A closed curve in a TQFT can support every anyon of the theory, not only the pure charges, and in the gravitational dictionary the additional sectors are those with $P_+ \neq P_-$, which are the sectors carried by the horizon of a rotating black hole. The boundary dressed region sees only the charges, and so it is adequate only when the content of the region is time reversal invariant in the sense of Sec.~\ref{sec:timesymmetric}. This suggests a second definition of a compact region which lifts this restriction.\footnote{See \cite{Wong:2025kpz} for a similar proposal of a local gravitating subregion.}

Alternatively, then, we could define the region $a$ by drawing the closed curve $\widetilde{\gamma}_a$ around its matter legs and factorizing the Hilbert space along $\widetilde{\gamma}_a$ directly, analogous to the two sided black hole example in Sec.~\ref{sec:btz} (see Fig.~\ref{fig:compactregionbulkcut}). For a topological tensor network built from a transformable group $G$, the edge modes required for such a factorization are the ones we explain in Appendix~\ref{app:ribbons}: the closed ribbon operators supported on $\widetilde{\gamma}_a$ are labeled by a flux $[t]$ together with a representation $\lambda$ of its centralizer, with label space the diagonal quotient \eqref{eq:diagonalquotient}, and we expect the edge modes along a closed cut to carry the same data. For gravity, the anyons supported on a closed loop would be the same as in the two-sided black hole of Sec.~\ref{sec:btz}.
So even on the disk, a compactly supported region can carry edge modes which are not present for any boundary anchored curve. The overall factorization would still be gauge invariant, because both the region and its complement are dressed to the cut $\widetilde{\gamma}_a$ by their edge modes, rather than to the boundary through a tube. We will call regions defined in this way ``compactly supported''.

We have not constructed the factorization map along a closed cut in this paper, and we list the ingredients required by that construction at the end of Appendix~\ref{app:ribbons}. In short, we would need the analogs of the nested algebras of Sec.~\ref{sec:algebras}, a measure on the label space \eqref{eq:diagonalquotient} for each of their traces, and the operator valued weights relating them. That being said, note that the Gaussian average of Sec.~\ref{sec:ensembledef} did not depend in any detailed way on the structure of the edge modes. It used only that the ensemble average implements a partial trace, \eqref{eq:firstmoment}, and that higher moments produce a sum over permutations, \eqref{eq:gaussianmoments}, both of which are statements about the matter legs rather than about the cut. We therefore expect the argument leading to \eqref{eq:finiteregionresult} to hold for compactly supported regions as well, under the same assumptions,\footnote{The analysis of generalized entanglement wedges \cite{Bousso:2023sya} in time dependent geometries suggests that some of the assumptions such as replica symmetry may fail for general compact subregions. One could imagine then using the ideas from one-shot holography \cite{Akers:2023fqr, Akers:2021fut, Akers:2020pmf} to perform a more refined analysis of the R\'enyi entropies of the tensor networks in that case.} so that
\begin{align}
	\mathbb{E}\left[S(\rho_a)\right] \approx \min_{\gamma \supset a}\left[H[\,p^{\gamma}] + \langle \hat{A}_{\gamma} \rangle_\rho + S_{\mathrm{bulk}}\!\left(\rho_{\mathrm{int}(\gamma)}\right)\right] \,, 
\end{align}
with the only difference being the label set over which the sectors run and the (quantum) group theory data which determines the spectrum of $\hat{A}_\gamma$ on it.
Following the proposal \eqref{eq:areainvariant} of Sec.~\ref{sec:doublemodel}, that spectrum should be $\ln(\mathcal{S}_1^{\,\xi})$ with $\xi$ running over all anyons, which in the dictionary of Sec.~\ref{sec:quantumgroups} is \eqref{eq:areaspectrum_VTQFT} with both $P_\pm$ free. Since this is exactly the spectrum which reproduced the horizon length of the rotating black hole in Sec.~\ref{sec:btz}, we expect $\hat{A}_\gamma$ to retain its interpretation as a length divided by $4G_N$ for compactly supported regions, including regions whose content is not time symmetric. Settling this requires the explicit closed cut factorization map, and we leave it for future work.

\section{Summary and Discussion} \label{sec:discussion}

In this paper, we discussed how to factorize the Hilbert space of tensor networks which prepare states in topological field theories with arbitrary transformable gauge groups $G$. When the tensor network has no matter legs, we found that the entropy of a boundary subregion $R$ is given by \eqref{eq:SrhoR_gen},
\begin{align}
	S(\rho_R) = H[\,p_\pi] + \langle \hat{A}\rangle_\rho + \int d\mu(\pi) q_\pi S(\rho_\pi) \,. \label{eq:disc1}
\end{align}
Here, $H[\,p_\pi] $ is the differential entropy which measures uncertainty about which sector $\pi$ the reduced state is in, $\langle \hat{A}\rangle_\rho$ is the expectation value of a state independent operator that we identify with the area operator in gravity, and the third term is an average entropy within each sector. 

When we include matter that has been set to a fixed state, we found the entropy formula \eqref{eq:Sgamma}
\begin{align}
	S(\rho_R[\gamma]) =  H[\,p^\gamma_\pi] + \langle \hat{A}_\gamma \rangle_\rho + S_{\mathrm{bulk}}(\rho_{\mathrm{int}(R \cup \gamma)}) \,, \label{eq:disc2}
\end{align}
where $\gamma$ denotes a fixed choice of cut through the tensor network, $\rho_R[\gamma]$ is the reduced state which has been factorized along $\gamma$, and the terms in the entropy formula have the same interpretation as in \eqref{eq:disc1}. 

When we took the wave function coefficients of the matter state to be Gaussian random variables, then the ensemble average of the entropy of $\rho_R$ over possible matter states instead takes the form \eqref{eq:mainresult}
\begin{align}
	\mathbb{E}\left[S(\overline{\rho}_R)\right] \approx \min_{\gamma \sim R}\left[H[\,p^\gamma_\pi] + \langle \hat{A}_\gamma \rangle_\rho + S_{\mathrm{bulk}}\!\left(\rho_{\mathrm{int}(R \cup \gamma)}\right)\right]  \,, \label{eq:disc3}
\end{align}
where $\gamma \sim R$ denotes a minimization over all boundary anchored curves which are homologous to the subregion $R$. 

Finally, we used this formalism to propose a definition of a finite region $a$ in the bulk (a subset of the matter legs of the tensor network), and found two possible answers for the entropy of this definition of subregion. If we did not average over the matter legs outside of the subregion $a$, we found that \eqref{eq:Sa_fixed}
\begin{align}
	S(\rho_a) = H[\,p^{\gamma_a}] + \langle \hat{A}_{\gamma_a} \rangle_\rho + S_{\mathrm{bulk}}\!\left(\rho_a\right) \,, \label{eq:disc4}
\end{align}
which roughly corresponds to the generalized entropy of the subregion $a$ itself, matching a proposal given in \cite{Jensen2023}. If instead we assumed that the wave function coefficients \emph{outside} the subregion $a$ are well-approximated by Gaussian random variables, we find \eqref{eq:finiteregionresult}
\begin{align}
	\mathbb{E}\left[S(\overline{\rho}_a)\right] \approx \min_{\gamma \supset a}\left[H[\,p^{\gamma}] + \langle \hat{A}_{\gamma} \rangle_\rho + S_{\mathrm{bulk}}\!\left(\rho_{\mathrm{int}(\gamma)}\right)\right] \,. \label{eq:disc5}
\end{align}
This roughly corresponds to the minimal generalized entropy among all bulk subregions which contain the original one, which matches the generalized entanglement wedge proposal \cite{Bousso:2022hlz,Bousso:2023sya}. 

The connection between topological entanglement entropy and holographic entanglement entropy has been anticipated in the literature \cite{Donnelly_2016,Lin:2017uzr,Jafferis:2019wkd,Wong:2022eiu,Mertens:2022ujr,Mertens:2025ydx,Mertens:2022aou,Chua:2023ios,Balasubramanian:2023dpj}. Specifically, in \cite{McGough:2013gka}, it was shown that the topological entanglement entropy should match the black hole entropy spectrum of 3D gravity, and \cite{Wong:2022eiu} gives another TQFT path integral perspective of this match. Furthermore, \cite{Carlip:1994gy} (see also \cite{Strominger:1997eq}) showed that counting edge modes near the horizon correctly captures the black hole entropy.
Our derivation differs from these results in a few important ways. First, we canonically quantized an analog of the 3D gravity Hilbert space and gave a method to compute the entanglement of arbitrary states in the Hilbert space. This provides a framework (within canonical quantization) which gives a more concrete justification of the identification of these quantities. Furthermore, we also showed \emph{why} tensor networks give results similar to gravity, at least in 3d: assuming the quantum group replacement works as expected, the 3D gravity Hilbert space can essentially be thought of as a string net, which has a tensor network description. Additionally, this tensor network description of the topological entanglement entropy provides a framework to compute this entropy for a compact region of spacetime. Finally, our entropy formula is in-principle well-defined even away from the saddle point approximation of a single dominant, semiclassical spacetime.

We emphasize, however, that even the hypothetical $\mathcal{U}_q(\sl(2,\R))$ tensor network would not precisely agree with 3D gravity. The phase space that the $\mathcal{U}_q(\sl(2,\R))$ tensor network quantizes agrees \emph{locally} with the phase space of canonical quantum gravity in three dimensions, and requires a global quotient by the mapping class group of the spatial slice $\Sigma$ to match completely. 
There is a subtlety associated with this quotient: Maloney and Witten showed in \cite{Maloney:2007ud} that the density of states of a single boundary Euclidean AdS spacetime receives negative contributions after taking the renormalized sum over the mapping class group of the boundary. While the context they studied is not quite the same as ours, their results show that one must be careful after gauging the mapping class group of $\Sigma$, for it is possible that the inner product on the physical Hilbert space might no longer be positive definite. 

If the inner product is not positive definite after summing over the mapping class group, then it would mean that 3D gravity is not a unitary theory by itself, and requires additional degrees of freedom to restore unitarity. It would be interesting to use the tensor network model to test various proposed resolutions to this puzzle, such as a sum over off-shell topologies \cite{Maxfield:2020ale}, and see which ones would restore positivity.
Indeed, we argued in \cite{Balasubramanian:2026ymu} that non-perturbative effects are required for a genuine holographic interpretation of the generalized entropy formulas we derived in the main text, so perhaps similar effects would cure any potential negativity as well.
Determining if the inner product of the physical Hilbert space remains positive definite after gauging the mapping class group is therefore an important step in checking if pure 3D gravity in Lorentzian AdS spacetimes is well-behaved on its own. It is interesting to speculate that after accounting for all of these features, the resulting tensor model may be equivalent to a canonical quantization of the three-dimensional path integral proposed in \cite{Jafferis:2025jle,Jafferis:2025yxt,Jafferis:2025vyp}.

To derive the entropy formulas \eqref{eq:disc3} and \eqref{eq:disc5}, we assumed that the matter state was taken to be randomly chosen with respect to a Gaussian distribution. This is not unusual in the tensor network literature \cite{Hayden_2016,Akers:2021pvd,Cheng:2022ori,Vasseur:2018gfy,Chirco:2017wgl,Qi:2022lbd,Chandra:2023dgq}, but we would like to understand the gravitational meaning of this behavior in more detail. One possibility is that although a smooth matter distribution in a semi-classical spacetime does not look very random, in the underlying quantum gravitational theory, such distributions are highly irregular. Indeed, there are many indications that the gravitational path integral behaves as if it has an inherently ``averaged'' \cite{Marolf:2020xie,Jafferis:2025vyp,Jafferis:2025yxt,Schlenker:2022dyo,Coleman:1988cy,Giddings:1988cx,Saad:2019lba,Chandra:2022bqq,Saad:2021rcu} or ``complex'' behavior \cite{Balasubramanian:2024lqk,Balasubramanian:2026azk,Couch:2018phr,Brown:2015lvg,Belin:2021bga,Stanford:2014jda,Rabinovici:2023yex,Cotler:2016fpe,Shenker:2014cwa,Maldacena:2015waa,Polchinski:2015cea,Perlmutter:2016pkf,Jensen:2016pah,Mezei:2016wfz,Yoshida:2017non,Hayden:2007cs}. 
It would be interesting to understand whether this assumption could be justified in more detail.

In the QES formula \eqref{eq:QESformula}, the minimization is over surfaces $\gamma$ which are \emph{homologous} to the boundary subregion $R$. On the disk with random matter we found the same constraint in \eqref{eq:disc3}: two cuts which enclose different sets of matter legs are not homotopic to one another, since neither can be deformed past a matter leg, but both are homologous to $R$, because $R \sqcup \gamma$ bounds the disk $\Sigma_R[\gamma]$ which defines the extended Hilbert space $\Hext{R}[\gamma]$. 
The homology condition affects both sides of the duality in AdS/CFT \cite{Haehl:2014zoa,Almheiri:2016blp,Papadodimas:2015jra}, and is known to be subtle in tensor network models of gravity \cite{Harlow:2016vwg}, so it would be interesting to check that it is homology, and not homotopy, which continues to control the minimization in more general topologies in the tensor model.

In this work, a crucial role was played by the edge modes introduced by the factorization maps we used to compute the entropy. These edge modes play a similar role to recently proposed observer models in quantum gravity \cite{Abdalla:2025gzn,Harlow:2025pvj}, quantum reference frames \cite{Fewster:2024pur}, and the embedding map of the corner symmetry proposal \cite{Ciambelli:2021vnn,Ciambelli:2021nmv,Freidel:2021cjp,Ciambelli:2022cfr,Ciambelli:2024qgi,Balasubramanian:2023dpj}. It would be interesting to understand these connections in more detail.

\paragraph{Acknowledgments:} We thank Xi Dong, Don Marolf, Wayne Weng, Chris Akers, Yimu Bao, Jon Sorce, Elba Alonso-Monsalve, Dan Sehayak, Leo Shaposhnik, Alexander Jahn, Wissam Chemissany, Sami Kaya, and Juan Maldacena for helpful discussions.
CC is supported by the National Science Foundation Graduate Research Fellowship under Grant No. DGE-2236662. 
CC was also supported in part by grant NSF PHY-2309135 to the Kavli Institute for Theoretical Physics (KITP). 
VB is supported in part by the DOE through DE-SC0013528 and QuantISED grant DE-SC0020360.

\appendix

\section{Closed ribbon operators}
\label{app:ribbons}

Every cut $\gamma$ used in the main text is a boundary anchored curve. The factorization maps of Sec.~\ref{sec:factorization}--\ref{sec:entropy_matter_random} were built accordingly: move 1 splits the single vertex of the reduced lattice, and the corner leg it produces supports the gauge invariant line operators which can end on the cut. For the rough boundary condition imposed at the cut in the main text, these line operators are labeled by a single representation $\pi \in \widehat{G}$. The operators which run along the cut $\gamma$, then, are generated by the anyons which can measure the total charge of the anyons ending on the cut. Because $\gamma$ ends on the smooth segment of the open boundary condition, the anyons which run along $\gamma$ are restricted to be the pure fluxes (which arrange themselves into the center $\mathcal{Z}(\mathcal{A}_R)$). However, an operator wrapping a closed curve $\gamma$ has no endpoints, so the boundary conditions do not constrain it. In this sense, gauge invariant operators supported on closed curves are more universal than boundary anchored line operators. The purpose of this appendix is to define these operators in the present model, and to determine what group theory data labels them.

Gauge invariant operators in topological tensor networks are technically not supported on lines $\gamma$: they are supported on \emph{ribbons} \cite{Kitaev:1997wr,Bombin_2008}. A ribbon is a thickened path through the edges of the lattice. Consequently, the gauge invariant operators in this appendix are also called ribbon operators. A ribbon operator acts on the legs along both of its sides: along one side it multiplies the group elements it crosses, a magnetic action which transports a flux along the path, and along the other it weights the configuration by a character, an electric action which transports a charge \cite{Bombin_2008,Delcamp:2016eya}. Ribbon operators were introduced for Kitaev's double model with $G$ finite \cite{Kitaev:1997wr}, where such an operator is labeled by a conjugacy class $[g]$ of $G$ together with an irreducible representation $\lambda$ of the centralizer $C_G(g)$, which respectively label the flux and the charge carried by the ribbon operator. These pairs label the anyons of the TQFT that the double model is equivalent to. 
In this Appendix, we will similarly construct ribbon operators which depend on the same data when $G$ is a more general transformable group (see Sec.~\ref{sec:themodel}). It would be interesting to demonstrate explicitly that the ribbons we construct exhaust the complete set of ribbon operators when $G$ is a general transformable group, as they do when $G$ is a finite group, but we will leave this analysis for future work.

\begin{figure}
	\centering
	\begin{tikzpicture}[scale=1.5,thick]
		\def\rbdy{0.075}      
		\def\rblk{0.05}       
		\def\ylab{-1.85}      
		\colorlet{latt}{black!50}
		\colorlet{matt}{blue}
		\colorlet{cutc}{red}
		
		\begin{scope}[shift={(0,0)}]
			\def\Ro{1.40}     
			\def\Ri{0.44}     
			\def\Rl{0.80}     
			\def\Rg{1.08}     
			\draw (0,0) circle (\Ro);
			\draw (0,0) circle (\Ri);
			\draw[latt,->-=0.125] (0,0) circle (\Rl);
			\foreach \A in {120,240,360}{                 
				\draw[latt,->-=0.55] (\A:\Ro) -- (\A:\Rl);
			}
			\foreach \A in {60,180,300}{                   
				\draw[latt,->-=0.55] (\A:\Ri) -- (\A:\Rl);
			}
			\foreach \A in {60,120,180,240,300,360}{       
				\fill[latt] (\A:\Rl) circle (\rblk);
			}
			\draw[cutc,very thick,->-=0.125] (0,0) circle (\Rg);
			\fill[cutc,opacity=0.15, even odd rule] (0,0) circle (\Rg) (0,0) circle (\Rl);
			\node[cutc] at (76:{\Rg+0.17}) {$\gamma$};
			\foreach \A in {120,240,360}{\filldraw[latt,fill=white] (\A:\Ro) circle (\rbdy);}
			\foreach \A in {60,180,300}{\filldraw[latt,fill=white] (\A:\Ri) circle (\rbdy);}
			\node at (0,0) {$\overline{R}$};
			\node at (10:{\Ro+0.24}) {$R$};
		\end{scope}
		
		\begin{scope}[shift={(3.95,-0.62)},scale=0.86]
			\def\Rl{1.05}     
			\def\Rg{1.28}     
			\def\yB{-0.12}    
			\def\yA{1.62}     
			\def\fanB{0.36}   
			\def\fanA{0.44}   
			\draw[latt,->-=0.125] (0,0) circle (\Rl);
			\foreach \A in {200,270,340}{
				\draw[latt,->-=0.55] ($(0,\yB)+(\A:\fanB)$) -- (0,\yB);
			}
			\draw[latt,->-=0.55] (0,\Rl) -- (0,\yB);
			\foreach \A in {45,90,135}{
				\draw[latt,->-=0.55] ($(0,\yA)+(\A:\fanA)$) -- (0,\yA);
			}
			\draw[latt,->-=0.55] (0,\Rl) -- (0,\yA);
			\draw[cutc,very thick,->-=0.125] (0,0) circle (\Rg);
			\fill[cutc,opacity=0.15, even odd rule] (0,0) circle (\Rg) (0,0) circle (\Rl);
			\fill[latt] (0,\yB) circle (\rblk);
			\fill[latt] (0,\yA) circle (\rblk);
			\fill[latt] (0,\Rl) circle (\rblk);
			\foreach \A in {200,270,340}{
				\filldraw[latt,fill=white] ($(0,\yB)+(\A:\fanB)$) circle (\rbdy);
			}
			\foreach \A in {45,90,135}{
				\filldraw[latt,fill=white] ($(0,\yA)+(\A:\fanA)$) circle (\rbdy);
			}
			\node[latt] at (160:{\Rl-0.20}) {$k$};
			\node[latt,anchor=east]       at (-0.10,{0.5*(\Rl+\yA)+0.04}) {$\ell_R$};
			\node[latt,anchor=west]       at (0.10,{0.5*(\Rl+\yB)}) {$\ell_{\overline{R}}$};
			\node[latt,anchor=north east] at (-0.10,{\Rl-0.07}) {$v$};
			\node[latt,anchor=west]       at (0.10,\yA) {$v_R$};
			\node[latt,anchor=east]       at (-0.10,{\yB+0.20}) {$v_{\overline{R}}$};
			\node[latt,anchor=south]      at (0,{\yA+\fanA+0.11}) {$\vec{g}_R$};
			\node[latt,anchor=north]      at (0,{\yB-\fanB-0.11}) {$\vec{g}_{\overline{R}}$};
			\node[cutc] at (44:{\Rg+0.20}) {$\gamma$};
		\end{scope}
		
		\node at (0,\ylab) {(a)};
		\node at (3.60,\ylab) {(b)};
	\end{tikzpicture}
	\caption{The cylinder $\Sigma_{0,2}$ and a closed ribbon $\gamma$ on it, shown in red. Lattice legs are gray and oriented, boundary vertices are white, bulk vertices are solid gray, and the cut induced by the ribbon is shown in red.
		(a) A lattice on the surface, drawn as an annulus with the $R$ boundary outside and the $\overline{R}$ boundary inside, with three marked points on each. Every boundary leg runs into a single loop of the lattice which wraps the non-contractible cycle. Unlike the lollipop plaquettes of Fig.~\ref{fig:reduced}, the region enclosed by that loop is left unshaded, because no magnetic constraint is imposed on it. The cut $\gamma$ is a closed curve isotopic to the loop, drawn offset from it.
		(b) The same surface after the moves of Sec.~\ref{sec:themodel} have collected the $R$ boundary legs into a single tree meeting the loop from outside, and the $\overline{R}$ boundary legs into a single tree meeting it from inside, so that the loop surrounds the $\overline{R}$ boundary. The loop carries $k$ and is based at the bulk vertex $v$; the legs $\ell_R$ and $\ell_{\overline{R}}$ join $v$ to the two tree vertices $v_R$ and $v_{\overline{R}}$; and the boundary legs carry $\vec{g}_R$ and $\vec{g}_{\overline{R}}$. In this presentation $\gamma$ encircles the loop and crosses exactly one leg, $\ell_R$, which is why a ribbon supported on $\gamma$ measures $k$ and deposits its flux on $\ell_R$.}
	\label{fig:cylinderanyons}
\end{figure}
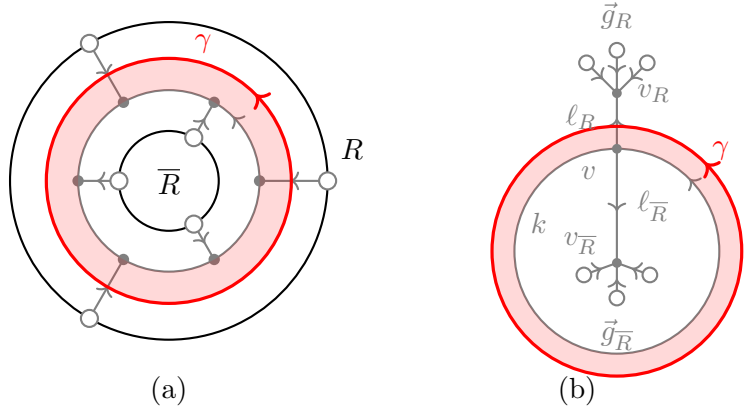

To define a closed ribbon operator, we need to work with a surface with a non-contractible cycle, so that such a closed loop exists. To this end, we work on the cylinder $\Sigma_{0,2}$, the simplest surface admitting a closed cut, presented by the lattice of Fig.~\ref{fig:cylinderanyons} \cite{Delcamp:2016eya}. Fig.~\ref{fig:cylinderanyons}(a) shows a lattice on the annulus (which is equivalent to the cylinder) in which every boundary leg runs straight into a single loop of the lattice wrapping the non-contractible cycle. Fig.~\ref{fig:cylinderanyons}(b) shows the same surface after the moves of Sec.~\ref{sec:themodel} have collected the boundary legs of each side into a tree, and it is the presentation we will work with. It has three bulk vertices: the vertex $v$ which the loop is based at, and the two vertices $v_R$ and $v_{\overline{R}}$ which the trees meet at. The loop carries $k$, the legs joining it to the two tree vertices carry $\ell_R$ and $\ell_{\overline{R}}$, and the boundary legs carry $\vec{g}_R$ and $\vec{g}_{\overline{R}}$. No magnetic constraint is imposed on $k$, because it wraps the non-contractible cycle. We write co-invariant basis states for $\Ha_{\mathrm{phys}}(\Sigma_{0,2})$ as
\begin{align}
	\dket{\vec{g}_R,\vec{g}_{\overline{R}},\ell_R,\ell_{\overline{R}},k} \,. \label{eq:cylbasis}
\end{align}

As in Sec.~\ref{sec:themodel}, a gauge transformation with parameter $u$ at the vertex $v$ multiplies the legs emanating from it on the left and conjugates the loop based there,
\begin{align}
	\ell_R \to u\, \ell_R \,, \qquad \ell_{\overline{R}} \to u\, \ell_{\overline{R}} \,, \qquad k \to u k u^{-1} \,, \label{eq:gaugev1}
\end{align}
leaving the boundary legs alone, while gauge transformations with parameters $u_R$ and $u_{\overline{R}}$ at the two tree vertices act on the far end of each leg by right multiplication,
\begin{align}
	\vec{g}_R \to \vec{g}_R\, u_R^{-1} \,, \quad \ell_R \to \ell_R\, u_R^{-1} \,, \qquad
	\vec{g}_{\overline{R}} \to \vec{g}_{\overline{R}}\, u_{\overline{R}}^{-1} \,, \quad \ell_{\overline{R}} \to \ell_{\overline{R}}\, u_{\overline{R}}^{-1} \,, \label{eq:gaugevR}
\end{align}
leaving $k$ alone. Co-invariance of \eqref{eq:cylbasis} means the state is unchanged under all three of these gauge transformations.

Because the boundary legs $\vec{g}_R$ and $\vec{g}_{\overline{R}}$ do not play a role in what follows, we will suppress them and write basis states as $\dket{\ell_R,\ell_{\overline{R}},k}$, restoring them only when we check that the ribbon operators we construct commute with the gauge transformations \eqref{eq:gaugevR}.

\subsection{Definition of the closed ribbon operator}

Let $[t]$ be a conjugacy class of $G$. We take its representative $t$ to be regular, so that by definition its centralizer $C_G(t) = T_B$ is the unique Cartan subgroup of $G$ which contains $t$. Then, let $\lambda$ be an irreducible unitary representation of that Cartan subgroup; since $T_B$ is abelian, $\lambda$ is a one dimensional character $\chi_\lambda$. Next, let
\begin{align}
	W(T_B) := N_G(T_B)/T_B
\end{align}
be the Weyl group of $T_B$, where $N_G(T_B)$ is its normalizer in $G$. For non-compact $G$ this is the \emph{real} Weyl group of that particular Cartan subgroup, and it can differ between Cartan subgroups which are not conjugate in $G$. For $G = \SL(2,\R)$, the hyperbolic Cartan subgroup has $W \cong \Z_2$, and the elliptic Cartan subgroup has a trivial Weyl group. We will use this data to define a closed ribbon operator $F_\gamma([t],\lambda)$ on the ribbon $\gamma$ shown in Fig.~\ref{fig:cylinderanyons}.

The remaining ingredients we need to define the closed ribbon operator are read off the state $\dket{\ell_R,\ell_{\overline{R}},k}$ itself. Given $k$, choose $x_0$ such that
\begin{align}
	x_0^{-1} k\, x_0 := k_t \in T_B \,, \qquad x_0 \in C_G(k) \backslash G / T_B \label{eq:x0def}\,.
\end{align}
In other words, choose a way of conjugating the holonomy $k$ into the same Cartan subgroup that houses the flux $[t]$. Such an $x_0$ exists precisely when the conjugacy class $[k]$ intersects the same Cartan subgroup $T_B$ as $[t]$; when the two conjugacy classes do not intersect the same Cartan subgroup, we define $F_\gamma([t],\lambda)\dket{\ell_R,\ell_{\overline{R}},k} = 0$. This means a closed ribbon operator only acts on states whose holonomy around the non-trivial cycle of the cylinder lies in the same Cartan subgroup as its flux: for example, if $G=\SL(2,\R)$, no closed ribbon connects the elliptic and hyperbolic sectors. Equivalently, the closed ribbon operators are block diagonal with respect to the $G$-conjugacy classes of Cartan subgroups.
We parameterize $x_0$ by the double coset $C_G(k) \backslash G / T_B$ to remove the redundancy in this parameterization: nothing below depends on a choice of representative for $x_0$ within this coset. 
Furthermore, there are finitely many choices of equivalence classes $x_0$ which satisfy \eqref{eq:x0def}, and they are all related by the action of the Weyl group as $x_w = x_0 w$, where $w \in W(T_B)$.

With these in hand, we can define
\begin{align}
	F_\gamma([t],\lambda) \dket{\ell_R,\ell_{\overline{R}},k} = \frac{1}{|W(T_B)|}\sum_{w \in W(T_B)} \chi_\lambda\bigl(w^{-1} k_t\, w\bigr)\; \dket{\, x_w\,t\, x_w^{-1}\, \ell_R,\; \ell_{\overline{R}},\; k} \,. \label{eq:ribbonweylsum}
\end{align}
This operator does two things at once, which is what makes it a general ribbon rather than a pure flux or a pure charge insertion. First, it deposits the flux on the $\ell_R$ leg, i.e., the one crossed by $\gamma$ in Fig.~\ref{fig:cylinderanyons}(b). The group element
\begin{align}
	f_w := x_w\,t\, x_w^{-1}\,, \label{eq:fluxelement}
\end{align}
which multiplies $\ell_R$, is the representative of the flux $[t]$ transported into the frame set by the holonomy $k$ of the co-invariant state. By a ``frame'' we mean the following. The holonomy $k$ picks out no preferred element of $T_B$ by itself; it is $x_0$, through \eqref{eq:x0def}, which identifies the centralizer of the holonomy with the Cartan subgroup housing the flux $[t]$. In other words, $x_0$ conjugates the Cartan subgroup $T_B$ containing $t$ as $C_G(k) = x_0 T_B x_0^{-1}$, and $f_w$ is the image of $w t w^{-1} \in T_B$ under that identification. In particular,
\begin{align}
	f_w \in C_G(k) \,, \label{eq:fluxcommutes}
\end{align}
so the flux deposited by a closed ribbon commutes with the holonomy it encircles. The sum over $W(T_B)$ then sums uniformly over the choices of such a frame.

Equation \eqref{eq:fluxcommutes} is also what makes \eqref{eq:ribbonweylsum} independent of the representative chosen for $x_0$ within its double coset, as claimed above. Right multiplication $x_0 \to x_0 \tau$ by $\tau \in T_B$ leaves $k_t$ and every $f_w$ untouched, since $T_B$ is abelian and $w t w^{-1} \in T_B$. Left multiplication $x_0 \to c\, x_0$ by $c \in C_G(k)$ leaves $k_t$ untouched and conjugates the flux, $f_w \to c f_w c^{-1}$; but by \eqref{eq:fluxcommutes}, both $c$ and $f_w$ lie in $C_G(k) = x_0 T_B x_0^{-1}$, which is abelian, so $f_w$ is unchanged as well.\footnote{This step is where we use that the holonomy $k$ is regular, so that $C_G(k)$ is the Cartan subgroup $x_0 T_B x_0^{-1}$ rather than something larger. We assume this alongside the regularity of the flux $t$; states with non-regular holonomy are discussed at the end of this appendix.}
Additionally, $F_\gamma$ weighs each term in the sum over the Weyl group by the character $\chi_\lambda$ evaluated on the holonomy, which is the charge measuring the flux it encircles. 

We will now demonstrate that $F_\gamma([t],\lambda)$ commutes with the gauge transformation~\eqref{eq:gaugev1}, and hence it is a gauge invariant operator. This implies that $F_\gamma([t],\lambda)$ has a definition directly on the physical Hilbert space, regardless of the choice of lattice $\Lambda$ which we use to present it.
The gauge transformations \eqref{eq:gaugevR} at $v_R$ and $v_{\overline{R}}$ act on the boundary legs and on the far end of $\ell_R$ and $\ell_{\overline{R}}$ by right multiplication. Since \eqref{eq:ribbonweylsum} leaves the boundary legs alone and multiplies $\ell_R$ from the left, $F_\gamma([t],\lambda)$ commutes with both of them immediately, and only gauge invariance at the vertex $v$ has to be checked.

At $v$, \eqref{eq:gaugev1} sends $k \to u k u^{-1}$, and the defining relation \eqref{eq:x0def} is then solved by $x_0 \to u x_0$, since $(ux_0)^{-1}(uku^{-1})(ux_0) = x_0^{-1}kx_0 = k_t$. The holonomy representative $k_t$ is therefore unchanged, so the characters $\chi_\lambda(w^{-1}k_tw)$ are unchanged, while the flux is conjugated, $f_w \to u f_w u^{-1}$. Since $\ell_R \to u \ell_R$, the leg variable of each output term goes to
\begin{align}
	f_w \ell_R \;\longrightarrow\; u f_w u^{-1}\, u \ell_R = u \bigl( f_w \ell_R \bigr) \,,
\end{align}
which is again exactly \eqref{eq:gaugev1} applied to the output state, while $\ell_{\overline{R}}$ and $k$ transform identically on both sides because $F_\gamma([t],\lambda)$ does not touch them. So $F_\gamma$ commutes with gauge transformations at $v$ as well, and therefore descends to an operator on the physical Hilbert space $\Ha_{\mathrm{phys}}(\Sigma_{0,2})$.

We can also show directly from \eqref{eq:ribbonweylsum} that the closed ribbon operators commute among themselves. Acting with $F_\gamma([t_2],\lambda_2)$ leaves $k$ untouched, so it leaves $x_0$ and $k_t$ untouched as well, and a second ribbon deposits its flux on $\ell_R$ by left multiplication using the same $x_0$. Both fluxes then lie in $C_G(k) = x_0 T_B x_0^{-1}$ by \eqref{eq:fluxcommutes}, which is abelian, and the characters contribute only scalars, so
\begin{align}
	F_\gamma([t_1],\lambda_1)\, F_\gamma([t_2],\lambda_2) = F_\gamma([t_2],\lambda_2)\, F_\gamma([t_1],\lambda_1) \,. \label{eq:ribbonsabelian}
\end{align}

Finally, one can show that \eqref{eq:ribbonweylsum} is equivalent to the more complicated expression\footnote{The group commutator is $[g,h] = ghg^{-1}h^{-1}$, and $\delta$ is the Dirac delta at the identity $e \in G$.}
\begin{align}
	F_\gamma([t],\lambda) \dket{\ell_R,\ell_{\overline{R}},k} = \int_{C_G(k) \backslash G / C_G(t)} \!\!\! d[x] \; \delta\bigl([x^{-1} k x,\,t]\bigr) \, \chi_\lambda(x^{-1} k x) \, \dket{\, k^{-1} x t x^{-1} k\, \ell_R,\; \ell_{\overline{R}},\; k} \,. \label{eq:ribbonop}
\end{align}
The delta function restricts $x$ to the locus where $x^{-1}kx \in C_G(t) = T_B$, which is exactly the set of solutions $x_0 \cdot N_G(T_B)$ of \eqref{eq:x0def}; the double coset then collapses this to $T_B \backslash N_G(T_B) / T_B = W(T_B)$, so the integral is the finite sum \eqref{eq:ribbonweylsum}. Integrating over the double coset rather than over $G$ is also what avoids the divergent factors $\mathrm{Vol}(C_G(t))$ and $\mathrm{Vol}(C_G(k))$, which is the  renormalized group averaging developed in \cite{Alonso-Monsalve:2025lvt}. Note also that on the support of the delta function, $x t x^{-1}$ commutes with $k$,\footnote{To see this, note that because $x^{-1} k x \in T_B$, it commutes with $t$. Therefore, conjugating both sides of that commutation relation by $x$, $x t x^{-1}$ commutes with $k$.} so the conjugation by $k$ in \eqref{eq:ribbonop} acts trivially and the two expressions deposit the same flux on the $\ell_R$ leg.

We record \eqref{eq:ribbonop} because it is what the systematic construction of ribbon operators produces \cite{Akers:2024wab,Balasubramanian:2025rcr}. For finite groups $G$, there is not a systematic classification of conjugacy classes in terms of Cartan subgroups and so forth, so the analog of \eqref{eq:ribbonop} is the best we can do. It is therefore somewhat surprising that for Lie groups the same expression collapses to something as simple as \eqref{eq:ribbonweylsum}. 

\subsection{What labels the ribbons?}

We conclude from the above construction that a closed ribbon in this model is labeled by a conjugacy class together with an irreducible representation of its centralizer, exactly as in the finite group double model. The extension to arbitrary transformable $G$ leads to the block decomposition by Cartan conjugacy class and the Weyl sum in \eqref{eq:ribbonweylsum}.\footnote{For compact groups $G$, there is a single Cartan subgroup.}

However, the two labels are not independent. Representing a regular conjugacy class $[t]$ by an element $t \in T_B$ makes $\lambda \in \widehat{T_B}$ a character, but $(t,\lambda)$ and $(wt,w\lambda)$ describe the same operator for every $w \in W(T_B)$, as \eqref{eq:ribbonweylsum} makes explicit.\footnote{The Weyl group acts on $T_B$ by conjugation, and hence on its characters by the pullback of this action, so that $\chi_\lambda(w^{-1} k_t w) = \chi_{w\lambda}(k_t)$.} The label space is therefore the \emph{diagonal} quotient
\begin{align}
	\bigl(T_B \times \widehat{T}_B\bigr) \big/ W(T_B) \,, \label{eq:diagonalquotient}
\end{align}
and not the product $\bigl(T_B/W\bigr) \times \bigl(\widehat{T}_B/W\bigr)$ of the separately reduced factors. Equation \eqref{eq:diagonalquotient} is therefore the correct label set for the regular anyons of the generalized double model when $G$ is a transformable Lie group.

Finally, we note that the boundary anchored ribbon operators are contained in this class of ribbon operators as a degenerate subset. If $[t] = [e]$ then $C_G(e) = G$, the label $\lambda$ is an irreducible representation $\pi$ of the whole group, and the pair is the pure charge $(e,\pi)$: precisely the label set of Sec.~\ref{sec:algebras}. The labels available to a closed cut therefore contain the ribbon operators we utilized in the main text as a special case. The construction above which used these labels assumed the flux to be regular, so strictly speaking it does not apply at $[t] = [e]$: the centralizer is no longer a Cartan subgroup, so no $T_B$, and hence no Weyl sum, is singled out by $t$. If one sets $t = e$ in \eqref{eq:ribbonweylsum} anyway, the flux \eqref{eq:fluxelement} becomes $f_w = e$, nothing is deposited on the $\ell_R$ leg, and the operator reduces to multiplication by $\chi_\pi(k_t)$, the character evaluated on the holonomy. That is a class function of the holonomy, which is precisely the form of the operators $\hat{C}$ that generate $\mathcal{Z}(\mathcal{A}_R)$ in Sec.~\ref{sec:algebras}, so taking the limit $[t] \to [e]$ reproduces these kinds of operators as well.

What does break down is the evaluation of that character at non-regular holonomy $k$: $\chi_\pi$ is a locally integrable function, but it diverges on the non-regular elements of $G$. For example, as $k_t \to e$, the character formally limits to the dimension $d_\pi$, which diverges for non-compact $G$. So the definition of the ribbon operators on states with non-regular holonomy, where the character blows up, has to be handled separately. However, because non-regular elements are measure zero within $G$, this may not be a problem, as this kind of divergence would not affect the matrix element of any physical operator. We leave a systematic treatment of states with non-regular holonomy to future work.

We expect \eqref{eq:diagonalquotient} to be the data carried by the edge modes of a factorization map along a closed cut, in the same way that the representations $\pi \in \widehat{G}$ label the edge modes introduced by the boundary anchored factorization maps of the main text. Constructing that map is the natural next step. To construct this map, one would need the analogs of the nested algebras $\mathcal{B}_R \supset \mathcal{A}_R \supset \mathcal{Z}(\mathcal{A}_R)$ for a closed cut, a measure on the label space \eqref{eq:diagonalquotient} defining a trace on each of them, and the operator valued weights relating those traces, following Sec.~\ref{sec:algebras}. As in Sec.~\ref{sec:algebras}, the mismatch in the measures for the analogs of $\mathcal{A}_R$ and $\mathcal{Z}(\mathcal{A}_R)$, generalized to include the full label set \eqref{eq:diagonalquotient}, is what would then fix the eigenvalues of the area operator along a closed curve. When $G$ is a finite group, these eigenvalues are known to be the logarithm of the ``quantum dimension'' of the anyons represented by the ribbon operators $F_\gamma([t],\lambda)$ \cite{Kitaev:2005dm,Levin_2005,Bonderson:2017osr}, so in principle there is no obstruction to determining this measure for the topological model of this paper. We leave this explicit determination for future work.

\newpage
\bibliographystyle{JHEP}
\bibliography{biblio}

\end{document}